\documentclass{elsarticle}

\usepackage{graphicx}

\usepackage{amssymb}
\usepackage{amsmath}
\usepackage{multirow}

\usepackage{lineno}

\begin{document}

\begin{frontmatter} 


\title{Construction and Performance\\ of the Barrel Electromagnetic Calorimeter\\ for the GlueX Experiment}

\author[uofr]{T.D.~Beattie}
\author[uofr]{A.M.~Foda}
\author[uofr]{C.L.~Henschel}
\author[uofr]{S.~Katsaganis}
\author[uofr]{S.T.~Krueger}
\author[uofr]{G.J.~Lolos}
\author[uofr]{Z.~Papandreou\corref{cor}}
\ead{zisis@uregina.ca}
\cortext[cor]{Corresponding authors: Tel.: +1 306 585 5379; +1 757 269 7625.}
\author[uofr]{E.L.~Plummer}
\author[uofr]{I.A.~Semenova}
\author[uofr]{A.Yu.~Semenov}
\author[jlab]{F.~Barbosa}
\author[jlab]{E.~Chudakov}
\author[jlab]{M.M.~Dalton}
\author[jlab]{D.~Lawrence}
\author[jlab]{Y.~Qiang\fnref{f1}}
\fntext[f1]{Current address: Toshiba Medical Research Institute USA, Inc., 706 N Deerpath Dr, Vernon Hills, IL 60061.}
\author[jlab]{N.~Sandoval}
\author[jlab]{E.S.~Smith\corref{cor}}
\ead{elton@jlab.org}
\author[jlab]{C.~Stanislav}
\author[jlab]{J.R.~Stevens\fnref{f2}}
\fntext[f2]{Current address: Department of Physics, College of William \& Mary, Williamsburg, VA 23187.}
\author[jlab]{S.~Taylor}
\author[jlab]{T.~Whitlatch}
\author[jlab]{B.~Zihlmann}
\author[cmu]{W.~Levine}
\author[cmu]{W.~McGinley}
\author[cmu]{C.A.~Meyer}
\author[cmu]{M.J.~Staib}
\author[uofa]{E.G.~Anassontzis}
\author[uofa]{C.~Kourkoumelis}
\author[uofa]{G.~Vasileiadis}
\author[uofa]{G.~Voulgaris}
\author[usm]{W.K.~Brooks}
\author[usm]{H.~Hakobyan}
\author[usm]{S.~Kuleshov}
\author[usm]{R.~Rojas}
\author[usm]{C.~Romero}
\author[usm]{O.~Soto}
\author[usm]{A.~Toro}
\author[usm]{I.~Vega}
\author[iu]{M.R.~Shepherd}
\address[uofr]{Department of Physics, University of Regina, Regina, Saskatchewan, Canada S4S 0A2}
\address[jlab]{Jefferson Laboratory, Newport News, Virginia 23606, USA}
\address[cmu]{Carnegie Mellon University, Pittsburgh, Pennsylvania 15213, USA}
\address[uofa]{National and Kapodistrian University of Athens, 15771 Athens, Greece}
\address[usm]{Universidad T\'ecnica Federico Santa Mar\'ia, Casilla 110-V  Valpara\'iso, Chile}
\address[iu]{Indiana University, Bloomington, Indiana 47405, USA}

\begin{abstract}
The barrel calorimeter is part of the new spectrometer installed in Hall D at Jefferson Lab for the GlueX experiment. 
The calorimeter was installed in 2013, commissioned in 2014 and has been operating routinely since early 2015.
The detector configuration, associated Monte Carlo simulations, calibration and operational performance are described herein.     
The calorimeter records the time and energy deposited by charged and neutral particles created by a multi-GeV photon beam. 
It is constructed as a lead and  scintillating-fiber calorimeter 
and read out with 3840 large-area silicon photomultiplier arrays.
Particles impinge on the detector over a wide range of angles, from normal incidence at 90 degrees down to 11.5 degrees, which defines a geometry that is fairly unique among calorimeters.  
The response of the calorimeter has been measured during a running experiment and performs as expected for electromagnetic showers below 2.5 GeV.
We characterize the performance of the BCAL using the energy resolution integrated over typical angular distributions for $\pi^0$ and $\eta$ production of
$\sigma_E/E$=5.2\%/$\sqrt{E(\rm{GeV})} \oplus$ 3.6\% and a timing resolution of  $\sigma$\,=\,150\,ps at 1\,GeV.

\end{abstract}   


\end{frontmatter}



\tableofcontents

\section{GlueX Detector  \label{sec:projectbackground} }

\begin{figure}[http]\centering
\includegraphics[width=0.7\textwidth]{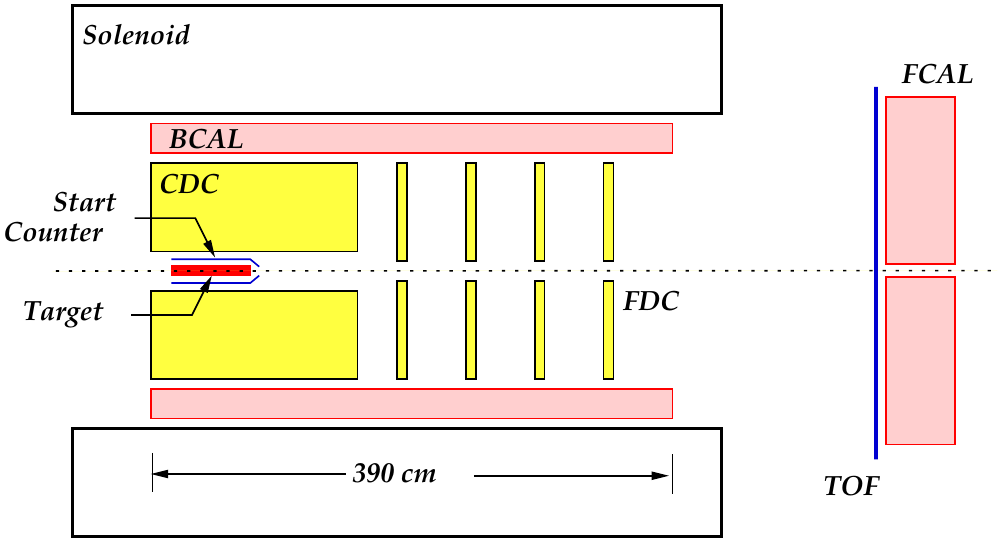}  
\caption{\label{fig:gluexsketch}          
  Sketch of GlueX detector.  The main systems of the detector are the Start Counter \cite{hdnote3064}, the Central Drift Chamber (CDC) \cite{VanHaarlem:2010yq} the Forward Drift Chamber (FDC) \cite{Pentchev2017281}, a scintillator-based Time of Flight (TOF) wall and a lead-glass Forward Calorimeter (FCAL) \cite{MORIYA201360}. The Barrel Calorimeter (BCAL) is sandwiched between the drift chambers and the inner radius of the solenoid.  (Color online)
  }   
\end{figure}

The primary motivation of the GlueX experiment is to search for and, ultimately, study the pattern of gluonic excitations in the meson spectra produced in $\gamma p$ collisions at 9~GeV~\cite{gluex-ref}.
Specifically, GlueX aims to study the properties of hybrid mesons --- particles where the gluonic field contributes directly to the $J^{PC}$ quantum numbers of the mesons~\cite{meyer:2015eta}. 
The design of the GlueX detector \cite{Ghoul:2015ifw} is based on a solenoidal magnet that surrounds all detectors in the central region, providing a magnetic field of about $2$~T along the direction of the photon beam, which impinges on a 
$30$~cm-long liquid hydrogen target.  A schematic of the detector including its major sub-detectors is given in Fig.\,\ref{fig:gluexsketch}.
The goal of GlueX calorimetry is to detect and to measure photons from the decays of $\pi^{0}$'s and $\eta$'s and other radiative decays of secondary hadrons.  The detector measures the energies and positions of the 
showers made by photons, as well as the timing of the hits. It also provides the timing of the hits caused by charged hadrons, allowing for time-of-flight particle identification.



\section{BCAL: Overview and Design \label{sec:overview}}

\begin{figure}[http]\centering
\includegraphics[scale=0.4]{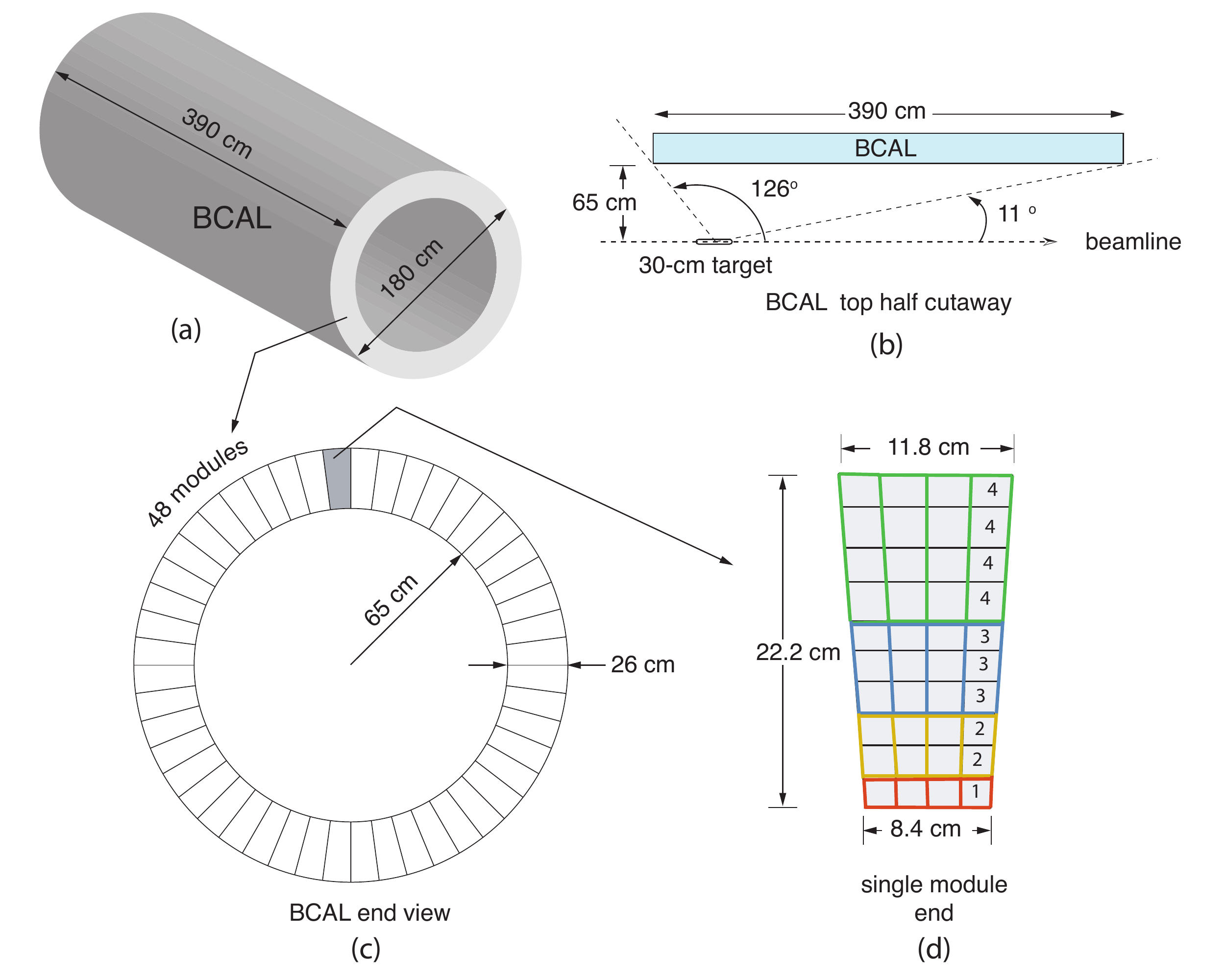}
\caption{\label{fig:bcalsketch}
  Sketch of the Barrel Calorimeter and readout. (a)~ A three-dimensional rendering of the BCAL; (b)~top-half cutaway (partial side view) of a~BCAL module showing its polar angle coverage and location with respect to the GlueX LH$_2$ target;  (c)~end view of the BCAL depicting all 48 azimuthal modules and (d)~wedge-shaped end view of a single module showing the location of light guides and sensors as well as the 1:2:3:4 readout summing scheme, described in Section~\ref{sec:electronics}. More details can be found in the text.  (Color online)
  }
\end{figure}

A practical solution to the requirements and constraints imposed by the experiment is a calorimeter based on lead-scintillating fiber sandwich technology.  The BCAL is modeled closely after the electromagnetic calorimeter built for the KLOE experiment at DA$\Phi$NE \cite{Adinolfi:2001sk,Adinolfi:2002zx,Adinolfi:2002jk}. 
The BCAL detects photon showers with energies between 0.05 GeV and several GeV, $11^{\circ}$--$126^{\circ}$ in polar angle, and $0^{\circ}$--$360^{\circ}$ in azimuthal angle. The containment of showers depends on the angle of particle incidence, with a thickness of $15.3$ radiation lengths for particles entering normal to the calorimeter face and reaching up to 67 radiation lengths at $14^{\circ}$. Geometrically, the BCAL consists of 48 optically isolated modules each with a trapezoidal cross section, forming a  390~cm-long cylindrical shell having inner and outer radii of 65~cm and 90~cm, respectively. The fibers run parallel to the cylindrical axis of the detector.  Schematics showing the geometry of the BCAL and readout segmentation are included in Fig.\,\ref{fig:bcalsketch}.

The light is collected via small light guides at each end of the module that is delivered to silicon photomultiplier (SiPM) light detectors, which were chosen due to their insensitivity to magnetic fields. Preamplifiers and summing circuits are situated near the light sensors and generate signals that are delivered to VXS electronics racks, conveniently located on the floor of the experimental hall.

The performance of the calorimeter is summarized by its ability to measure the energy, position and timing of electromagnetic showers. In general, the energy resolution of an electromagnetic calorimeter is expressed in the form:
\begin{equation}
 \frac{\sigma_E}{E} = \frac{a}{\sqrt{E\mbox{(GeV)}}} \oplus b \oplus 
\frac{c'}{E\mbox{(GeV)}},
\label{eq:Eresolution}
\end{equation} 
where the symbol $\oplus$ means that the quantities are added in quadrature. The $a/\sqrt{E}$ term contains the combined effect of sampling fluctuations and photoelectron statistics, with the former dominating the resolution.  This is commonly referred to as the stochastic term. The constant term, $b$ in Eq.~\ref{eq:Eresolution}, originates from sources with uncertainties that scale with energy. These sources can be  mechanical imperfections, material defects, segment-to-segment calibration variations, 
instability with time and shower leakage. The term $c'/E$ results from noise and pileup in high-rate environments. Measurements of an early full-length prototype\footnote{The matrix material of the prototype was very similar to the final detector. 
However, the readout used standard photomultiplier tubes and a uniform segmentation, which could result in a different energy dependence for the resolution.}
using a photon beam \cite{Leverington2008327} provided expectations for the various contributions to the resolution: $a\approx 5.4\%\sqrt{\rm{GeV}}$, $b \approx 2.3\%$ and $c'$ negligible.

The azimuthal angular resolution is dominated by the readout granularity in azimuth, which is about 2 cm. The polar angular resolution depends on the position resolution along the length of the barrel ($z$), which is determined by measuring the time difference between signal arrivals at the upstream and downstream faces of the barrel. The resolution in $z$ depends on the resolution of half the time difference, which may be parameterized as
\begin{equation}
\sigma_t = \frac{c}{\sqrt{E\mbox{(GeV)}}} + d,   
\label{eq:Tresolution}
\end{equation} 
where $c$ and $d$ are empirical constants.
At 1~GeV, the prototypes obtained a timing resolution of $\sigma_t \approx 200$~ps, which leads to a resolution in $z$ of $\sigma_z \approx 3$~cm.  The timing resolution is also important for measuring flight times from the target, which are used to help with charged particle identification.

In the following sections we provide details of the individual components that were used in the construction of the BCAL and then describe the performance of the calorimeter during the experiment.


\section{Major Components}
\label{sec:majorcomponents}


\subsection{Fiber Selection}
\label{sec:fiberselection}

Kuraray SCSF-78MJ double-clad, blue-green fibers\footnote{Kuraray Plastic Scintillating Fibers (\url{http://www.kuraray.com/products/plastic/psf.html}).} \cite{bcal-2} were selected for their high light output and long attenuation length ($\sim$ 4 m).
The light yield affects the energy and timing resolutions as well as the detection threshold.  
Over three quarters of a million fibers were used in the construction of the BCAL.  
Fiber testing and the fabrication of the BCAL modules were carried out at the University of Regina.
Acceptance testing of the fibers was based on measuring the response of about 0.5\% of the fibers, which were selected uniformly from each of the
manufactured batches. These fibers were evaluated by employing a 373-nm UV LED to stimulate the fibers along their length and reading out the light using a spectrophotometer and a photodiode in order to extract the spectral response and the attenuation length, respectively~\cite{Baulin201348}.  Single exponential fits to the spectral response at 100-280~cm distance from the light source yielded an average bulk attenuation length and standard deviation of $(387\pm26)$~cm.  Double-exponential fits to the spectral response over the entire 4-m length also allowed the extraction of long and short attenuation-length components at $(486\pm54)$~cm and $(75\pm22)$~cm, respectively.  The relative strength of the short to long 
attenuation lengths was determined to be 0.3$\pm$0.08. 

The quality of these fibers was evaluated further by exciting the fibers at their mid point using a $^{90}$Sr source in order to determine the light yield using a calibrated photomultiplier (PMT)\footnote{Hamamatsu R329-02 with a standard progressive voltage divider.} attached to one end of the tested fiber.  
The average number of photoelectrons from this sample was $9.2\pm 0.6$ at a source distance of 200~cm from the PMT~\cite{Beattie2014245, BeattieIEEE}.  


\subsection{Matrix Construction}
\label{sec:matrixConstruction}

\begin{figure}[tph]\centering
\includegraphics[width=0.25\linewidth]{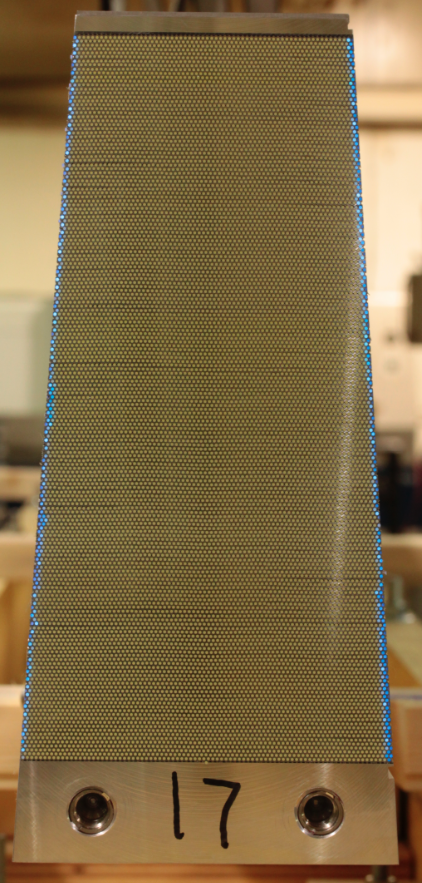}
\includegraphics[width=0.25\linewidth]{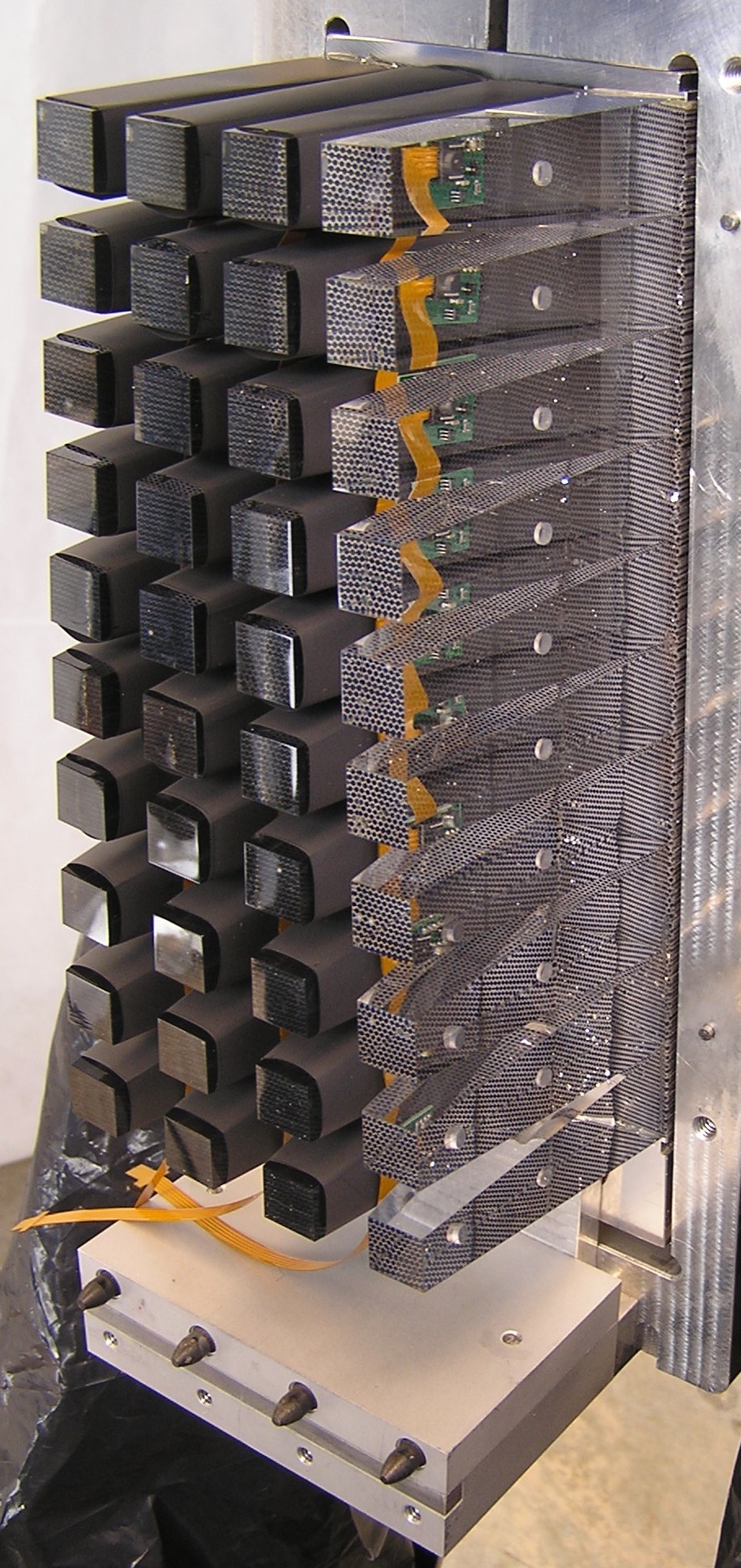}
\caption{\label{fig:ModPhotos}
  Left) Photo of the machined end of the lead-scintillator matrix. Individual fibers can be seen between the lead corrugated sheets. The matrix lies between the the thin Al plate on top and the thick Al base plate at the bottom.  
  Right) Arrangement of light guides glued to the face of the matrix. Three columns are already covered with black covers. Visible are LED boards showing through the last column as well as the location of the side pockets for
  the LEDs. (Color online)
  }
\end{figure}

Each BCAL module consists of 185 layers of corrugated lead sheets that are 0.5 mm thick. They are interleaved with 184 layers of 1-mm  fibers bonded to the 0.5-mm grooves in the lead sheets using BC-600 optical cement.\footnote{St. Gobain Crystals \& Detectors, Hiram, OH 44234, USA (\url{http://www.bicron.com}).} 
During construction, the overhead fluorescent lights were covered with yellow UV-absorbing film (TA-81-XSR\footnote{Window Film Systems, London, ON, Canada (\url{https://www.windowfilmsystems.com}).}) to protect the fibers from UV exposure~\cite{Baulin201348}. Custom devices were built for the construction: a plastic deformation device ({\em swager}) and  two electro-pneumatic presses. The detailed construction procedure is described elsewhere~\cite{hdnote3164} and is only recounted briefly herein. Photos of one completed module are shown in Fig.\,\ref{fig:ModPhotos} and the main properties of the calorimeter are listed in Table\,\ref{tab:bcalproperties}  \cite{hdnote840}.

\begin{table}[ht!]\centering
\caption{Summary of BCAL properties. 
\label{tab:bcalproperties}}
\begin{tabular}{|l|c|}
\hline
\hline
 Property                         &   Value \\ \hline 
Number of modules                 &  48  \\  
Module length                        &  390~cm  \\  
Module inner/outer widths          &  84.0~mm/118.3~mm  \\ 
Lead-scintillator matrix thickness   &  221.9~mm  \\ 
Inner/outer Al plates thickness      &  8~mm/31.75~mm          \\ 
Module azimuthal bite             & $7.5^\circ$  \\
Total number of fibers                &  685000  \\ 
Lead sheet thickness              & 0.5~mm  \\ 
Kuraray SCSF-78MJ multi-clad fiber                  & 1.0~mm   \\ 
Fiber pitch radial/lateral                 & 1.22~mm/1.35~mm   \\ 
Weight fractions (\% Pb:SF:Glue)      & 86.1: 10.5: 3.4  \\ 
Effective density               &  4.88 g/cm$^3$  \\ 
Effective Radiation Length     & 1.45 cm \\
Effective Moli\`{e}re radius  & 3.63 cm \\
Effective Atomic Weight & 71.4 \\
Effective Atomic Number & 179.9 \\ 
Sampling fraction               & 0.095 \\
Total weight                            &  28~t   \\   \hline
\hline
\end{tabular}
\end{table}

%
%
The swager was constructed to produce lead sheets, which were 4 m long, with 0.5-mm groves along their length, spaced 1~mm apart.  
It consisted of two motor-driven steel drums with adjustable speed settings that would rotate in a manner to draw in a lead sheet and extrude it from the opposite side with corrugated top and bottom surfaces. 

The presses were used to compress the matrix and squeeze out excess glue.
 Each press was based on steel tubing, 5 m in length, with an aluminum plate on top, accurately machined to flatness.  These were welded to a steel table frame which was bolted to the lab floor and also leveled. The upper section of each press was comprised of aluminum frames through which 20 pistons with rams were fitted. That section was reclined and the pistons retracted when the matrix layers were being built.  


An aluminum plate formed the base of each module.  These plates were machined to have a length of 4.2~m and a thickness of $1 \frac{3}{4}$".  To ensure that the base plates remained flat during matrix construction, two 4''-wide, 1''-thick steel rails were bolted to the underside of each base plate along its length and outer edges; the rails were removed before each module was installed in its final, barrel configuration.  All necessary features on the base plates (a center-line groove to place the first guide fiber, bolt holes, slots, etc.)\ were machined prior to the matrix build.  
Sixteen, equally-spaced points were marked on each of the long edges of the base plate to be used as reference points in checking the accumulating matrix build height after each press sequence was completed. The surface of the base plate was abraded using 80~grit sand paper to improve  bonding characteristics with the first lead sheet, which was attached to the base plate using industrial epoxy\footnote{Araldite 2011 epoxy in a ratio of 5:4 (resin:hardner).}.  
Optical cement BC-600\footnote{St. Gobain Crystals \& Detectors, Hiram, OH 44234, USA (\url{https://www.bicron.com}) in a ratio of 100:28 (resin:hardner).} was applied to the top surface of each lead sheet using paint brushes.
A presorted fiber bundle --- prepared on a table covered with a copper sheet so as to ground for static electricity --- was laid on the epoxy starting at the middle of the module and gradually working towards the ends, ensuring that each groove contained a fiber.  
A lead sheet was then rolled onto the fibers and checked for alignment. The working time for the epoxy was reached after about 10 layers at which time the activity was suspended for the day\footnote{The viscosity of the epoxy was 
monitored by a sample placed right after its mixing into a clear plastic cup. }.
Subsequently, a teflon sheet and  a 1'' aluminum plate would be placed on top on the lead sheet.  Then, the upper section of the press was brought to its vertical position and the rams deployed gradually in pairs from 
the middle out, in order to press and squeeze out excess epoxy from the matrix ends.  
The pressure was applied continuously for approximately 20~hours before the arms were loosened and the upper section and rams retracted to the reclined position to continue layer stacking the next day.  The height of the matrix was checked (using a digital depth gauge) and recorded along the $2\times 16$ reference points on the base plate to ensure an even build in terms of height off the table, i.e. the thickness of the matrix.  
After 185 layers were completed, a machined 8~mm thick aluminum plate was bonded to the top surface of each module using industrial epoxy, ensuring a flat surface within 1~mm along the entire top surface spanning $10 \times 400$~cm$^2$.

By constructing two modules in parallel on the two presses the construction of 48 production modules and one spare were completed within 2.5~years.  
Completed modules were shipped to an industrial firm\footnote{Ross Machine Shop, 40 Kress St., Regina, SK, Canada.} where they were machined to their final dimensions subject to tolerance specifications.  The modules' end surfaces were polished using a tool with sets of 4 cutter blades and a fifth ``wiper'' blade.
The modules were shipped in sets of four to Jefferson Lab using refrigerator trucks with their interior trailer temperature set at $15^{\circ}$\,C for their 3,300~km journey, in order to minimize possible thermal stresses on the matrix particularly when shipped from Regina during winter.

%


\subsection{Light Collection \label{sec:lightcollection}}
The ends of the calorimeter wedge-shaped modules are polished faces of the lead and scintillating fiber matrix. Due to the cylindrical geometry of the BCAL, the edges of each module are tapered at an angle of 3.75$^\circ$. Each wedge is divided into four equal azimuthal slices, or columns, and each column is covered by 10 different light guide shapes (see Fig.\,\ref{fig:ModPhotos}). 

The geometry of the readout is determined by the 40 light guides that are used
to collect the light from the fibers and guide it down to the light sensor. 
The input cross sections of the 
light guides (against the module) increase radially outward, with the innermost light guide having approximate dimensions of 21$\times$21 mm$^2$ and the outermost guide approximately
27$\times$25 mm$^2$. The output faces of all light guides have the same area of 13$\times$13 mm$^2$ that matches the area of the SiPM light sensor. As expected, the Monte Carlo-calculated light collection varies with radial
depth of the calorimeter, starting at 75\% for the innermost layer and decreasing to about 48\% for the outermost layer. We note that the collection efficiency is relatively high because the 
trapping angle inside the fibers is 26.7$^\circ$, which leads to a narrow input angular range into the light guide. 
Details of the geometry and calculations of acceptance can be found in 
Ref.\,\cite{hdnote1784}. The length of the light guides is 8 cm, which was chosen to ensure at least one reflection of light rays to randomize the position of light at the output relative to the input surface. 

A total of 4000 light guides were fabricated by the Universidad T\'ecnica Federico Santa Mar\'ia to exacting dimensions to match the end faces of the calorimeter wedges.
A cylindrical pocket (4-mm diameter, 2-mm deep) was drilled into one side of the light guide, approximately half way down the length to accommodate a light emitting diode (LED) to inject light to monitor the response of
individual sensors. The LED was directed toward the opposing (far) side of the BCAL module producing signals on both sides of the module whenever it fired. The light sensor at the far end of the module received
about twice as much light as the near sensor but with large variations between channels. 
Each light guide was covered with black Tedlar wrapping\footnote{Tests were made with Tedlar, acid-free paper, aluminized mylar and no wrapping. They all had similar light collection, which is expected if the collection is due to total internal reflection.} to prevent light from scattering into adjacent light guides. Each light guide was glued onto the face of the BCAL wedge using NOA 87
UV-curing glue.\footnote{Norland Products, \url{https://www.norlandprod.com/adhesives/noa87.html}.}
The SiPM sensors are placed at the output face of the light guide across a 0.5 mm air gap. About 30\% of the light is lost through the gap, but  practical considerations led us to install the photosensors 
on the electronic package that could easily be detached from the BCAL modules.


\subsection{The Photo Sensors  \label{sec:photosensors}  }

The selection of the optimum light sensor for a given detector depends on many factors including the properties of the sensors, geometrical constraints, cost, safety, and ease of use.  Our selection of SiPMs as the light sensors for the BCAL was driven largely by their insensitivity to magnetic fields, which is about 1~T at the sensor location. 
We selected the Hamamatsu S12045(X) Multi-Pixel Photon Counter (MPPC) array \footnote{Hamamatsu Corporation, Bridgewater, NJ 08807, USA \\ (\url{http://sales.hamamatsu.com/en/home.php)}.}, 
which is a $4\times4$ array of $3\times3$ mm$^2$ tiles \cite{hdnote2913}.
Each tile is composed of 3600 50$\times$50$\mu$m$^2$ pixels. 
The array has 20 pins: four bias pins to power individual rows
and 16 outputs for the signals of each of the tiles. The individual outputs were used during acceptance testing to check the signals from individual tiles. However, for installation in the detector all four bias pins were connected to a single input and the 16 output signals were joined to generate a single summed output for the array. 
 
\begin{figure}[ph]
\includegraphics[height=12cm,clip=true]{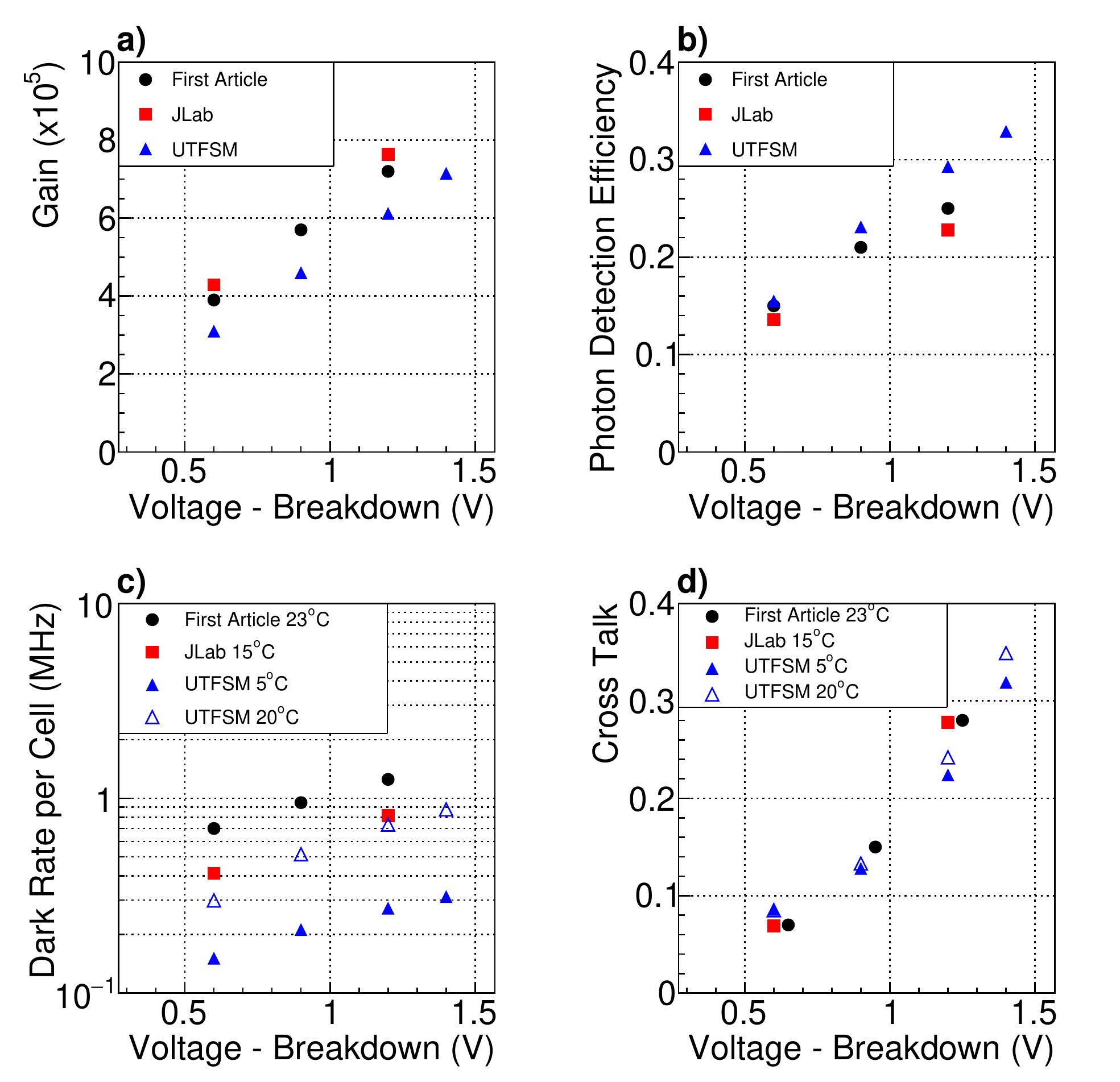}
\caption{Measurements of the first-article samples (black circles) \cite{Barbosa2012100,hdnote1777}, production samples at JLab (red squares) and production 
samples at UTFSM (triangles) \cite{soto,Soto201489} 
of four basic SiPM parameters as a function of the voltage over breakdown. 
a) gain, b) photon detection efficiency, c) dark rate per tile (the dark rate for the array is 16 times higher) and d)  cross talk determined from deviations of the single-pixel distributions from a pure
Poisson function. As long as the voltage over breakdown is kept constant,
the dark rate is the only parameter that has a significant temperature dependence. The nominal operating voltage for the GlueX experiment is 1.4 V above breakdown. (Color online)
\label{fig:plot_sipm_parms}}
\end{figure}

The SiPM arrays, here referred to as SiPMs, were tested extensively before acceptance. Four thousand units were purchased and 3840 are installed in the detector. The testing of the individual outputs, where individual single-pixel peaks were discernible, was divided
between two collaborating institutions: 30\% at JLab \cite{Barbosa2012100} and 70\% at the Universidad T\'ecnica Federico Santa Mar\'ia (UTFSM) \cite{soto,Soto201489}. The main operational  parameters are graphed in Fig.\,\ref{fig:plot_sipm_parms}.\footnote{
For details of the determination of each of the parameters shown, please refer to the cited literature, especially references \cite{Soto201489} and \cite{Qiang2013234}.}
As shown in Fig.\,\ref{fig:plot_sipm_parms}a, there is an absolute gain difference in the normalization of the measurements between JLab and UTFSM, but they show consistent behavior as a function of voltage above breakdown (overbias) \cite{hdnote2910}. 
A small number of arrays were tested at the University of Regina \cite{BeattieIEEE} but using a summed-output mode, where the
single-pixel peaks cannot be distinguished. The average photon detection efficiency (PDE) for those 11 SiPM units was measured to be (28$\pm$2(stat)$\pm$2(syst))\% at an overbias of  0.9~V and at 460~nm wavelength. The average PDE of $(24\pm2)$\%  for 3000 units measured by UTFSM at the same overbias~\cite{soto,Soto201489} was extracted at a wavelength of 518~nm.  The PDE varies as a function of wavelength, with it being larger by a factor of 1.11 in going from 518~nm to 460~nm according to the manufacturer.  This correction increases the UTFSM number to $\sim$27\% at 460~nm, consistent with the Regina results.

Our nominal operating point is 1.4 V over the breakdown voltage, which was selected to reduce the effect of readout thresholds. Even at this relatively high overbias, the noise level is dominated by fluctuations in the electronics baseline and not by single
pixel noise. The readout threshold should be set as low as possible and depends both on the width of the pedestal as well as baseline drifts, which have a small temperature dependence. We note that the optimal overbias considering properties of the SiPM alone is between 0.9 and 1.2 V (see for example Section 6 of Ref.\,\cite{Soto201489}). 


The gain of the SiPM depends on the voltage above the breakdown voltage, V$_{br}$, which is about 70~V for these sensors. The breakdown voltage has been determined to be a linear function of temperature over a broad temperature range \cite{Lightfoot:2008im} with a slope between 50 and 60~mV/$^\circ$C.  This imposes practical constraints on the operation of SiPMs if the gain is to be kept constant during operation. Our strategy is two-fold. First, we stabilize the temperature within practical limits ($\pm$ 2$^\circ$C) 
using a chilled water system. Two chillers are installed, one to cool the upstream and the other the downstream end of the BCAL. The water coolant of the chillers is circulated via manifolds through copper pipes that are in thermal contact with an aluminum plate that cools the back
side of the forty SiPMs on each module face using silicon pads. Second, we stabilize the gain 
using a custom circuit by adjusting the bias voltage using a negative coefficient thermistor, which 
is in contact with the cooling plate.  The temperature on each cooling plate is measured and recorded. The chillers are set to 5$^\circ$C to reduce the single-pixel noise. This maintains the cooling plate temperature at about 7.5$^\circ$C. 
The container housing for the SiPMs and
preamplifier circuits is flushed with dry nitrogen gas to prevent condensation. The temperature of the nitrogen inside the container is 
about 21$^\circ$C, 1--5\% relative humidity, and dew points of less than --20$^\circ$C. 

Silicon detectors generally deteriorate after exposure to radiation, especially neutrons, and SiPMs are no exception \cite{Qiang2013234}.
We estimate that the dark rate per array will increase from about 16 MHz to 100 MHz, which is our specified maximum rate, 
after approximately 7 years of operation at high intensity 
production physics running in Hall D and operating at a temperature of 5$^\circ$C with periodic annealing periods at 40$^\circ$C. To date, we have accumulated a couple of months of running
at low intensity, so we do not yet expect to observe any effects due to radiation.


\section{Electronics}
\label{sec:electronics}

Detailed Monte Carlo simulations were performed to investigate the optimal readout segmentation for electromagnetic shower reconstruction, with the aim of balancing performance against the costs of electronics channels.  
It was found that the fine granularity provided by the readout of all  SiPM light sensors individually does not provide significant advantages in determining either the energy or position of showers. 
In particular, the detailed measurement of the longitudinal profile does not help determine
the shower parameters and in fact adding thresholds for each readout can lead to missing small energy depositions. 
The simulations demonstrated that a summing scheme along the radial direction is a good match to the shower development and retains  the essential information regarding shower parameters.
The innermost layer is read out individually as the first conversions are important for reconstructing the position of impact and could help to discriminate between photon and neutron initiated showers. 
Increasing the segmentation in the outer layers of the BCAL does not yield any considerable advantage in performance. 
A summary of the number of instrumented channels is given in Table\,\ref{tab:bcalcounts}. 

Following the preamp stage, the outputs of each array are summed by columns in a 1:2:3:4 scheme as indicated in  Fig.\,\ref{fig:bcalsketch}d (see Ref.\,\cite{hdnote1795} for details of the summing circuit). This scheme consists in
no summing for the beginning of the shower in layer~1, adding two sensors for layer~2, three sensors for layer~3 and four sensors to capture the tail of showers in layer~4.
We refer to these summed outputs as readout cells.
This reduced the number of channels from a single module side with 40 SiPMs down to 16 signals delivered to flash Analog-To-Digital Converters (FADCs). 
In total, 96 electronics packages were assembled each containing 
40 SiPMs, their preamplifiers and summing circuits, the temperature compensation system and cooling plate described previously, and all associated connectors;  each package bolts onto the aluminum base plates at the end of a module. 

\begin{figure}[tp]\centering
\includegraphics[width=0.5\linewidth,clip=true,bb=60 550  325 750]{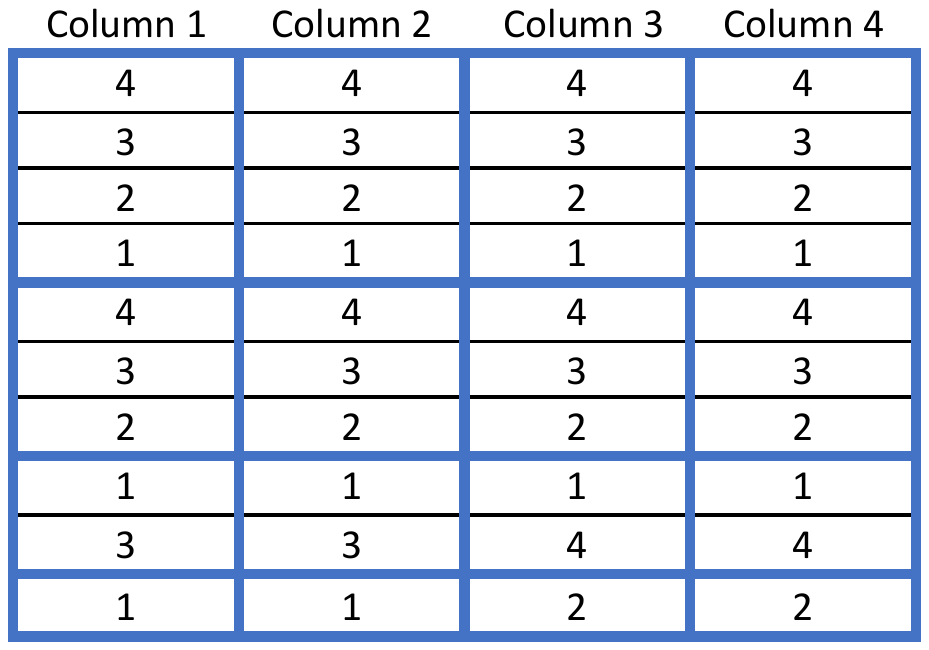}
\caption{\label{fig:biasdistribution}
  Bias distribution shown for the SiPMs in the 1:2:3:4 summing readout configuration for one end of one module. The numbers indicate the bias line feeding each SiPM and the summing groups are indicated by the thick lines. By powering a single bias line and pulsing
  one column, it is possible to select a specific SiPM within a summing group.
  }
\end{figure}

Voltage levels for the package are provided using commercial Wiener MPOD crates.\footnote{WIENER, Power Electronics GmbH, Linde 18, D - 51399 Burscheid Germany.
The low voltage for the circuitry is generated by MPOD MPV8008 modules and the bias voltages for the SiPMs are generated using ISEG EHS 201P-F-K modules, Bautzner Landstr. 23,  01454 Radeberg, Germany.}
Power is distributed to ten SiPMs at a time along rows and different summing groups (see Fig.\,\ref{fig:biasdistribution}), 
which allows selected groups of SiPMs to be powered and readout during special runs, such as when using the LED pulser system, which has four `strings' of LEDs that run along the columns orthogonal to the bias distribution lines. 

SiPMs were sorted according to their breakdown voltages first and then 
their bias voltage was further individualized by setting trim resistors on each SiPM assembly. 
The supply voltage to the ten SiPMs was set to the average required for the ten sensors after adjusting for the trim 
resistors. This delivers an operational voltage to within 10~mV of the calculated value, which corresponds to less than a percent change in gain for that channel. 

\begin{figure}[tp]\centering
\includegraphics[height=5cm,clip=true]{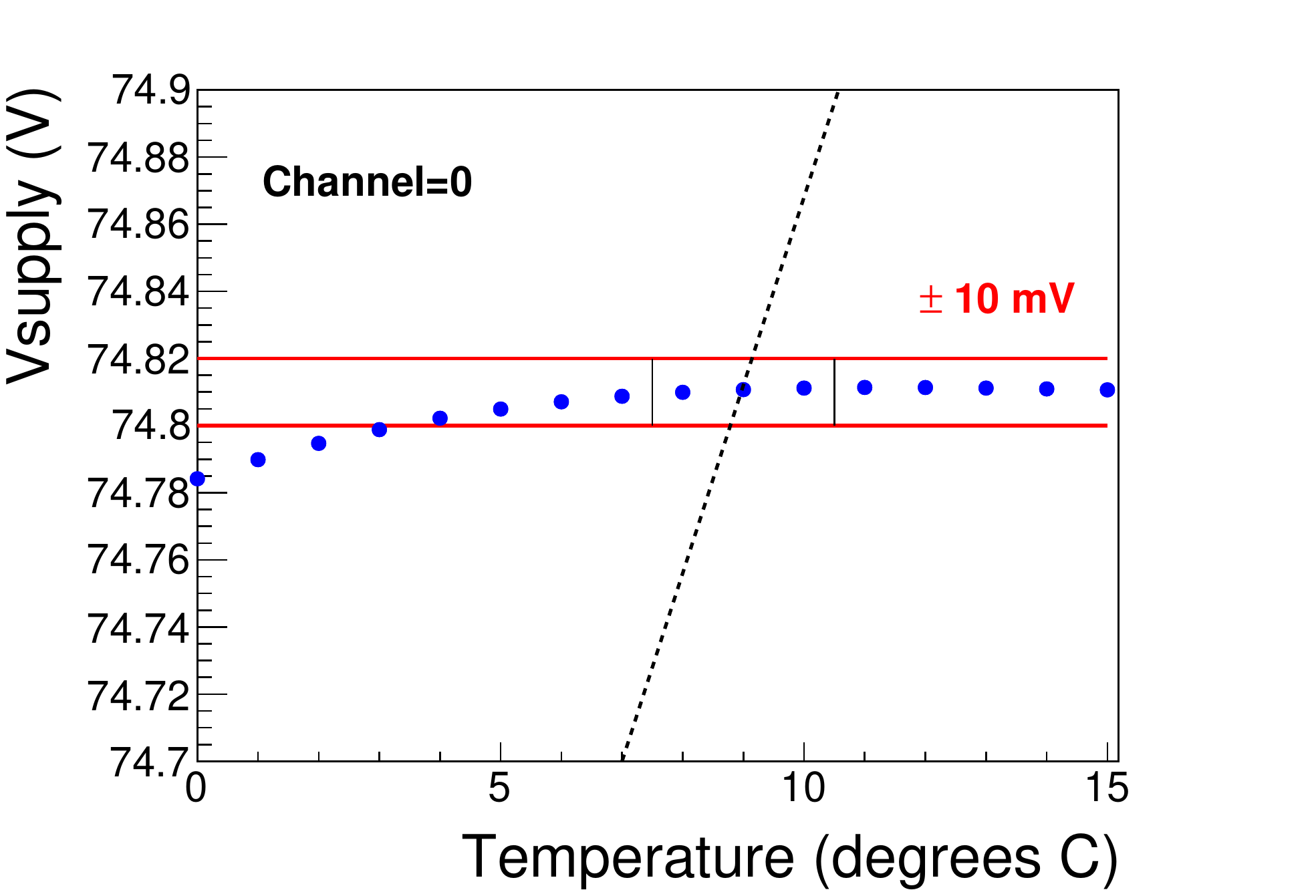}
\caption{\label{fig:plot_Tdependence1}
  Voltage to set a particular SiPM to the desired gain as a function of temperature.  With the chiller temperature set at 5$^\circ$C, the temperatures at the cooling plates ranged from  7.5$^\circ$C to 10.5$^\circ$C. The average temperature was 7.6$^\circ$C. The solid lines
  indicate the range of voltages that correspond to gain changes of about 1\%. The dotted line indicates the temperature dependence without the compensating circuit. (Color online)
  }  
\end{figure}

The sorting and operational bias voltage used for each sensor was determined from the operational voltages on the Hamamatsu factory data sheets, specified at 25$^\circ$C.\footnote{The Hamamatsu specification sheets 
provide the recommended operating voltage for a nominal gain of $7.5\times 10^5$, although our measurements indicate lower gains (Fig. 4a). We determined that this operational voltage on average corresponds to 0.9 V above breakdown; 
to obtain our setting at an overvoltage of 1.4 V, we added 0.5 V and then adjusted for temperature.}
  These values were checked with our own measurements and found to be  reliable and consistent. The voltage settings at other temperatures were adjusted assuming that the breakdown voltage changes by 56~mV/$^\circ$C.\footnote{This slope is the nominal 
value from Hamamatsu and our own measurements show some variation among sensors. Variations of this slope for sensors within a single summing group could increase the constant term of the energy resolution.} As mentioned earlier, the impact of the sensitivity of the breakdown voltage to temperature is minimized with a circuit that attempts to compensate for this effect. The response of this circuit is demonstrated in Fig.~\ref{fig:plot_Tdependence1} where the calculated supply voltage is shown to have very little temperature dependence over the range of temperatures that are maintained by the chiller cooling loop. The supply voltage setting is determined for each group of 10 SiPMs at the beginning of each physics run period by calculating its value at the measured temperature of their cooling plate. 

Custom readout electronics are mounted in standard VXS crates and include 
JLab 12-bit 250~MHz flash ADCs (FADCs) \cite{hdnote1022}, discriminators and F1 TDCs \cite{hdnote1021}.
The output signals from the summing circuits are delivered to the readout racks via RG-58 coaxial cables (55 ft).\footnote{Belden 9907 coaxial cable, thinnet 10Base2 Ethernet.}
The signals from the inner two summing layers (1 and 2) from both the upstream and downstream ends of each module are plugged into a single 16-channel FADC. The outer two summing layers (3 and 4) of each module, upstream and downstream, are plugged into a second FADC. Signals from both ends of each module are plugged into the same FADC module in order to make their sum easily available to the trigger. 
The summed signal compensates at first order for attenuation and provides a relatively uniform measure of the energy over the length of the module.  
Additionally, the signals from the inner three layers\footnote{The fourth layer is not instrumented with TDCs as that layer usually detects a small amount of energy from the showers and 
thus does not contribute significantly to the photostatistics that govern the timing resolution of the detector.} (12 signals per side) are amplified by a factor of five and transmitted to the electronic racks via 55-ft-long RG-58 cables (refer to 
footnote 19)
for input to discriminators. The discriminated outputs are used as inputs to the pipeline F1 TDCs.

Before final construction, a shorter prototype module (58-cm in length) was tested under the Hall B tagging spectrometer~\cite{Sober2000263}.  The response to electrons of 0.6 GeV at 90$^\circ$, 1.2 GeV at 20$^\circ$, and 1.9 GeV at 5$^\circ$ was measured.  Based on the results of these tests, the gain in the pre-amplifer circuits was reduced by a factor of 2.5, the copper thickness of the ground layers in the PCB boards was increased to control unacceptable ``ringing'' and noise, and the board layout was optimized.
These tests also established the amount of the cooling power required and confirmed the need for a dry-nitrogen environment to prevent water condensation.  Incorporating all these improvements resulted in an effective signal-to-noise ratio better than 400 for a 2V signal.


\begin{table}[ht!]\centering
\caption{BCAL channel counts. 
\label{tab:bcalcounts}}
\begin{tabular}{|l|c|}
\hline
\hline
Item                          & Quantity  \\ \hline 
 Trapezoidal Light Guides  & 3840   \\  
 Hamamatsu 144 mm$^2$ S12045 SiPM & 3840 \\ 
FADCs combined radially 1:2:3:4  &1536  \\  
TDCs combined radially 1:2:3     & 1152  \\  
Blue LED Model SMS1105BWC, Bivar, Inc. & 3840 \\     
\hline
\hline
\end{tabular}
\end{table}







\section{Assembly and Installation \label{sec:installation}}

\begin{figure}[thp]\centering
\includegraphics[scale=0.125]{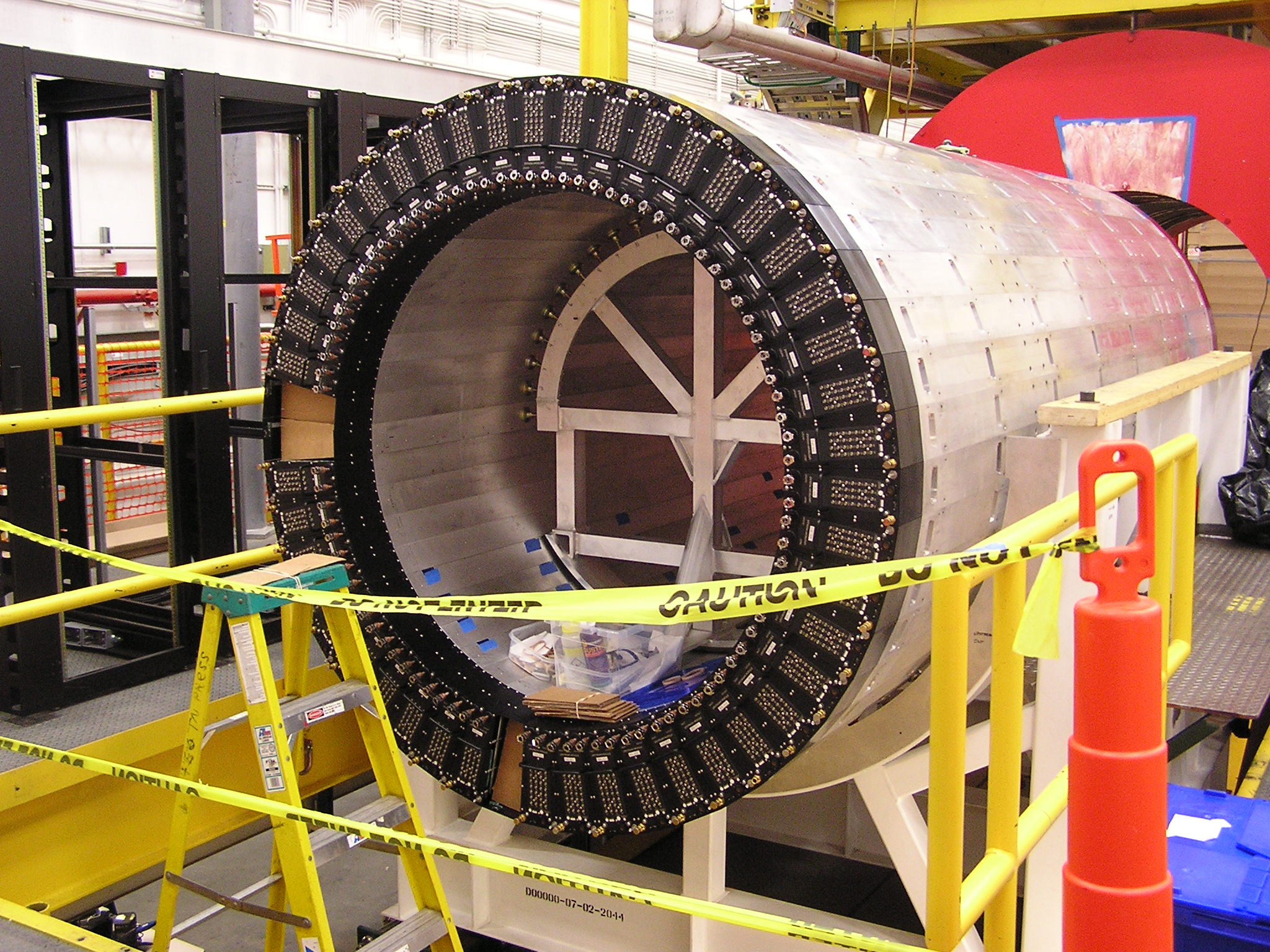}
\caption{\label{fig:bcalinstalled}
  BCAL assembly before insertion into the bore of the magnet (red yoke in the top right corner). Most of the electronic packages (black) have been mounted on the end of each module but cables are not yet connected. Also visible inside  the BCAL is a temporary fixture to support the upper modules during assembly. (Color online)
  }
\end{figure}
The lower half of the BCAL was assembled on a custom-built cradle, starting from the bottom-most, central position, and was then hoisted to the upstream GlueX detector platform since the overhead crane could not lift the entire BCAL (almost 30 tons).  The upper half of the BCAL was installed there, module by module, using a temporary adjustable support frame positioned on top of the bottom modules.   The machining tolerances of the modules 
required some small shims to match the overall cylindrical shape.  As each  module was positioned on the cradle, bolts were used to attach it to its neighbor though pockets machined in their base plates.  Once assembled, the full barrel was self-supporting and the internal frame could be removed. The nearly assembled BCAL, including the electronic packages at both ends, is shown in Fig.~\ref{fig:bcalinstalled}.  
The BCAL was then manually pulled into the solenoid on greased stainless steel rails using hand-operated winch and chains, and subsequently the cable connections, nitrogen gas and cooling lines were installed.


\section{Fine-grained Monte Carlo \label{sec:finegrainedMC}}

Two different Monte Carlo simulations, each containing a fine-grained representation of the detector, were used to optimize the calorimeter design and investigate its response to particle interactions.  Both  simulations contained a detailed geometric description of the module matrix, including the corrugated lead, fibers, glue, and base plates. The simulations were compared to each other at various checkpoints and agreed to within the estimated uncertainties.

A GEANT3 based simulation~\cite{Stamatis_thesis},  comprising a single BCAL module, was used to study the  shower longitudinal distribution,  sampling fraction,  energy leakage from the module, and the expected  energy resolution.  Simulations were carried out for photons, with energies between 50-1000~MeV, incident on the BCAL at $12^{\circ}$, $14^{\circ}$, $45^{\circ}$, and $90^{\circ}$.
During physics experiments, small angles receive the majority of flux incident on the BCAL.
Data at $12^{\circ}$ were used to study energy leakage from the downstream end of the calorimeter.  
The $14^{\circ}$ angle represents the direction of greatest material thickness for a photon from the target.
Simulations at $45^{\circ}$ were chosen as an intermediate angle and normal incidence is a standard for validating such simulations.

The second simulation, which modeled five adjacent modules, was based on FLUKA~\cite{Irina-Fluka}. It was employed to study the  dynamic range of the signal amplitude and to optimize the radial (i.e. depth) segmentation of the readout.  
The energy depositions were examined all the way down to the detector threshold and so FLUKA was used as it is more accurate at low energy.

The fine-grained GEANT3 Monte Carlo was used to determine average quantities for the BCAL matrix.  In order to minimize computational resources required for simulation, the standard GlueX Monte Carlo (\textsc{hdgeant}) has the BCAL implemented as single homogeneous material using these average properties.  The effective parameters used in \textsc{hdgeant} are listed in Table\,\ref{tab:bcalproperties}.

\subsection{Longitudinal Energy Deposition Profile}

The parametrized longitudinal development of the electromagnetic shower describes the shower reasonably well using the gamma distribution. If the depth in the material is expressed in units of radiation length,  ${t={x}/{X_0}}$, and the energy of the incident particle is $E_0$ then the profile is given by Ref.\,\cite{Agashe:2014kda}:
\begin{equation}
\frac{dE}{dt}=E_0\,b\,\frac{(bt)^{a-1}\,e^{-bt}}{\Gamma(a)}   \label{eq:EMprofile}
\end{equation}
The parameters \textit{a} and \textit{b} depend on the nature of the incident particle and the type of the absorbing material. The shower maximum, $t_{max}$, is equal to $(a-1)/b$. This equation generally fits the simulated data well.  

\begin{figure}[htbp]\centering
 \includegraphics[scale=0.3,angle=0.,trim=0mm 15mm 0mm 0mm]{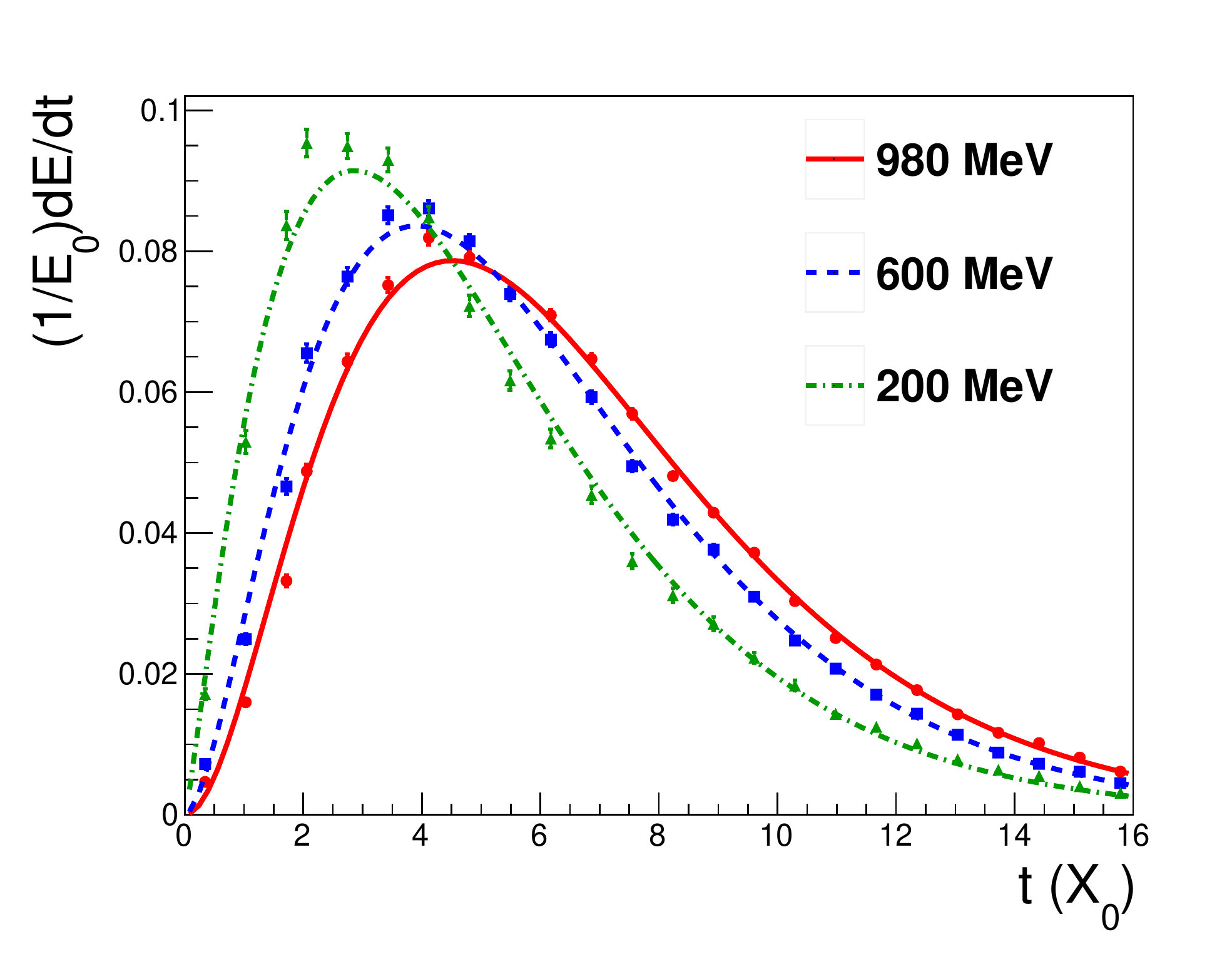}
\caption{\label{fig:longprofile}
  Longitudinal shower profile at 
  $90^{\circ}$ for incident photons of 200~MeV, 600~MeV and 980~MeV.  The data points correspond to the Monte Carlo results and the curves are fits using Equation~\ref{eq:EMprofile}.  }
  \end{figure}

A sample of the generated and fitted longitudinal profiles is shown in Fig.\,\ref{fig:longprofile}, for photons incident at
$90^{\circ}$.  The distribution is roughly independent of the incident angle except for small changes due to leakage. At shallower angles the shower develops to greater depth and the entire shower is practically captured.  
At $14^{\circ}$, where the apparent BCAL is at its thickest, low-energy particles deposit almost all of their energy in the first layer. This effect motivates the non-uniform 1:2:3:4 readout segmentation.

\subsection{Expected Energy Resolution}
\begin{figure}[htbp]\centering
 \includegraphics[width=0.5\linewidth]{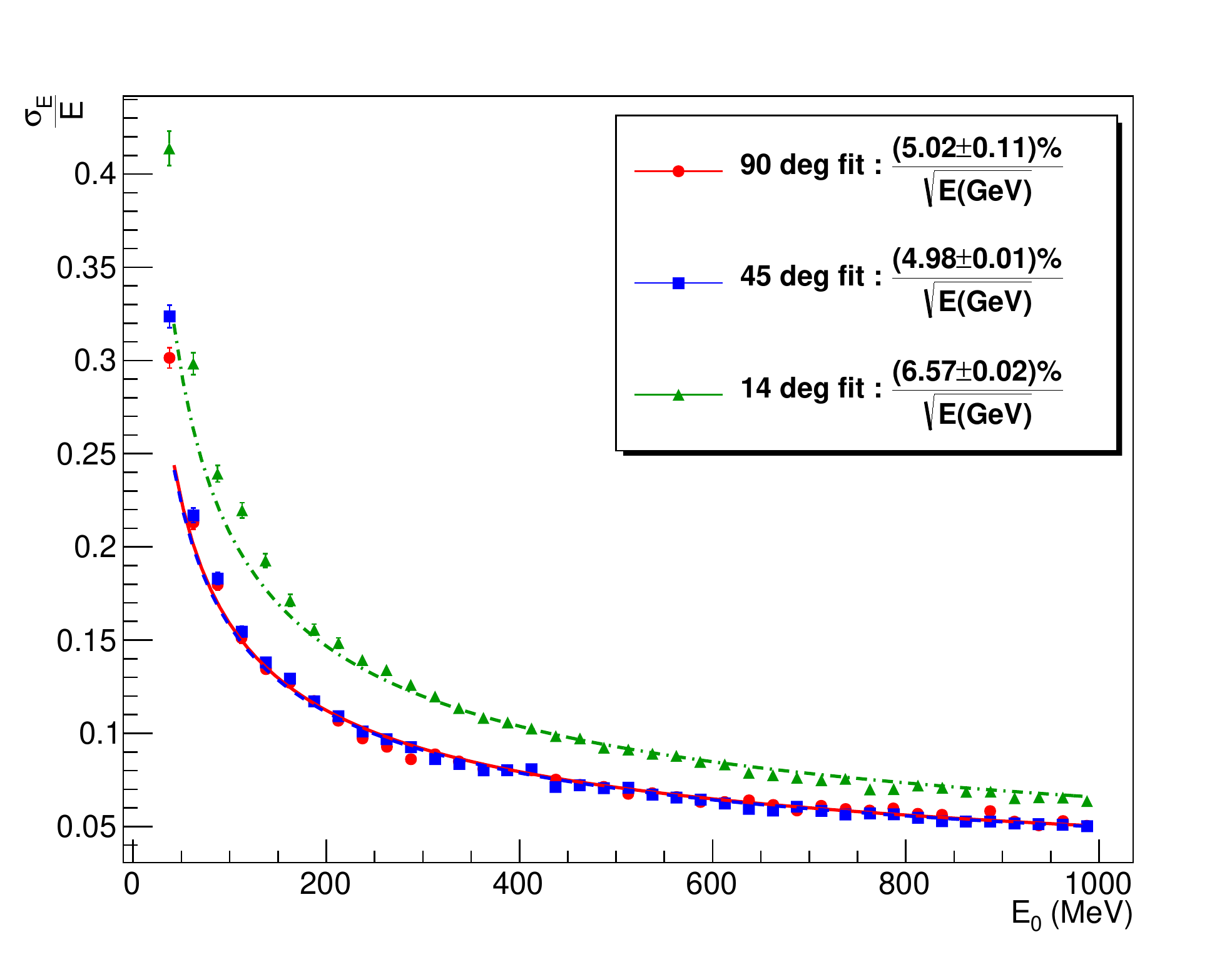}
\caption{\label{fig:eressim}
  Energy resolution as a function of incident photon energy at $14^{\circ}$, $45^{\circ}$ and $90^{\circ}$.The points are simulated data whereas the lines represent fits to the data using Equation~\ref{eq:resolutionfit}.
  }
\end{figure}

As shown in Equation~\ref{eq:Eresolution}, the energy resolution can be given as the sum of three terms in quadrature. If a high signal-to-noise ratio is achieved, the noise term is usually negligible.  The simulations did not include noise from the photosensors or the electronics and therefore the noise term, $c'$, was set to zero, whereas the constant term, $b$, was extracted from the fits and found to be consistent with zero within errors.  The  simulated resolution is shown in Fig.~\ref{fig:eressim} for photon angles of incidence of $14^{\circ}$, $45^{\circ}$ and $90^{\circ}$, including the results of the fits for the statistical term, $a$. The  function used to fit the graphs is:
\begin{equation}   
\label{eq:resolutionfit}
 \frac{\sigma_E}{E}\,=\,\frac{\textit{a}}{\sqrt{E(GeV)}}\,\oplus\,\textit{b}.   
\end{equation}
    
For photons at normal incidence, these simulations are in agreement with earlier ones~\cite{gx842}.   The curves at $45^\circ$  and $90^\circ$ are nearly identical.  At $45^\circ$ the module is able to contain more of the electromagnetic shower as compared to $90^\circ$, although the difference is small because even for 1~GeV photons the energy leakage out the back is only $\approx 3$\% at $90^\circ$ and $\approx 0.5$\% at $45^\circ$ (see Fig.~\ref{fig:energyleakage}). The better photostatistics at $45^\circ$ is likely the 
reason for the improved uncertainly in the fit.
The resolution degrades at $14^\circ$ and smaller impact angles because of the increased energy leakage out the front face of the module (albedo) as seen in  Fig.~\ref{fig:energyleakage}.
The energy resolution extracted from photon-beam data is described below in Section~\ref{sec:energyres}.
  
\subsection{Energy Leakage}

The distribution of energy leakage out of the BCAL is qualitatively different from most calorimeters because many particles are incident at relatively shallow angles.  The generation of secondary particles in the lead depends on the incident photon energy and has a stochastic behavior.  Since the sampling thickness effectively increases at shallow angles, the effective threshold for generation of secondaries is affected and this can complicate the interpretation of simulated results.

\begin{figure}[tbp]
 \centering
 \includegraphics[scale=0.195]{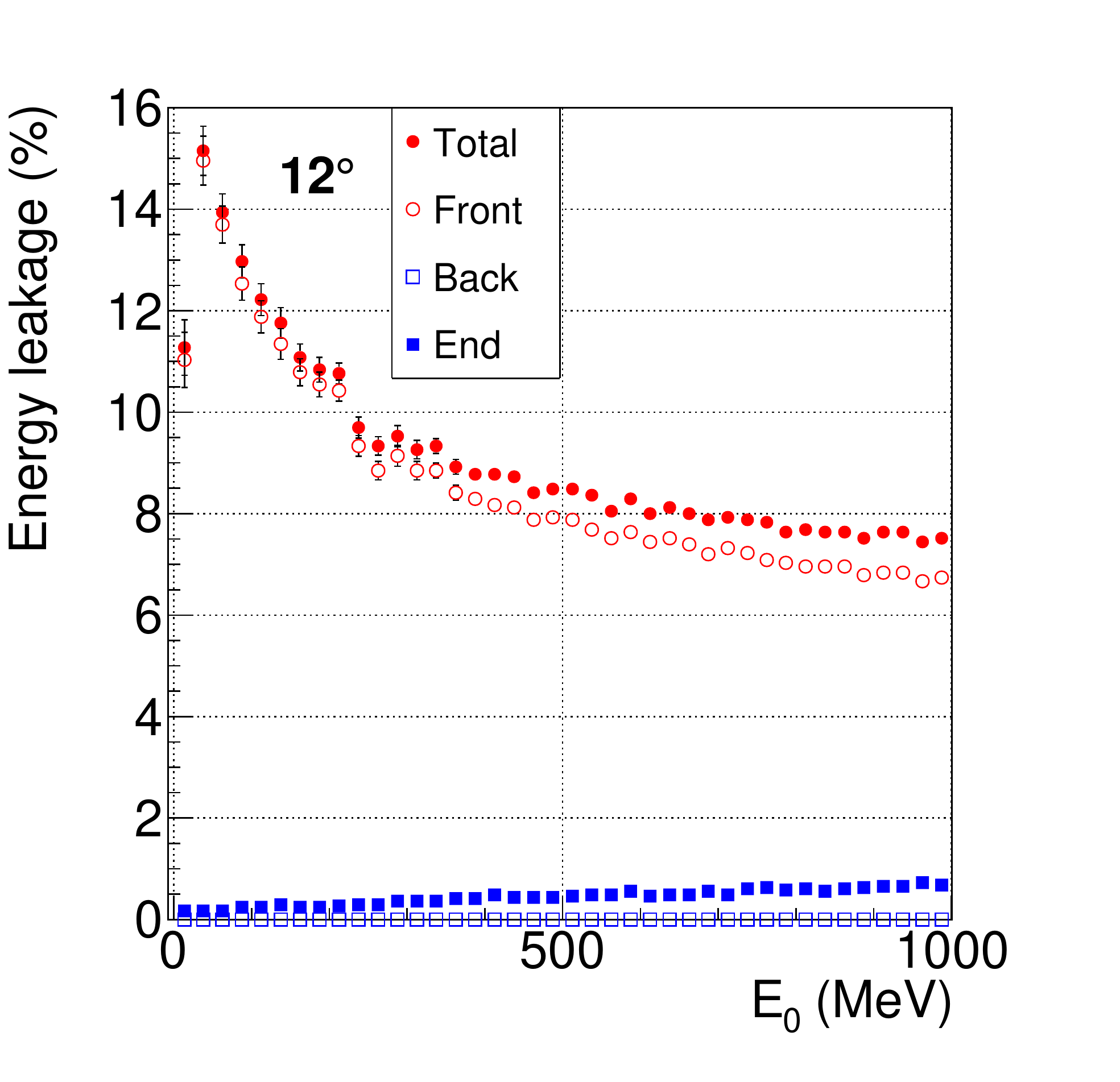}
 \includegraphics[scale=0.195]{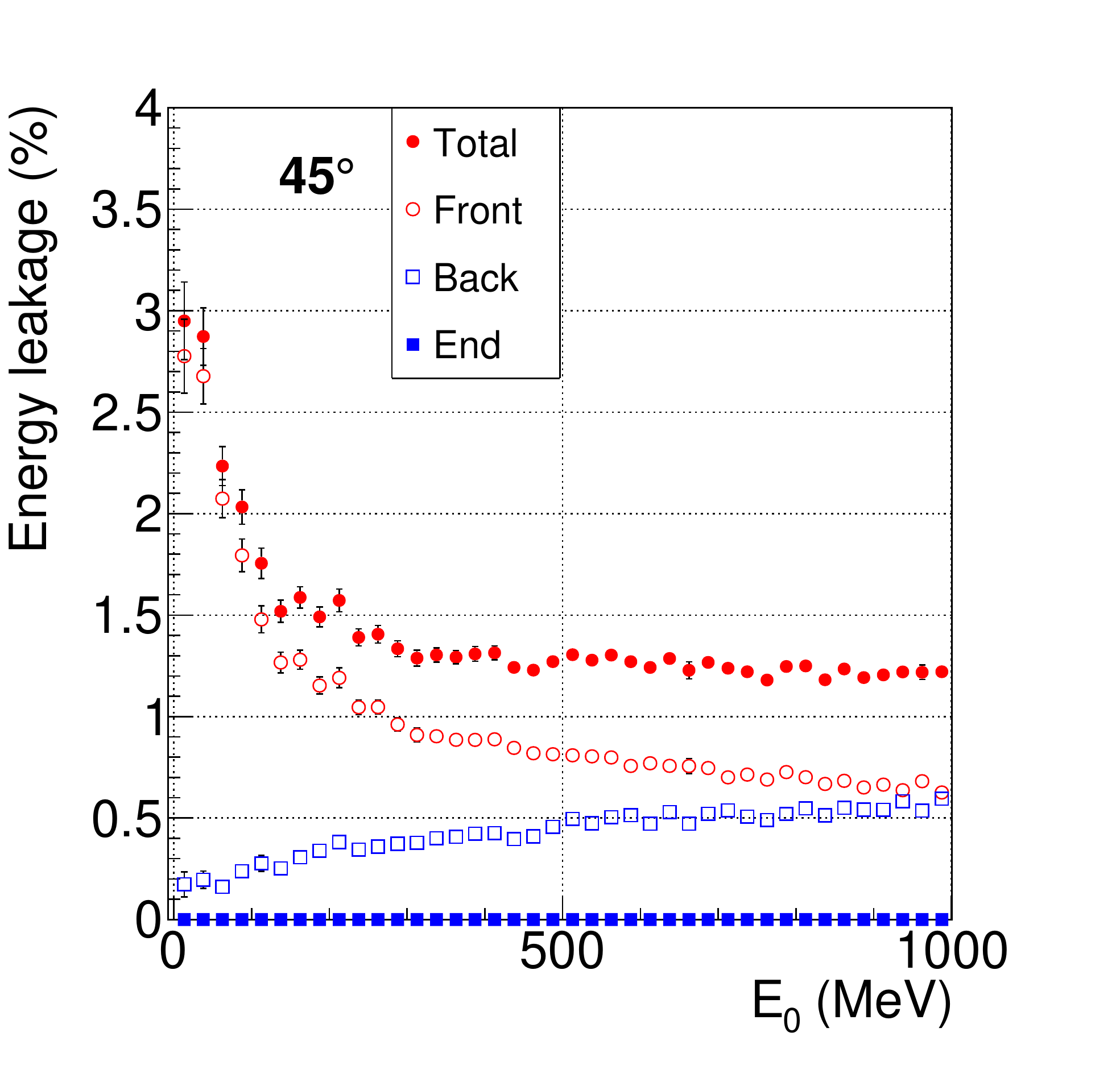}    
 \includegraphics[scale=0.195]{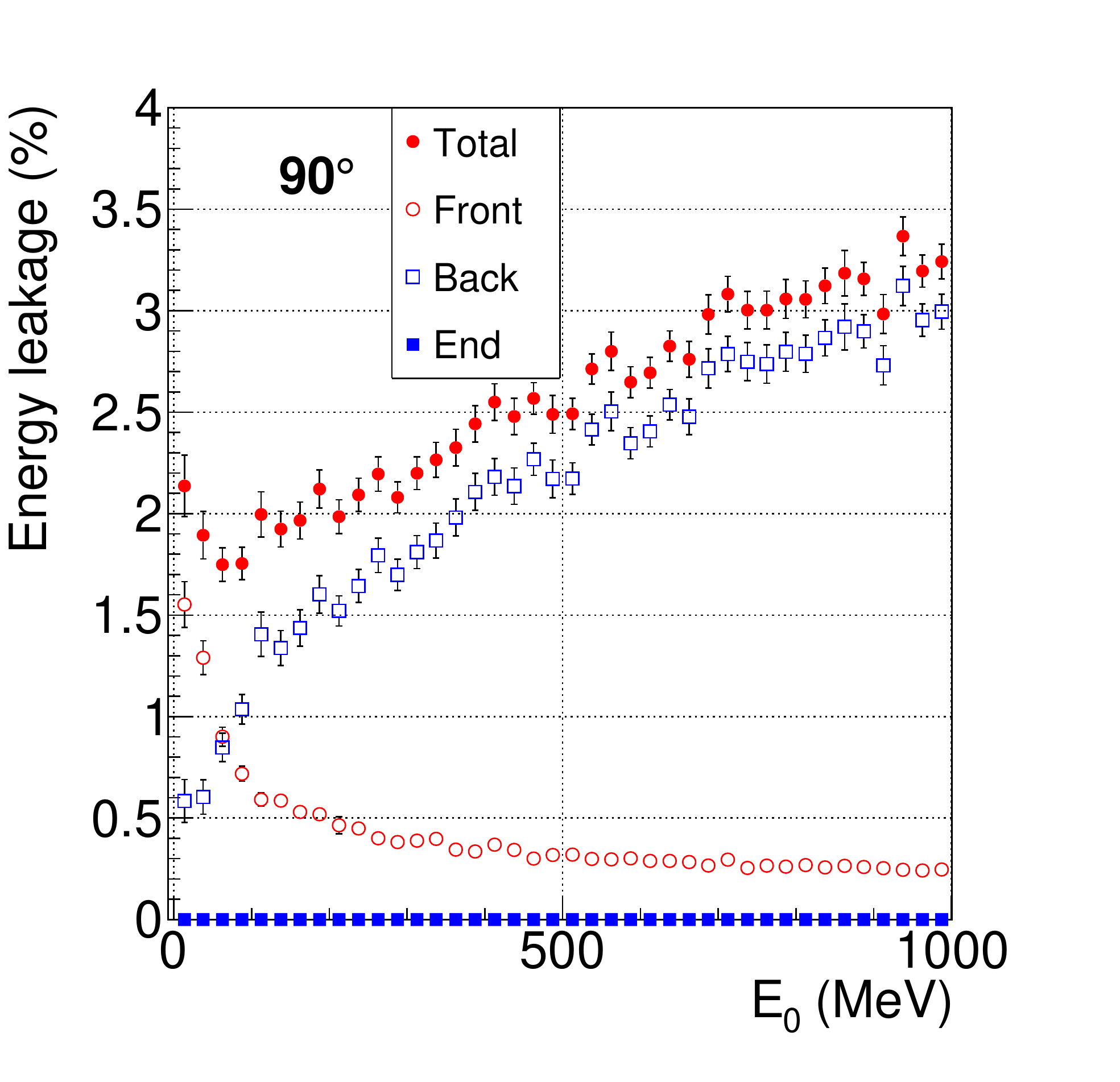}
 \caption{
 Simulated energy leakage (in percent) versus incident photon energy for photons incident at $12^\circ$, $45^\circ$ and $90^\circ$. The polar angles correspond to photon angles from the center of the GlueX target. 
 Front and back refer to the inner and outer surfaces of the module, respectively, with respect to the radial direction.  
Leakage from the ends refers to the energy leaking out the surfaces of the module along the direction of the beam, where the SiPMs are located.   
 The leakage out the ends is consistent with zero in all cases except at very forward angles ($12^\circ$).
 }
\label{fig:energyleakage}  
\end{figure}

The fractional energy leakage out of the calorimeter volume is shown in Fig.~\ref{fig:energyleakage}. 
For incident photon angles of $90^\circ$, irrecoverable energy leakage occurs mostly out the back of the module. This leakage increases monotonically with energy, following the evolution of $t_{max}$, which means that the shower develops further inside the module along the radial direction. 
There is little albedo at $90^\circ$, but it increases with decreasing angle and dominates the leakage at small angles ($12^\circ$-$14^\circ$).   It also tends to be a larger fraction at lower energies.  The albedo affects the reconstruction of showers in the calorimeter as it may create additional hits in the detector.

\subsection{Sampling Fraction}

\begin{figure}[tp]
 \centering
 \includegraphics[width=0.50\linewidth]{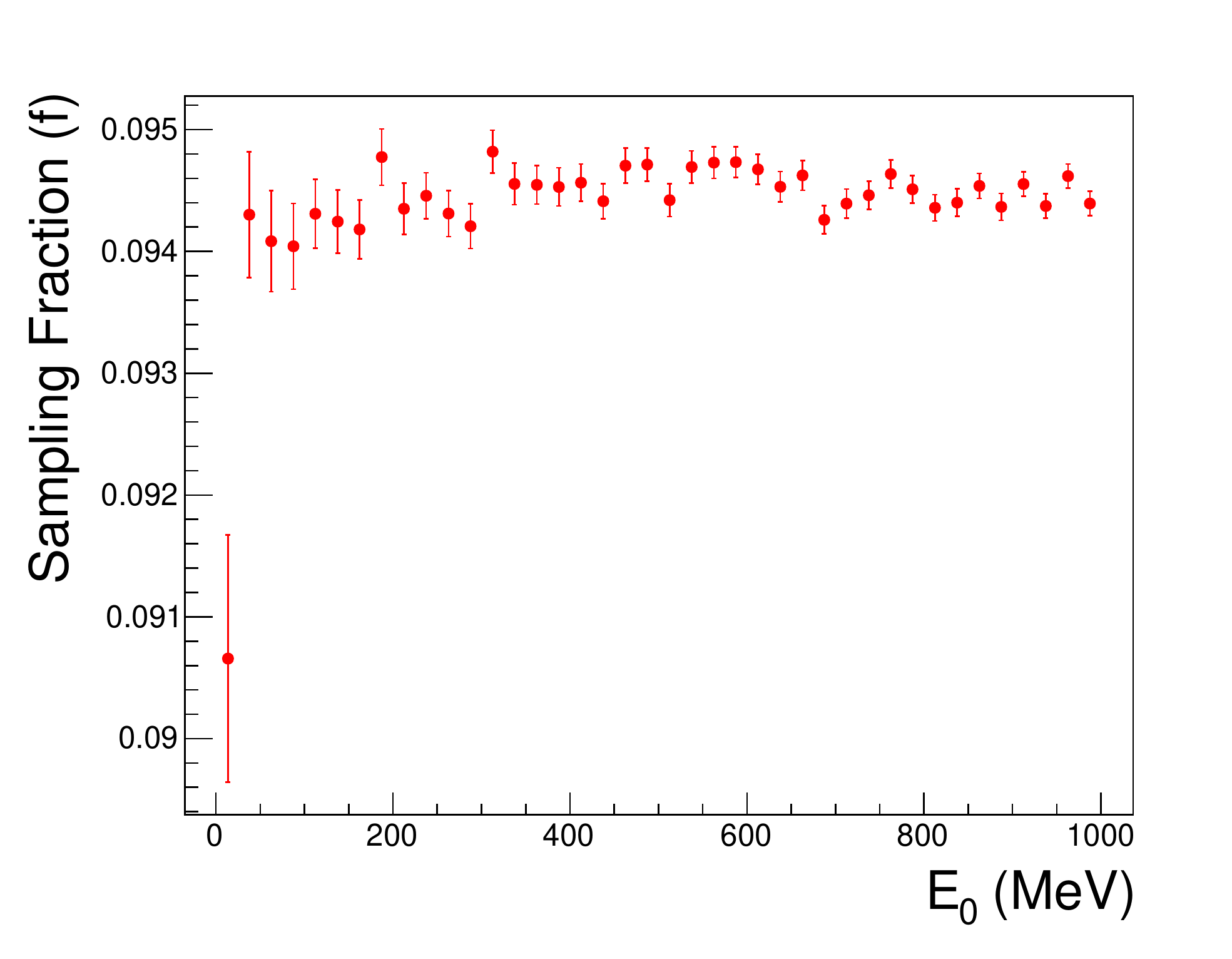}	  
\caption{Sampling fraction versus incident photon energy at  $90^\circ$ incidence. The sampling fraction is approximately constant across the full, simulated energy spectrum.}
\label{fig:fracdep}
\end{figure}

The sampling fraction is defined as the fraction of the energy that is deposited in the sampling material (i.e. scintillating fibers) over the total energy deposited in the module: 
\begin{equation}
 \textit{f}=\frac{E_{\rm scifi}}{E_{\rm mod}}
\end{equation}
The sampling fraction, as determined by simulation, is plotted as a function of the incident photon energy in Fig.~\ref{fig:fracdep} for  $90^\circ$, but is nearly identical for other angles.  The sampling fraction is approximately constant at $\approx$9.45\% but with some non-linearities below 50 MeV. 
To properly account for the sampling fraction, only the energy deposited in the fiber core is considered part of the sensitive volume. Therefore the material had  to be properly modeled including the two layers of cladding of the fibers. 

\subsection{Threshold Calibration using Michel Electrons}
Cosmic-ray muons that stop in the BCAL decay at rest via $\mu \rightarrow e  \nu  \overline{\nu} $ with a characteristic lifetime of 2.2 $\mu$s. The product electrons, known as ``Michel electrons,'' have a well-known energy distribution 
with a sharp peak at $52.8\,MeV$ (half the rest mass of a muon). This reaction can be tagged using stopping tracks and looking for the decay signals in the cell where the muons come to rest. For this test we took data with a relatively
long readout window of 511 samples (2044 ns), so that delayed signals could be recorded in the data stream.  The exponential time distribution of decay signals relative 
to stopped tracks was fit in the range between 0.4 and 1.8 $\mu$s and found to be $\tau\,=\,$2.21$\pm$0.02\,$\mu$s, consistent with the muon lifetime. The energies of the decay signals were then compared to a Monte Carlo that 
seeded muons above the
detector and let them decay inside the BCAL volume \cite{hdnote3097}. The Monte Carlo and data samples were compared and the readout threshold for a cell was determined to be 2.2 MeV. 
The measured energy depositions in each layer were compared to the detailed simulation and matched closely in layers 2 and 3. Some discrepancies between data and simulation were found in layers 1 and 4, but they are
on the outer edges of the BCAL where it is possible that either the muon or the electron were not adequately contained in the detector.  

 \begin{figure}[tbp]
\begin{center}
\hspace*{-0.4in}
\includegraphics[width=0.7\linewidth, angle=0.]{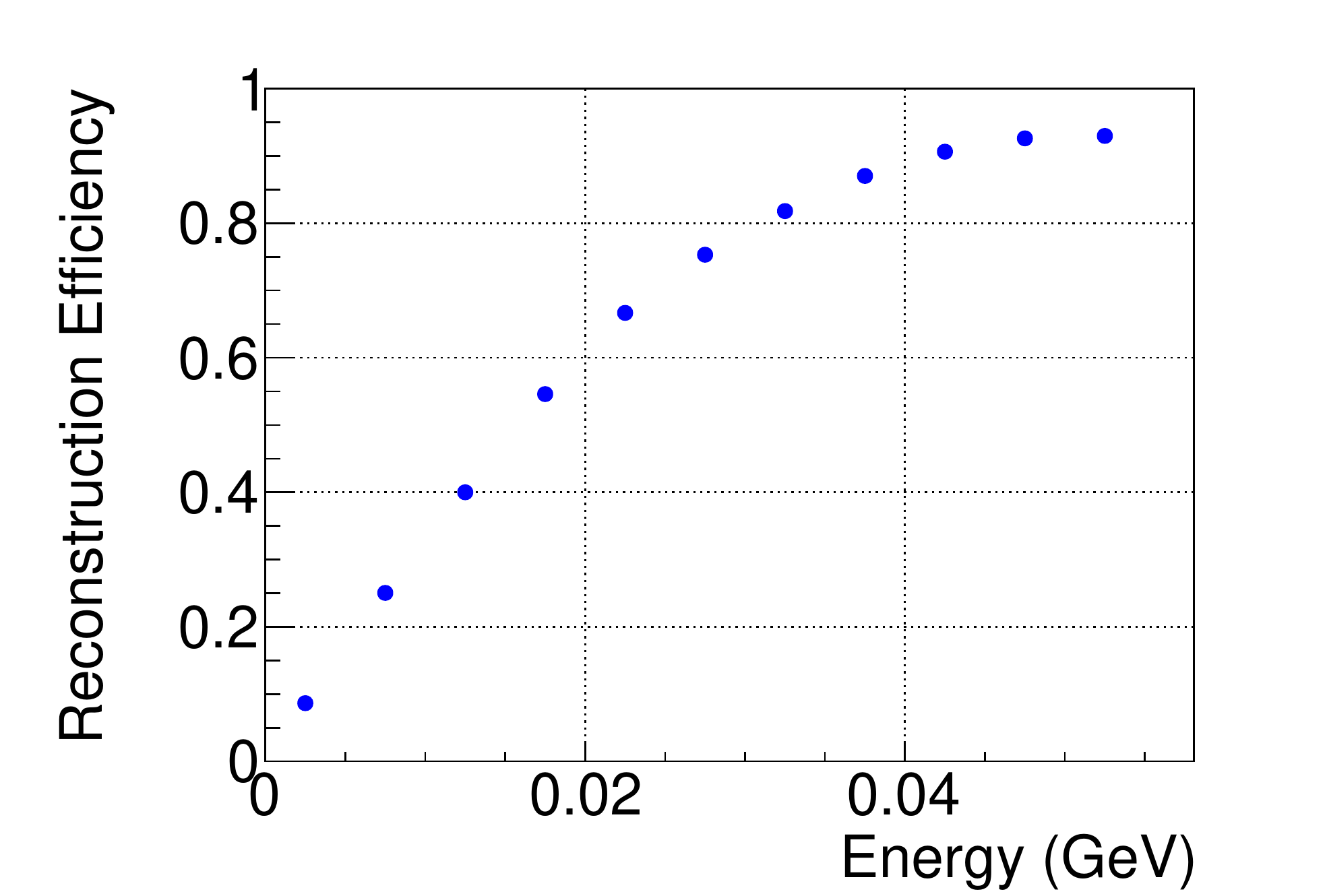}
\caption{\label{fig:Michel_efficiency}
Threshold for the reconstruction efficiency in the BCAL predicted by Monte Carlo, which was matched to the energy distribution of electrons from muon decays.}
\end{center}
\end{figure} 

These Michel electrons present a clean sample of low-energy electrons with energies that span the expected threshold range of the BCAL. Therefore, we have used this study to determine the effective reconstruction
thresholds. Using the same parameters and thresholds for the Monte Carlo that best matched our data of Michel electrons, we determined the detection  efficiency for low-energy electrons, which is shown in Fig.\,\ref{fig:Michel_efficiency}. 
The efficiency reaches a plateau between 40 and 50 MeV.

\subsection{Shower Distribution in the BCAL \label{sec:showerdistribution}}

\begin{figure}[tbp]\centering
 \includegraphics[width=0.8\linewidth]{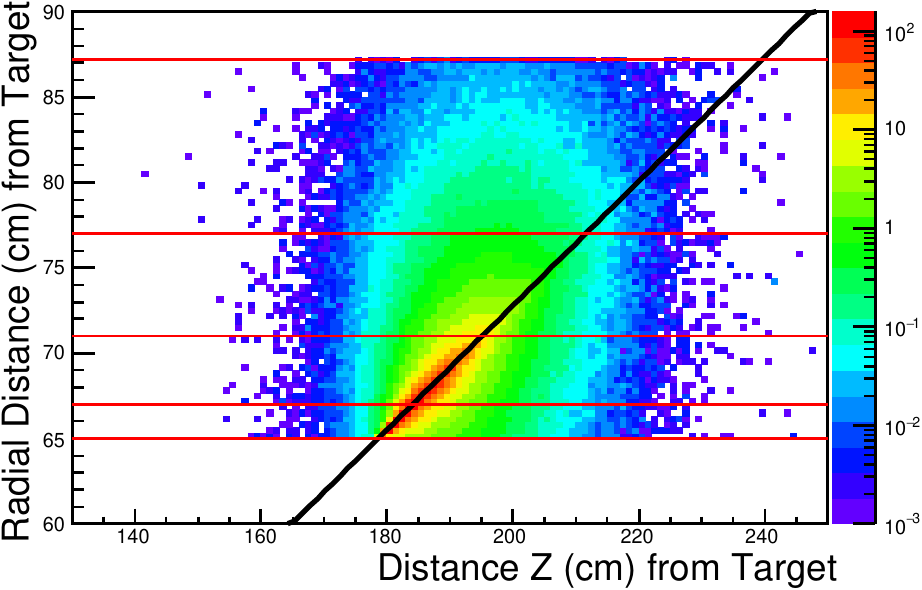}
\caption{\label{fig:Shower_curvature}
  Average distribution of the energy deposition in the BCAL for a  1.2 GeV shower corresponding to a photon incident at 20$^\circ$.  The horizontal lines indicate the approximate location of the radial readout  layers. (Color online)
  }   
\end{figure}

\begin{figure}[tbp]\centering
 \includegraphics[scale=0.64,angle=0.]{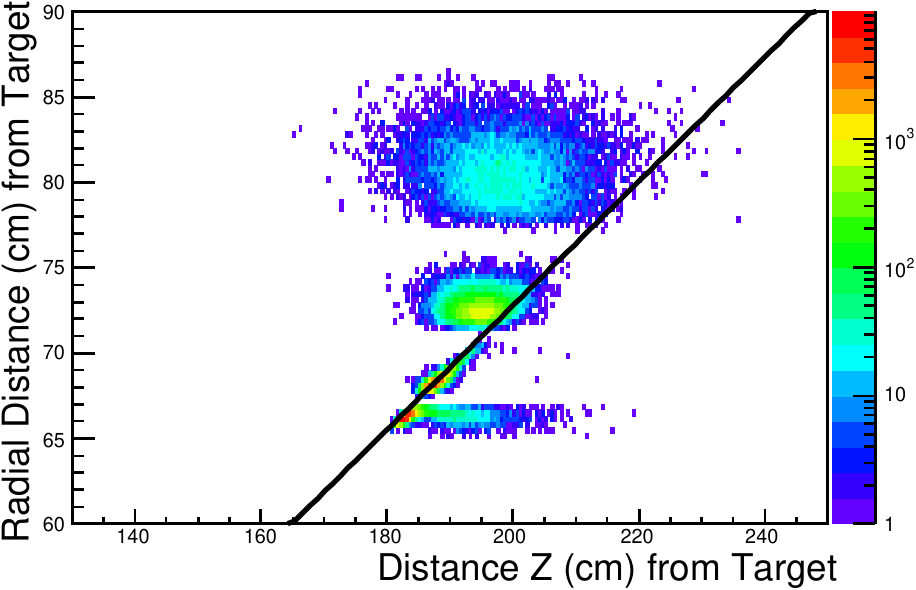}
 \includegraphics[scale=0.64,angle=0.]{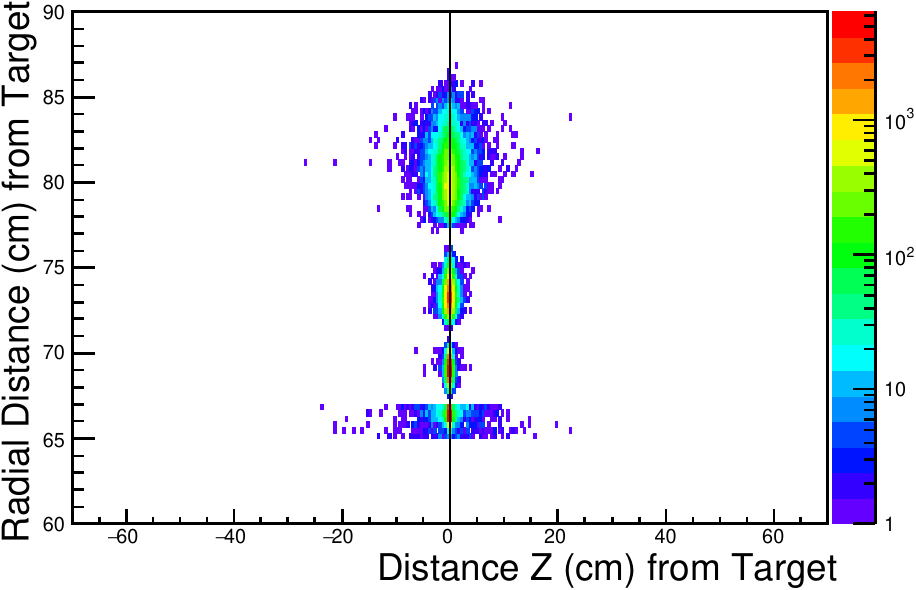}
\caption{\label{fig:showerCurvature}
Centroids of energy depositions for showers produced by photons at 1.2~GeV and 20$^\circ$ (left) and 0.5 GeV and 90$^\circ$ (right). At forward angles, the readout segmentation results in the centroid in each layer to be offset from the direction of the incident photon.The shower is distributed symmetrically around the initial photon trajectory only at normal incidence.   (Color online)
  }
\end{figure}

The majority of showers in the BCAL are created by particles impinging at shallow angles. For example, particles coming from the LH$_2$ target and hitting the center of the BCAL correspond to an incident angle of about 25$^\circ$. This configuration leads to some surprising features in the distribution of energy within the layers of the calorimeter. 
Centroids of shower depositions are not located along the trajectory of incident photon, as can be seen in Fig.\,\ref{fig:Shower_curvature}. The reason is that the photon shower consists of a high-energy-deposition part located along the direction of the incident photon and an almost isotropic (conical) halo. The centroid of the halo is sliced (sampled) by layer readout cells and is visibly shifted from the direction of the sloped incident photon. The centroid of the first layer is shifted downstream from
the photon direction while the outer layers are shifted upstream. Only the centroid in Layer 2 falls approximately along the direction of the incident photon. This behavior is exemplified in Fig.\,\ref{fig:showerCurvature} where we show the 
reconstructed position of energy depositions in various layers for a couple of selected shower energies and angles. 
This effect impacts the determination of the position of the shower in various layers and is addressed by using empirically determined constants in the reconstruction (see Section\,\ref{sec:postime}).

\section{Raw Data Processing \label{sec:raw_data}}
In this section we expand on the format of the raw data, which is read out from the digitizers. For a description of the electronics, see Section\,\ref{sec:electronics}. 
 
\subsection{FADC Raw Data \label{sec:adcwaveforms}}
\begin{figure}[htbp]\centering
 \includegraphics[width=0.6\linewidth]{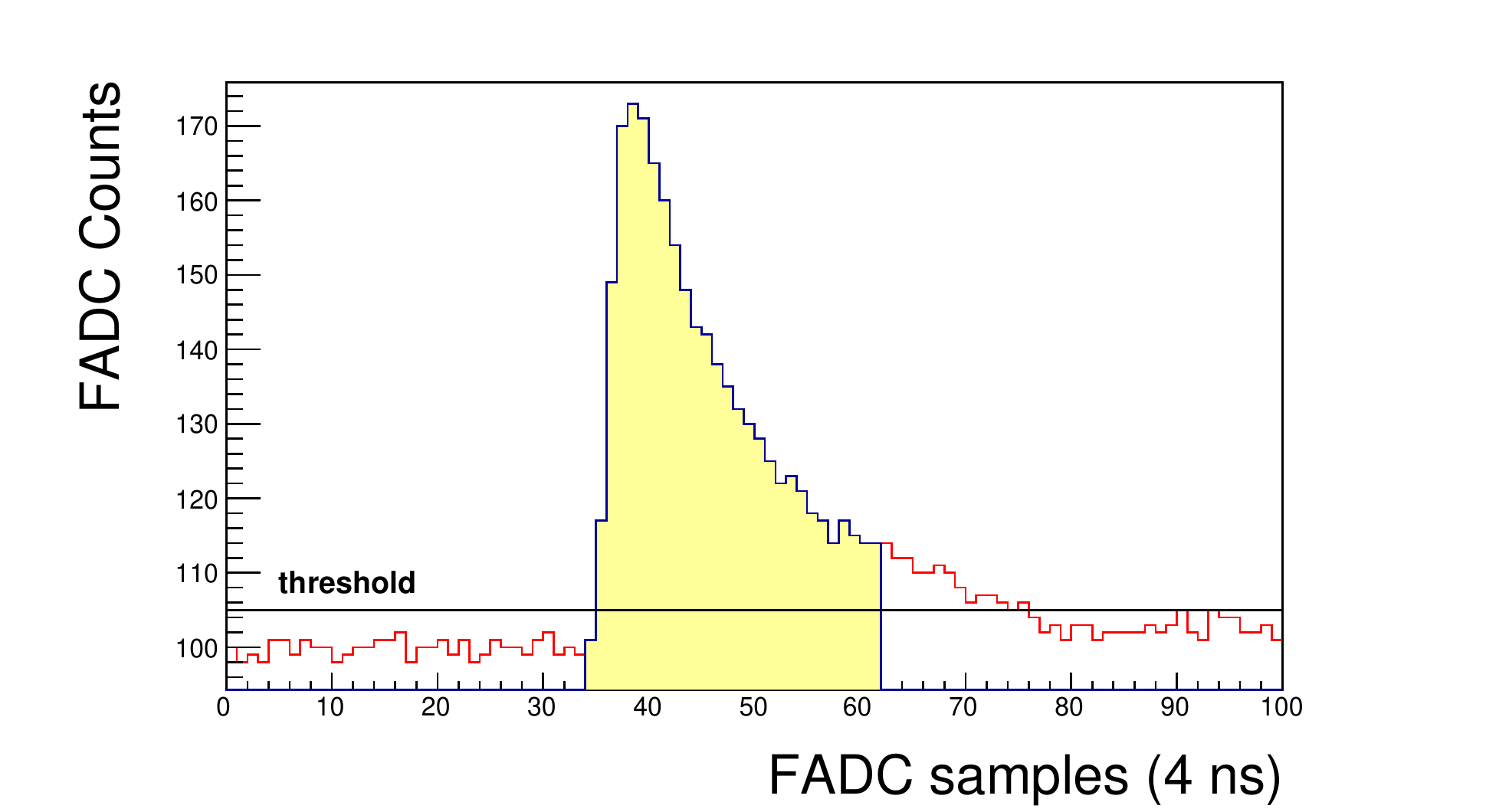}
\caption{\label{fig:Plot_waveform10}
 Typical FADC waveform corresponding to modest energy deposition. The baselines were adjusted a few times per running period to a value of 100 counts. The 
 readout threshold was set to 105 counts, where each counts corresponds to 0.4882  mV. The filled area in the histogram
 corresponds to the integration window.
 (Color online)
  }
\end{figure}

The JLab 250 MHz flash ADCs \cite{hdnote1022} sample each BCAL channel every 4 ns and generate raw waveforms as shown for a $\sim$20 MeV pulse in Fig.\,\ref{fig:Plot_waveform10}. 
Each BCAL waveform consists of 100 samples 
 (400 ns), which are available for further processing by the firmware upon a trigger signal if there is a threshold crossing. The firmware computes several derived features of the pulse: pedestal, peak value, integral over a selected window, and
 time of the half-way point on the leading edge. At most one pulse is extracted from each readout window. These  pulse features constitute the raw data that is read out from the FADC. 
 Optionally, the full waveforms can be read out for diagnostic purposes
 and to check the firmware output against the offline emulation of the parameter extraction. We record full waveforms in less than about 1\% of the production runs. The pedestal is determined for each channel event-by-event, appropriately scaled, and then subtracted from the peak and integral to obtain signals proportional to the energy deposited in the calorimeter. 
 Pulses are identified by the first sample that exceeds a threshold, 
 currently set to 105 counts. Typical pedestal widths are $\sigma\sim$1.2-1.3 counts. The integral is determined using a fixed number of samples relative to the threshold crossing, 
 which was determined by maximizing the ratio of signal to pedestal noise. 
 The integration window begins one sample before the threshold time (NSB) and extends to 26 samples after the threshold time (NSA). See Fig.\,\ref{fig:Plot_waveform10}.
 The algorithm that determines the time of the pulse is pulse-height independent and therefore no time-walk correction is required
 for the FADC times.
 
 We note that the FADCs also provide information to the trigger, which at the moment computes a sum over all BCAL signals. The inputs to the trigger consist of the sums of each FADC card. Each card corresponds
 to upstream and downstream signals from half of one module (16 channels), either the two inner layers, or the two outer layers. 
 The trigger pulses are computed in a similar way as for data pulses, but
 with different parameters. The threshold used is 120 counts, NSB=3 and NSA=19. The higher threshold reduces the sensitivity of the trigger to noise. 
 
 Each FADC sample digitizes a voltage between 0 and 2 V with a 12 bit ADC (0 to 4095). Voltages above this value are truncated to 4096 and the integral uses these truncated samples in the
 summation. A corrected sum can be estimated offline with some reduced accuracy assuming the pulse shapes remain unchanged except for a scale factor.
 
 \subsection{TDC Raw Data}
 The outputs of the three inner layers of the BCAL cells are connected to custom JLab leading-edge (LE) discriminators \cite{hdnote2511}, which feed the JLab F1 Time-To-Digital Converter (TDC) \cite{hdnote1021}. 
 The discriminator thresholds are typically set to about 35 mV. However, we find that there is considerable variation in the baseline relative to the threshold from one channel to another, so the thresholds have been adjusted
 channel by channel.  The pulse times are recorded relative to the trigger in a 12-bit word. Multiple hits, up to eight, may be recorded per channel per event, but are culled at a later time by comparison to FADC times. The nominal least count is configured to 58\,ps.
 



\subsection{Combining Raw Data}
The raw ADC and TDC data from individual channels are combined to produce calibrated energy, time and position values for reconstructed showers.  The raw data from both ends of a BCAL cell are combined into ``points,'' where the time difference is used to determine the position along the BCAL in $z$, while the time sum determines the time of energy deposition.  The energy of a point is the sum of the energy from the ends corrected for attenuation along the scintillating fibers.  These points are combined into ``clusters'' of points that 
collect all the energy deposited by a single particle incident in the BCAL.  The algorithms are tuned to minimize splitting a single physical cluster into multiple logical clusters while also avoiding combining two neighboring physical clusters into a single logical cluster.  Once the clusters are determined, a nonlinear energy correction is applied to convert them into ``showers.''  The determination of the calibration constants and the algorithms used are discussed in detail in the next section.

\section{Calibration \label{sec:calibration}}

The calibration proceeds in the following order.  First we remove hardware time offsets in the ADCs and TDCs.  
The TDC times are then corrected for time-walk using ADC information.  The positions of the hits, $z$, are calibrated using a sample of pion-enriched negatively charged tracks, comparing the drift chamber track position projected to the BCAL to the time difference between the two ends of the BCAL.  For the same sample of tracks, the point times are calibrated by projecting the time sum of the two ends back to the target using the momentum and trajectory of the track and comparing it to the appropriate accelerator bunch time.  The attenuation within the scintillating fibers and the ratio of the gains between each end of a BCAL cell are then extracted by comparing the size of the signal at each end with the $z$ position, calculated from the time difference.  The overall gain for each BCAL cell is then determined in an iterative procedure by minimizing the width of the reconstructed $\pi^0$ invariant mass distribution from its $2\gamma$ decay.
 These steps are detailed in the sections that follow.



\begin{figure}[t]\centering
 \includegraphics[width=0.7\textwidth]{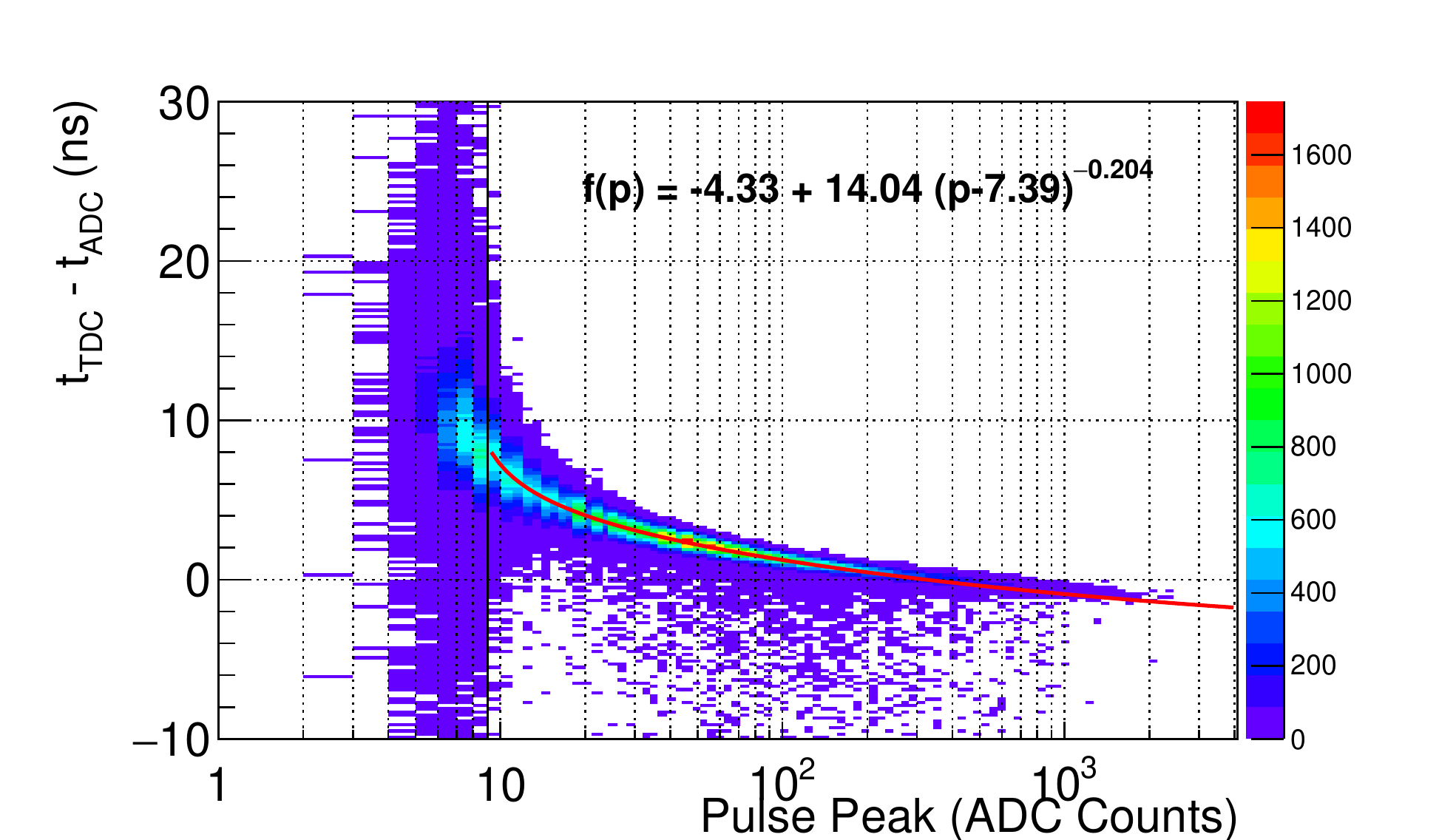}
\caption{\label{fig:timewalk}
Example of the fit of the measured leading-edge discriminator time as a function of the FADC pulse peak value
for module 1, layer 1, sector 1, upstream. The fit corresponds to the time-walk correction function. (Color online)
}  
\end{figure}

\subsection{Timing Calibration \label{sec:timecalib}}

The TDC data represent the time of the threshold crossing from the JLab leading-edge discriminators. Therefore, for accurate time applications, this value must be corrected for time-walk, 
namely the variation of the measured time with pulse amplitude. The corrected time for each cell, $t_w$,
is given by
\begin{eqnarray}
t_w & = & t - f_w(p), \label{eq:timewalk}
\end{eqnarray}
where $f_w(p)$ is the time-walk correction function, assumed to be only a function of the pulse peak height, $p$, and $t$ is the uncorrected time.
An empirical form for the time-walk correction is chosen based on previous experience \cite{clasnote2002007} to be
\begin{eqnarray}
f_w(p) & = & c_0 + {c_1 (p - c_2) ^{c_3}}, \label{eq:timewalkfunc}   
\end{eqnarray} 
where $c_0,\,c_1$, $c_2$ and $c_3$ are time-walk-correction constants.  This functional form has sufficient parameters to fit the data over the full dynamic range
and reduces to the commonly used inverse-square-root 
dependence when $c_3=-0.5$. The correction and therefore the TDC time determination is only valid for pulses above the threshold, approximately 9 FADC counts.  The offsets 
$c_0$ are used to match the TDC to the FADC times. The time-walk correction constants are obtained by minimizing the difference between $t_w$ and the measured FADC time.
An example 
of the measured time in one TDC as a function of its FADC pulse peak value is shown in  Fig.\,\ref{fig:timewalk} together with a fit of Eq.\,\ref{eq:timewalkfunc} to the data, along with the
fit results.

\subsection{Position Determination from Time Measurements \label{sec:postime}}

The precise determination of the effective speed of light in the BCAL fibers is crucial for the position reconstruction of the particles in the BCAL.  For a point source, the measured times are expected to be linear functions of the position along the module, with a proportionality constant given by the effective speed of light propagation in the fibers, $v_{\rm eff}$:
\begin{eqnarray}
z & = & \frac{1}{2}\,v_{\rm eff} \left[ (t^u_{w} - t^d_{w} )  + (t^u_{0} - t^d_{0}) \right] \nonumber \\
    & \equiv & \frac{1}{2}\,v_{\rm eff} \left[ t_{\rm diff} + (t^u_{0} - t^d_{0}) \right] ,     \label{eq:ztime}
\end{eqnarray}
where $t_w$ represent the time-walk corrected TDC values or the ADC times, which need no walk correction. For the outermost
layer, only FADC times are available. 
We define the difference of upstream and downstream times as $t_{\rm diff}$.
The constant time offsets $t_{0}^{u(d)}$ are determined such that $z=0$ corresponds to interactions at the center of the module.

   
\begin{figure}[tbp]
\centering
\includegraphics[scale=0.7]{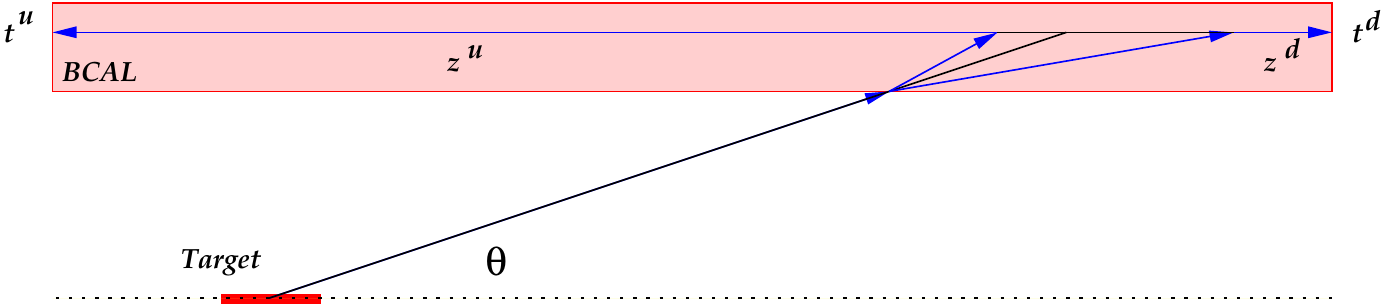}
\caption{\label{fig:shower_in_bcal} Toy model of the shower development inside the BCAL and associated timing signals.}
\end{figure}

In order to calibrate the correspondence between the measured BCAL times and the location of particle interactions, we use a sample of charged tracks that point to the BCAL and we determine 
the impact position $z$ using the tracking chambers. The sample includes a mixture of non-interacting tracks as well as hadrons that interact in the BCAL and generate hadronic showers. For this ensemble, we find
that Eq.\,\ref{eq:ztime} has to be modified due to the fact that showers spread out as they develop through different layers. The upstream and downstream boundaries of the energy flow will affect the measured
times of upstream and downstream sensors.

\begin{figure}[tph]
\centering
\includegraphics[width=0.48\textwidth]{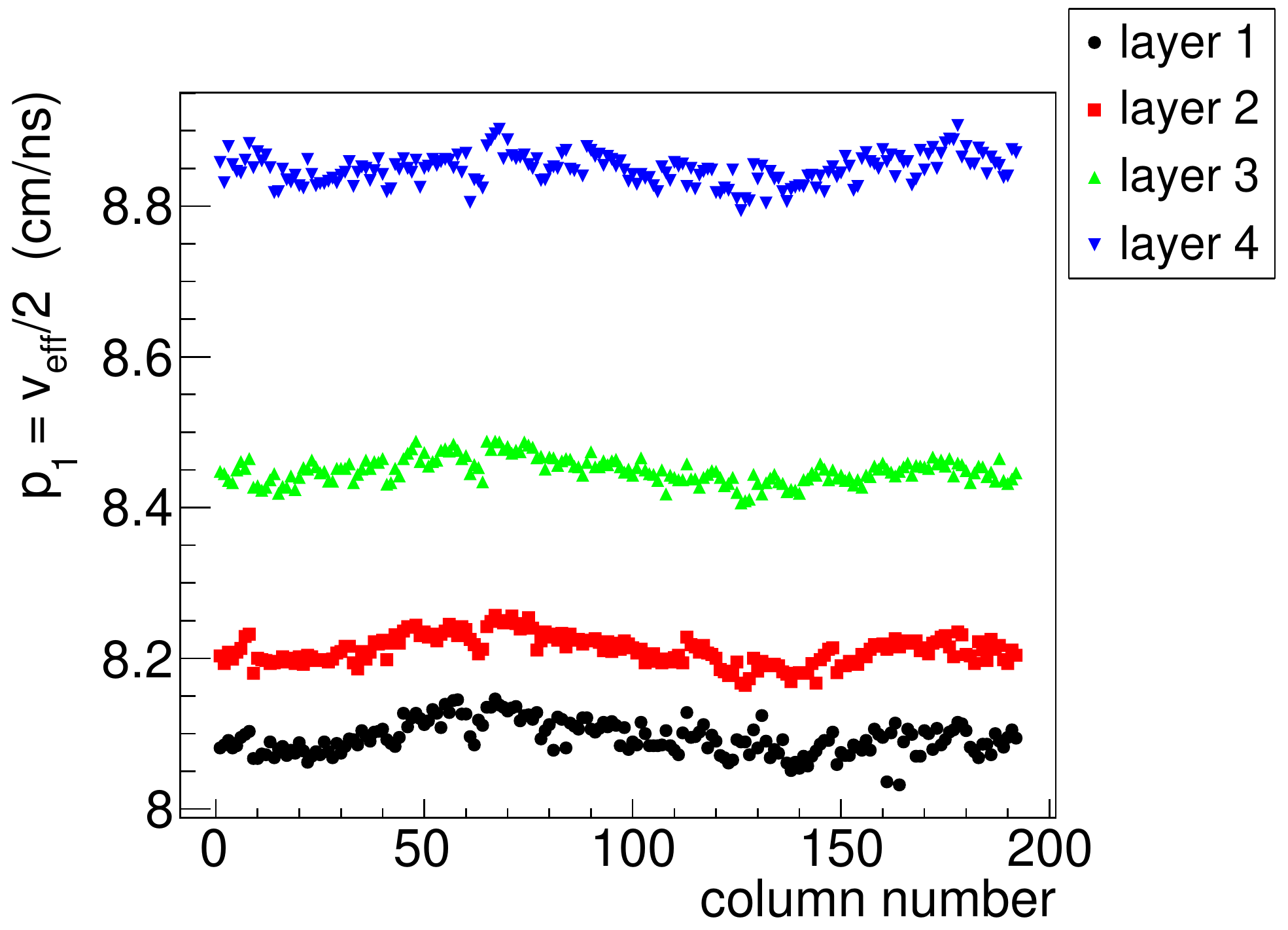}   
\includegraphics[width=0.48\textwidth]{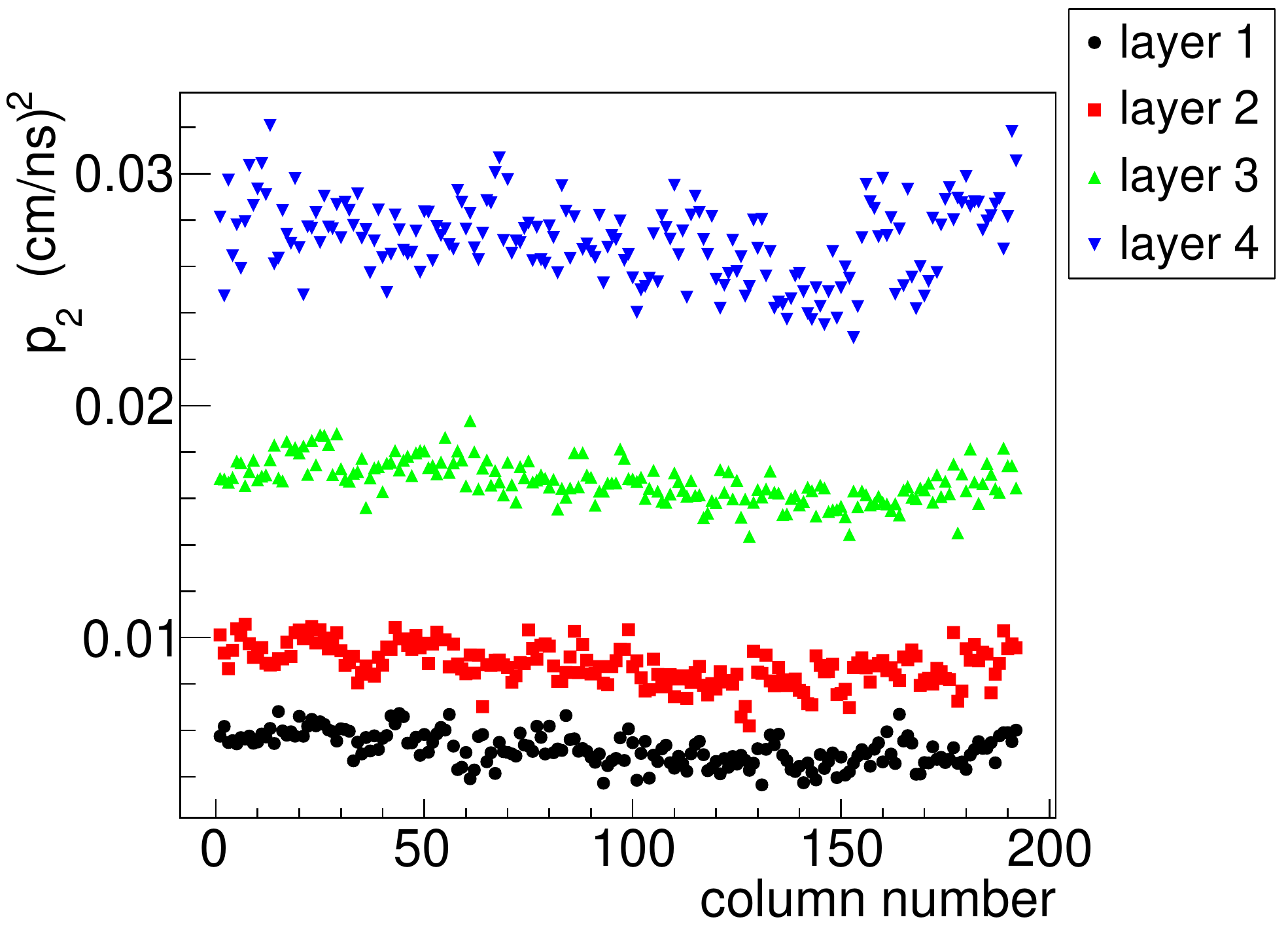}
\caption{\label{fig:linear_vs_quadratic_production_runs}   
Linear term in Eq.\,\ref{eq:ztime2}, $p_1$, representing half the effective speed of light for each BCAL channel (left) and a quadratic term, $p_2$,  (right).}
\end{figure}

Figure~\ref{fig:shower_in_bcal} illustrates the shower 
development of a particle that hits the BCAL surface. As the shower propagates through each layer the shower width grows larger. The timing signal that is recorded at the ends of each module
comes from the edges of the shower. This leads to a smaller effective length that the light has to travel before it reaches the ends of the module and this effect becomes stronger as the particle moves to 
subsequent layers, resulting in the layer-dependent values for the effective speed of light. This effect also depends on the polar angle $\theta$, which in turn introduces a $z$-dependence that 
requires introducing a quadratic term into Eq.\,\ref{eq:ztime}: 
\begin{eqnarray}
z & = & p_0 + p_1 \, (t_{\rm diff}) + p_2\, (t_{\rm diff})^2.   
\label{eq:ztime2} 
\end{eqnarray} 
We find that this form matches our data and is used to calibrate the determination of position from the time difference. The calibration constants are layer dependent as expected from the sketch in Fig.\,\ref{fig:shower_in_bcal}
and from Monte Carlo simulations. We extracted the effective speed of light for each channel, which is
represented by the linear coefficient, $v_{\rm eff} = 2 p_1$. The parameters determined by this procedure are shown in 
Fig.\,\ref{fig:linear_vs_quadratic_production_runs}. We do observe a small azimuthal asymmetry (represented by column number in the plot) of the parameters, but this dependence is attributed to our track reconstruction, since
the BCAL is azimuthally symmetric. 

\subsection{Attenuation Length and Gain Ratios}
\begin{figure}[tbp]
\centering
\includegraphics[width=0.48\linewidth]{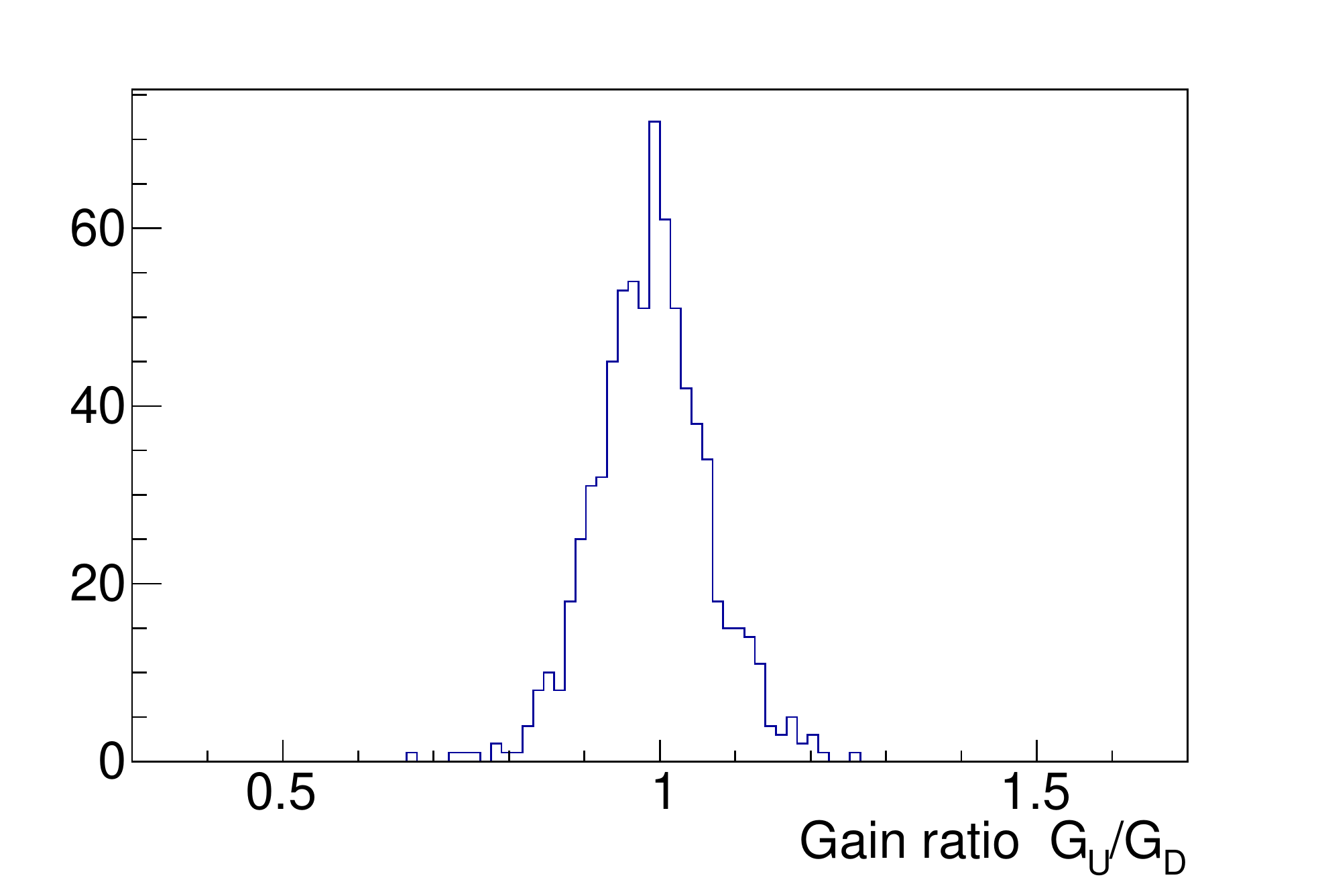}
\includegraphics[width=0.48\linewidth]{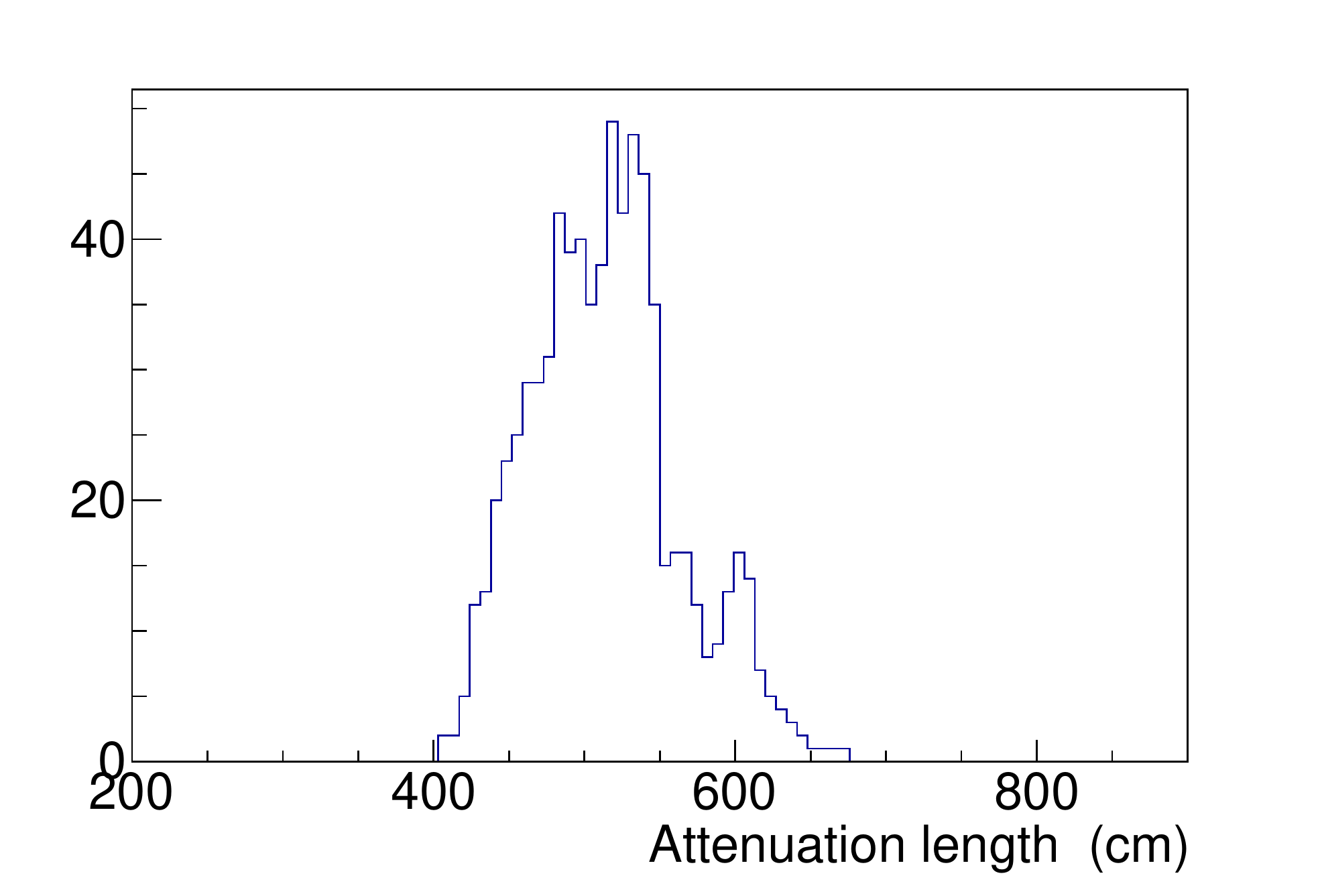}  
\caption{\label{fig:Lambda_GainRatios} 
Left) Distribution of measured upstream to downstream gain ratios in each cell. Right) Distribution of measured attenuation lengths in each cell. 
}
\end{figure}

We briefly outline some of the formalism needed to define the parameters that characterize the calorimeter.
Given a signal $A$ recorded in one of the FADCs, one may compute the amount of energy deposited at a position $z$  along the length of the module relative to the center as
\begin{eqnarray}
E_{U} & = &  g_{U}\, e^{(L/2+z)/\lambda} \, A_{U} \label{eq:upE} \\ 
E_{D} & = & g_{D}  \, e^{(L/2-z)/\lambda} \,  A_{D} \label{eq:downE} \\  
\overline{E} & = & \textstyle{\frac{1}{2}} (E_U + E_D).
\end{eqnarray}
The two measurements of the energy are obtained by using signals from the upstream and downstream ends separately, but each depends on the attenuation length, $\lambda$, the position of energy deposition, $z$,
and the gain of that channel, $1/g$. $L$ is the length of the module and the subscripts are used to indicate upstream or downstream ends. The usual estimator of the energy is given
by the average value $\overline{E}$. For the case of charged particles, one may determine the location of impact by 
tracking the particle to the calorimeter module. The position of energy depositions registered in the upstream and downstream sensors can be
determined from the timing information of the hits (see Section\,\ref{sec:postime}).
Using a sample of hits that illuminate the length of the BCAL, one may use the following equation (obtained by dividing Eq.\,\ref{eq:upE} by Eq.\,\ref{eq:downE},
which should be equal aside from fluctuations)  to determine the attenuation length $\lambda$ 
\begin{eqnarray}
\ln \left[\frac{A_{D}}{A_{U}} \right]  & = & \frac{2}{\lambda} z  - \ln \left[\frac{g_{D}}{g_{U}} \right],  \label{eq:zeq}
\end{eqnarray}
where the log of the ratio of pulse heights is a linear function of the position $z$ with a slope equal to $(\frac{2}{\lambda})$. The intercept 
can be used to determine the ratio of upstream to downstream gain factors $g_{U}/g_{D}$. The distribution of measured parameters is
plotted in Fig.\,\ref{fig:Lambda_GainRatios}. 

The measured attenuation lengths are systematically about 25\% longer than those determined from individual fibers \cite{Baulin201348}.
This is, at least in part, because the signals include reflections from the opposite end.
We expect a 10\% spread in attenuation lengths due to measured variation of the fibers themselves.
In addition, we find that the attenuation length systematically increases with layer number. 
This effect is expected for showers that propagate into deeper layers (see Fig.\,\ref{fig:shower_in_bcal}) and generate an extended source inside the module.

 \subsection{Shower Reconstruction \label{sec:showerreconstruction}}
 
 A particle incident on the BCAL will deposit energy in several neighboring BCAL cells. We have developed a clustering algorithm to identify which cells\footnote{Most of the energy is contained in double-ended hits, although single-ended hits are
 added to the total energy sum but are not used to compute the position or time of the cluster.}
  share the energy deposited by a single particle incident on the BCAL. The first step of the clustering algorithm is to search through all of the cells that contain energy deposition and find the highest energy cell. The highest energy cell is used to seed the first cluster if it is above the cluster threshold of 15 MeV.\footnote{The noise level is below the
 readout threshold of 2.2 MeV.}  Cells with deposited energy that are
in geometrical proximity to the cluster seed are added to the cluster based on how close and how much energy was deposited relative to the total cluster energy. Each time a cell is added to the cluster, the cluster's centroid position, time and energy are updated and that cell is removed from the available pool for other clusters. This procedure is repeated until all of the remaining hit cells are exhausted. In the next step, the clustering algorithm searches for the highest energy cell in the event that hasn't been included into a cluster already. If the highest energy cell remaining is above the cluster seed threshold then it will form a new cluster and the procedure is iterated.
 This will continue on until there are no more cells in the event that are above the cluster seed threshold.

After all clusters have been formed from all the hit cells, the clustering algorithm will check that each cell was placed in its  most appropriate cluster. The cell's most appropriate cluster is identified based on the cell's proximity and energy relative to its original cluster and all other clusters in the event, which may cause that cell to be reassigned. Typical sizes of clusters are 6$^\circ$ in the azimuthal angle, $\phi$, and 4$^\circ$ in the polar angle, $\theta$.
After this procedure is finished, the clustering algorithm will try to merge clusters that are very close together into a single cluster.

Once we have gathered cells into clusters, we combine the information from these cells to form ``showers." The energy of the shower is the sum of all of the energy deposited in each of the cells corresponding to that shower with
a non-linear correction (see Section\,\ref{sec:nonlinear}) applied to the entire cluster. The shower time is determined as the energy-square-weighted average of all the cell times in the shower. The distance from the target center to the shower centroid position and the shower's polar angle position with respect to the center of the target are also energy-square-weighted averages of the cell's respective quantities. The $\phi$ position of the shower is just the energy-weighted average of the $\phi$ position of each cell. The weighting scheme is empirical, but motivated by the fact that cells should contribute based on the size of the signal (energy) and based on the measurement accuracy that generally improves with energy except 
for the determination the azimuthal angle $\phi$.
The shower information is used to reconstruct a particle's four-vector to be used for physics analyses. 

\subsection{Gain Determination}

\subsubsection{Gain Coefficients from Cosmic Rays \label{sec:cosmicgains}}

\begin{figure}[tbp]\centering
 \includegraphics[width=0.4\linewidth]{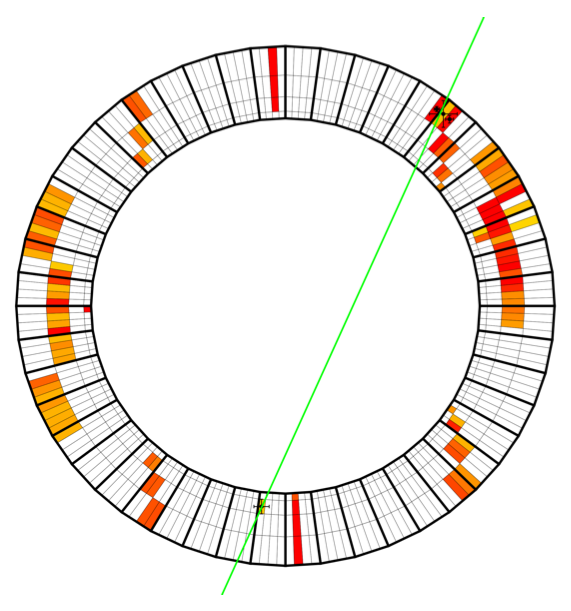}
 \includegraphics[width=0.4\linewidth]{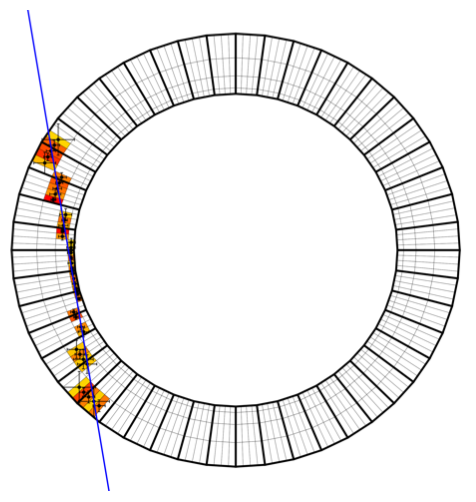}
\caption{\label{fig:cosmicsTracks}
  Cosmics events showing a ``noisy'' event (left) and ``clean'' event that fits the track profile (right).  The latter candidate events were used to extract the initial seed for the SiPM gain coefficients, which were refined subsequently by employing a $\pi^0$ width minimization algorithm. (Color online)
  }
\end{figure}

A starting point for the extraction of the gain factors was obtained from the energy deposition of cosmic-ray tracks traversing the detector with the magnetic field off. These depositions were constrained using fits to the cosmic track profile in the BCAL, the latter obtained by developing an algorithm to process self-triggering events in the BCAL.  The algorithm identified candidate tracks and rejected poorly reconstructed ones.  A sample of both kinds is shown in Fig.~\ref{fig:cosmicsTracks}.  The coefficients for all 1536 FADC readout channels were extracted and clustered around 0.042\,MeV/(integral ADC count).  Gain coefficients were determined in this way to a typical accuracy of 20\% for layers 1-3 and about 30\% for layer 4. This accuracy was sufficient as a starting point for the $\pi^0$ width minimization scheme as described below. We also note that very few $\pi^0$ showers
penetrate into layer 4, so the cosmic-ray method  is used to obtain the gain calibrations for this layer.




\subsubsection{Gain Calibration using $\pi^0$'s \label{sec:gaincalibration}}
The determination of the gains of individual channels is obtained using a high-statistic sample of $\pi^{0}\rightarrow\gamma\gamma$ using the full range of incident photons between 3 and 12 GeV. Assuming that two detected photons come from the decay of a single particle, then the decaying particle's invariant mass squared is derived from the energy-momentum relation:
\begin{equation}
m^{2}=E^{2}-|\vec{p}|^{2}.
\label{eq:EpRelation}
\end{equation}
This can be rewritten in a form that contains only the reconstructed shower quantities:
\begin{equation}
m^{2}= 2E_{1} E_{2} (1-\cos\psi).
\label{eq:invMass}
\end{equation}
For the $\pi^{0}\rightarrow\gamma\gamma$ decay, $E_1$ and $E_2$ are the energies of two photon showers in the BCAL and $\psi$ is the angle between them in the plane defined by their momentum vectors. The angle is determined using the reconstructed shower positions,  obtained from BCAL time measurements (Section\,\ref{sec:postime}),  and the vertex as determined from the charged tracks in the event.

\begin{figure}[tbp]\centering
\includegraphics[width=0.9\textwidth]{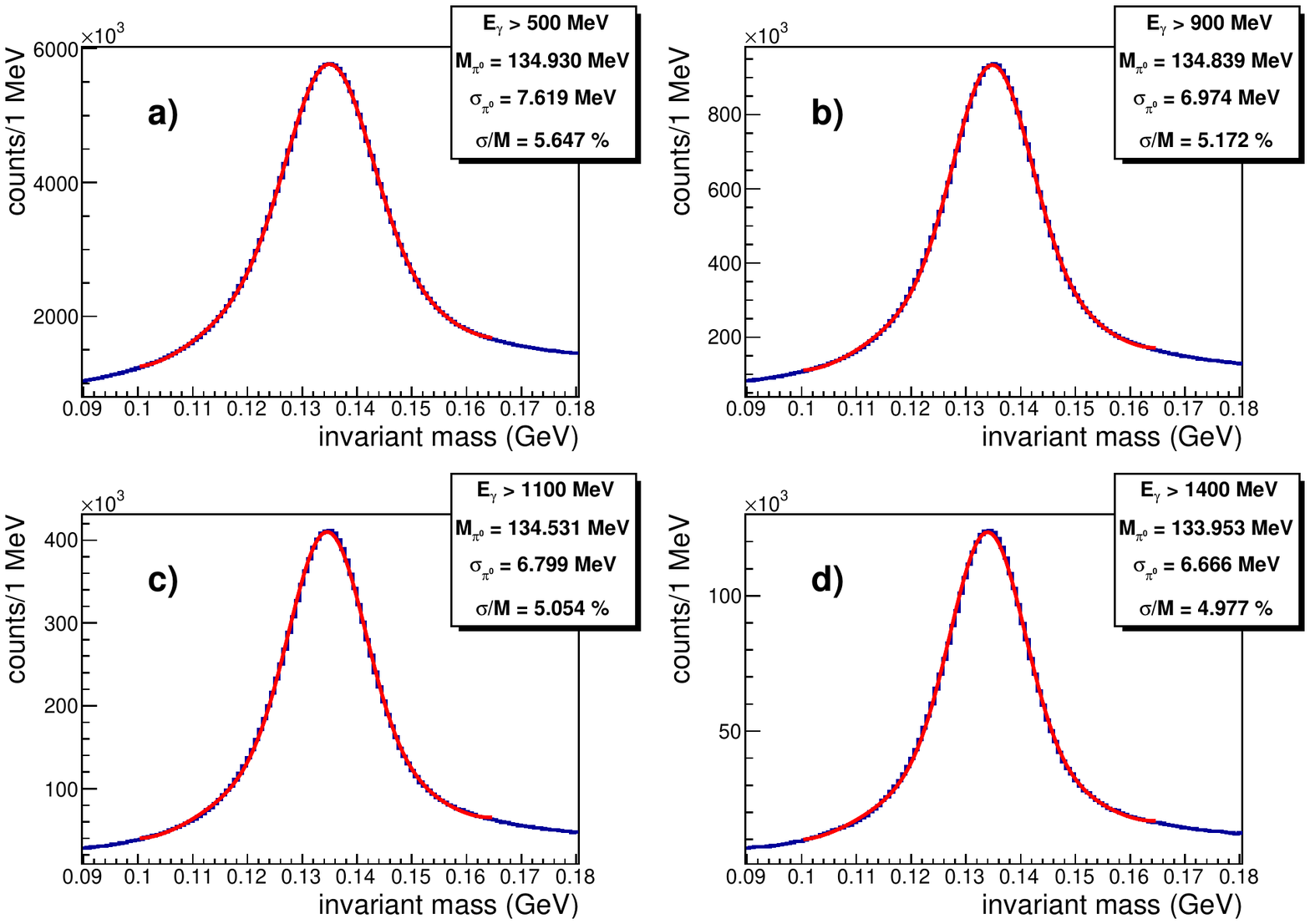}
\caption{\label{fig:pi0_distributions}
  Two-photon mass distributions for four different energy cuts after completing the calibration procedure.
 }   
\end{figure}

We selected events that have at least two charged particles and two photons that reconstruct close to the $\pi^0$ mass. The charged particle tracks were required to be able to determine the event vertex position. Since the BCAL has an angular coverage from $11^{o} < \theta < 126^{o}$ with respect to the target position, the $\pi^{0}$ mass distribution peak and width depend heavily on the location of the interaction. We reconstruct the four-momentum vectors of the photons fixing the event vertex position as their point of origin. 
The gain constants were determined for periods that had uniform experimental conditions and collected sufficient statistics to determine all gain constants. 
These calibration periods typically lasted several weeks.
We adjusted the minimum shower energy ($\sim$\,0.3\,GeV) that was included into the calibration sample in order to ensure we had a 
clean $\pi^0$ sample. This resulted in a signal-to-background ratio of greater than 4 to 1. Typical distributions of the two-photon invariant mass spectrum after the calibration converged are shown in Fig.\,\ref{fig:pi0_distributions}.  

\begin{figure}[tbh]\centering
 \includegraphics[scale=0.3,angle=0.]{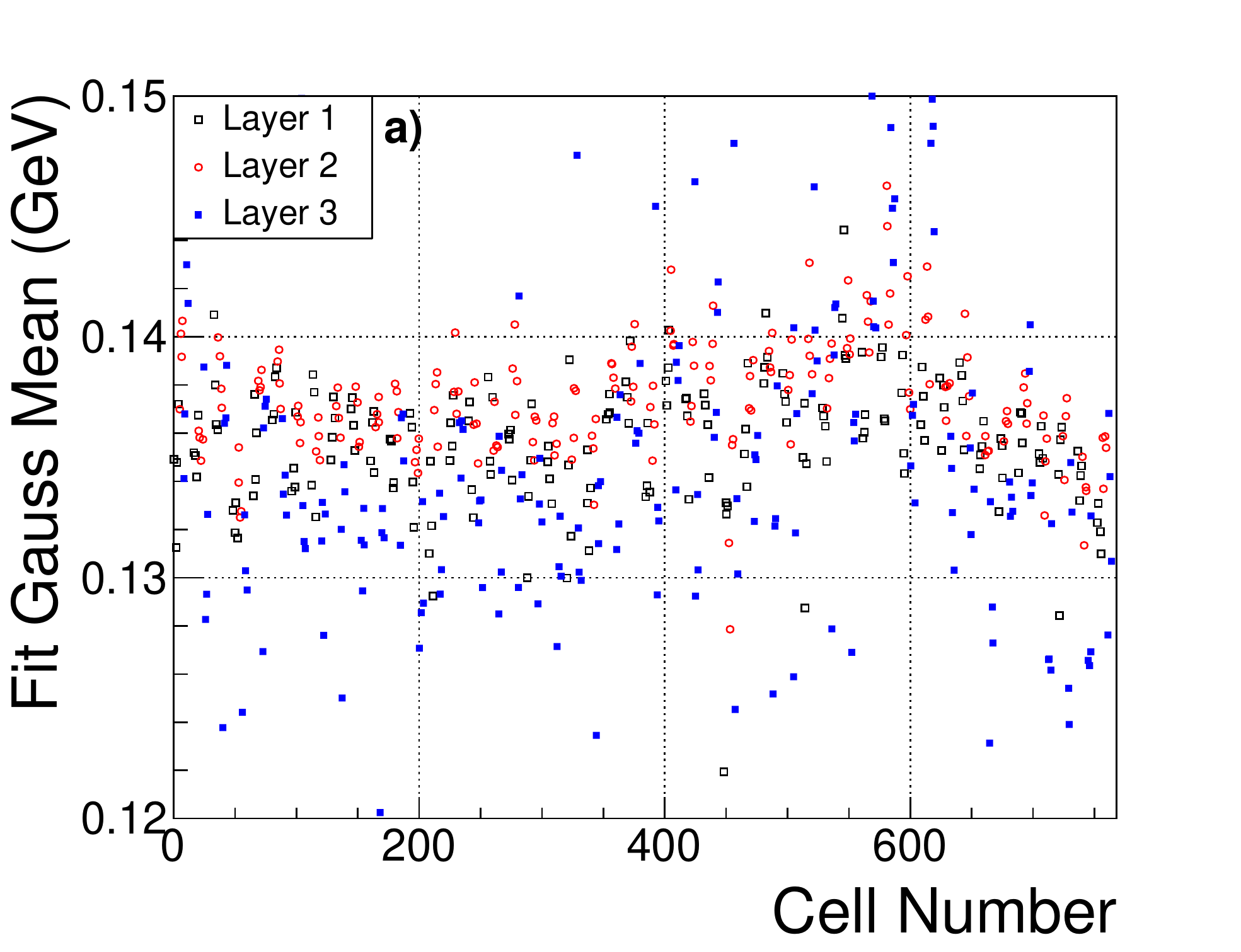}     
 \includegraphics[scale=0.3,angle=0.]{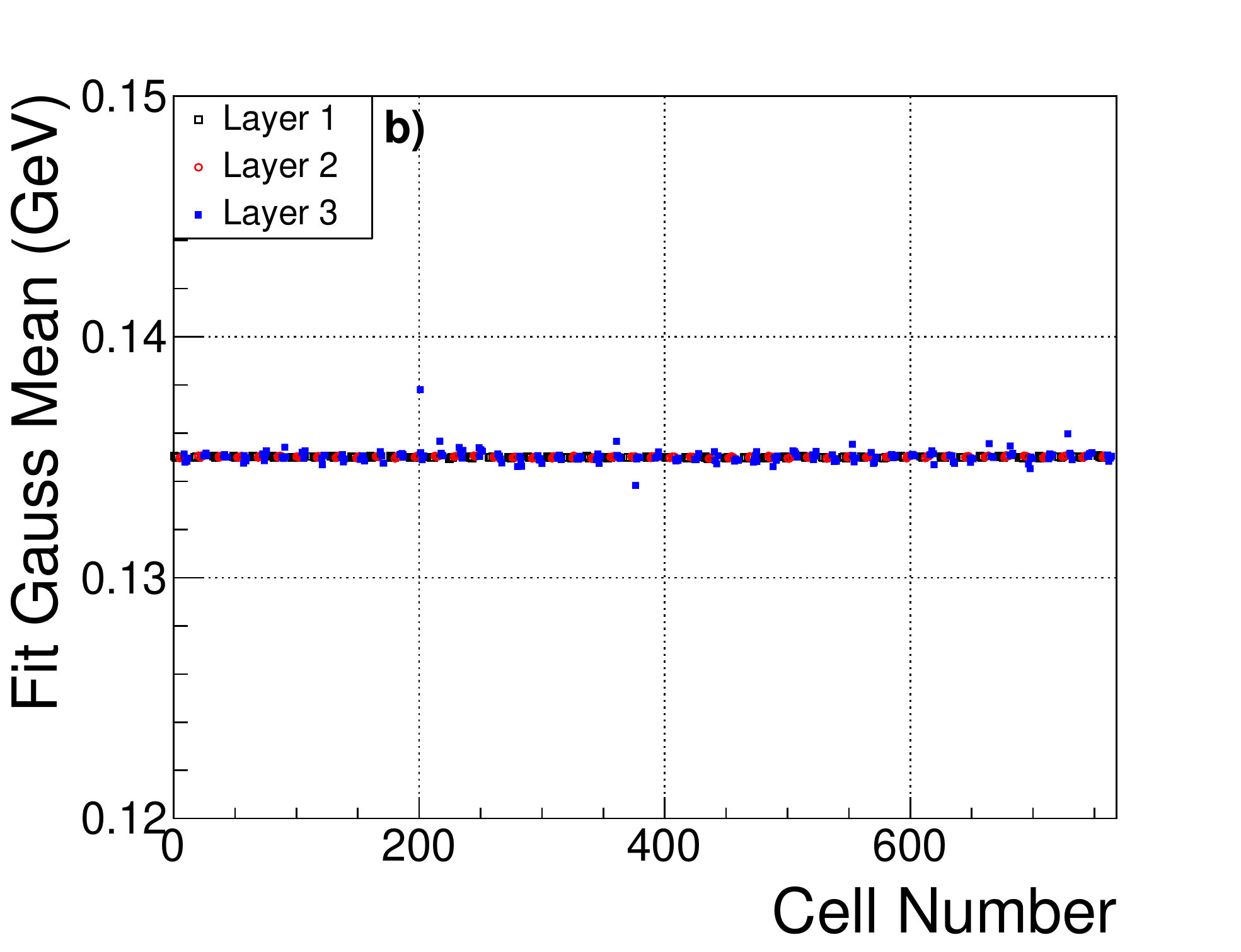}
\caption{\label{fig:newGainCalibCompare} $\pi^{0}$ fit mean values as a function of cell number a) before and b) after applying the calibration procedure. (Color online)}   
\end{figure}

The calibration is an iterative method that adjusts individual cells only when they contribute significantly to the shower energy. For layers 1 and 2 we compute the two-photon invariant mass when a particular cell 
contributes more than half of the total energy of its shower. 
We histogram all such events and fit the mass distribution with a Gaussian for each cell. We then compute the ratio of the known $\pi^{0}$ mass and the fitted $\pi^{0}$ peak position and apply this ratio to the gain of the corresponding cell. 
This procedure moves the reconstructed $\pi^{0}$ mass peak position closer to the known $\pi^{0}$ mass and is applied iteratively until the reconstructed $\pi^{0}$ peak positions are all aligned with the known $\pi^{0}$ mass for each cell. 
This method is less efficient for the third layer of the BCAL, where we relax the minimal cell contribution to 40\% of the total shower energy in order to acquire sufficient statistics for reliable fits. 
For the fourth layer where $\pi^{0}$ decays 
don't deposit sufficient energy in a single cell, we use the cosmic-ray determination of the gain (Section\,\ref{sec:cosmicgains}). The fitted peaks of the $\pi^0$ masses are shown in Fig.\,\ref{fig:newGainCalibCompare}
before and after the calibration procedure.  We note that the light output of the fibers varies by about 7\% \cite{Baulin201348}.
Following the final iteration, the calibrated $\pi^0$ mass peaks are less than 1~MeV away from their true value.

\subsubsection{Non-linear Corrections \label{sec:nonlinear}}

\begin{figure}[tbp]\centering   
 \includegraphics[scale=0.2,angle=0.,width=0.45\textwidth]{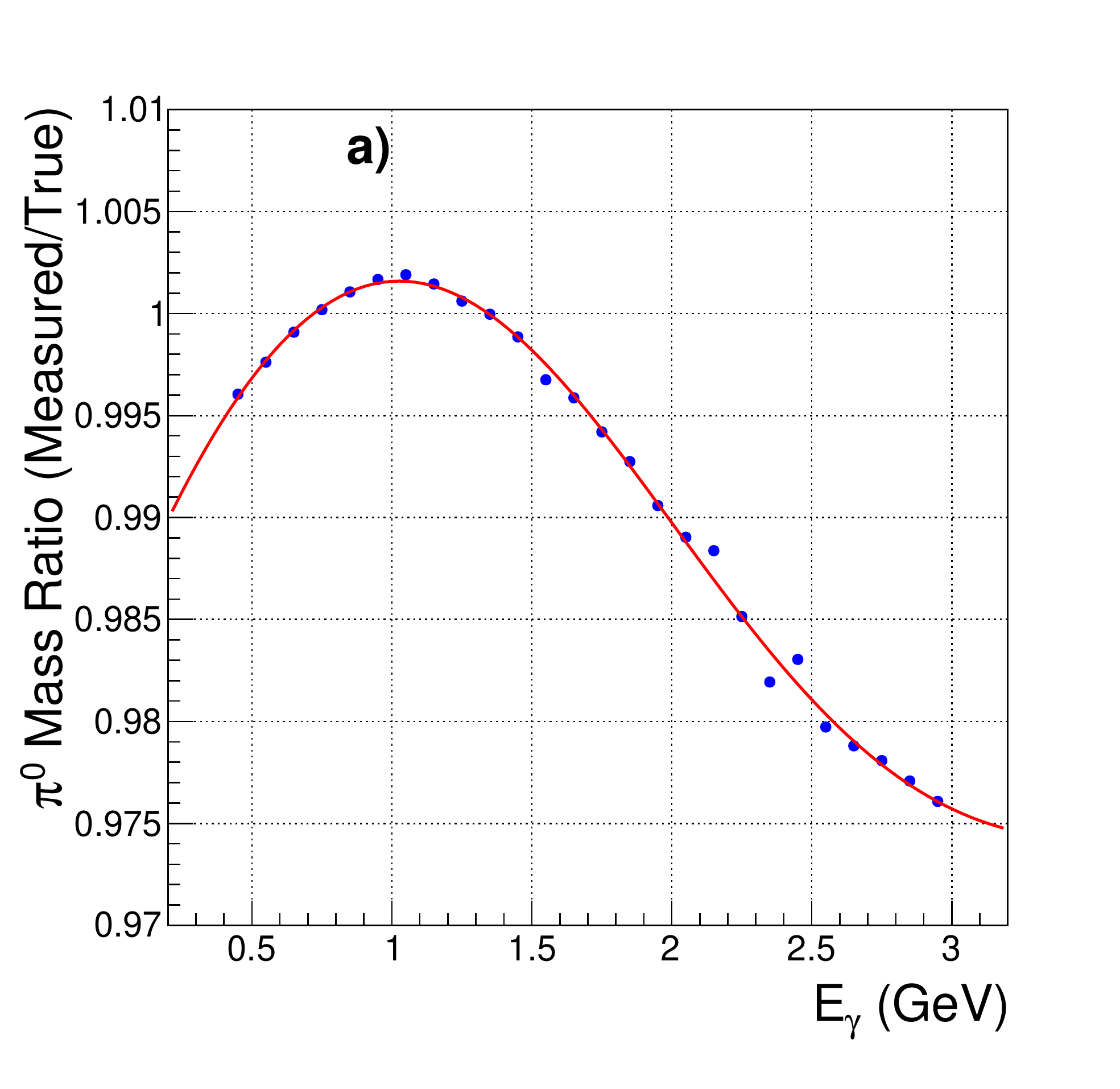}   
 \includegraphics[scale=0.2,angle=0.,width=0.45\textwidth]{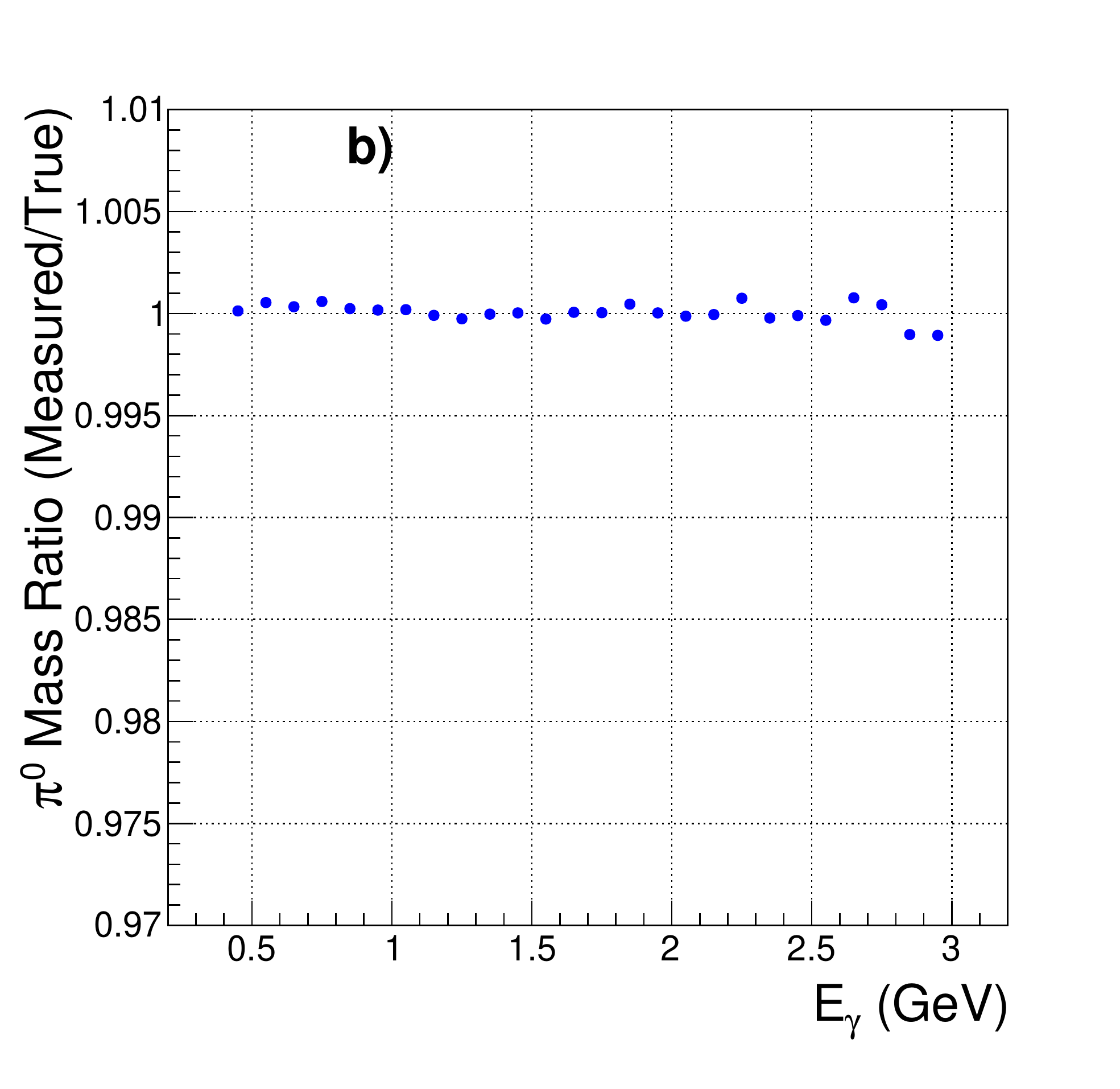}   
\caption{ \label{fig:nonlinb4correct}
a) The Gaussian mean fit to the $\pi^{0}$ mass distribution peak divided by the known $\pi^{0}$ mass value as a function of the 2$\gamma$ average energy when both showers are within 100 MeV of each other. The red line represents the function that we use to correct the reconstructed single photon shower energy. b) Ratio of the corrected $\pi^0$ mass distribution peak to the true value after applying the correction. (Color online)}
\end{figure}

After the gain calibration is complete, non-linearities are observed as an energy dependence of the reconstructed $\pi^0$ mass. At low energy the non-linearity is due to threshold effects 
and at high energies it arises from saturation, either in the SiPMs or the FADCs, or both.
The pixelation of the SiPMs is an inherent source of non-linearity (see Section\,\ref{sec:photosensors}) since each pixel outputs only one pulse that does not change with the number of incident photons. 
We estimate that a pulse with a peak height of 4000 counts ($\sim$2\,GeV) would fire about 20000 pixels. We have not made any explicit correction for this effect in our data. 
Saturated pulses in the FADCs have been corrected by assuming a constant pulse shape to normalize the saturated pulse integral 
(see Section\,\ref{sec:adcwaveforms}). Therefore some non-linearities remain and again we use a $\pi^{0}$ sample to determine an average value of this correction. We use the change of 
the  $\pi^{0}\rightarrow\gamma\gamma$ invariant mass to make a correction to each single photon shower, therefore we require that both showers be within 100~MeV of each other. The relative mean as a function of energy and the function that we use to correct the reconstructed single photon shower energy are shown in  Fig.\,\ref{fig:nonlinb4correct}. 
The measured $\pi^0$ widths, after all corrections, are tabulated in Table\,\ref{tab:pi0Widths}, and for photons above 1.1 GeV the measured $\pi^0$ width is 6.8 MeV.
\begin{table}[!ht]
\centering
\caption[pi0Widths]{Widths extracted from a gaussian fit of the $\pi^{0}$ mass distribution for various shower energy cuts after the gain calibration was completed.
}
\begin{tabular}[\linewidth]{|c|c|c|c|}
\hline
\hline
Shower Energy Cut (MeV)		&$\pi^{0}$ Gaussian width	 (MeV)	  	\\ \hline
E $>$ 500 	&7.62  \\ \hline
E $>$ 900     &6.97  \\ \hline
E $>$ 1100   &6.80  \\ \hline
E $>$ 1400  &6.67  \\ \hline
\hline
\end{tabular}
\label{tab:pi0Widths}
\end{table}

\subsection{Layer Efficiencies}
The efficiency for observing a signal in a given layer due to a charged track hitting the BCAL was studied with a sample of charged tracks. The event selection required a track candidate with sufficient 
number of hits in the drift chambers 
to reconstruct a track with a good figure-of-merit and a matched shower in the BCAL. We note that the sample includes non-interacting tracks 
(e.g. muons) as well as particles that produce hadronic showers inside the BCAL.
The efficiency of a given layer was checked only when energy was observed in a layer beyond
the one in question. For example layer 2 was checked only when there were hits in either layer 3 or 4 to ensure that the shower penetrated beyond layer 2. A match in the layer under study was obtained when a hit was observed
in the layer within 3 columns of the extrapolated track position to that layer. 
The computed efficiencies include a small correction for accidental hits. 
 
\begin{figure}[thp]\centering
\includegraphics[width=0.9\textwidth]{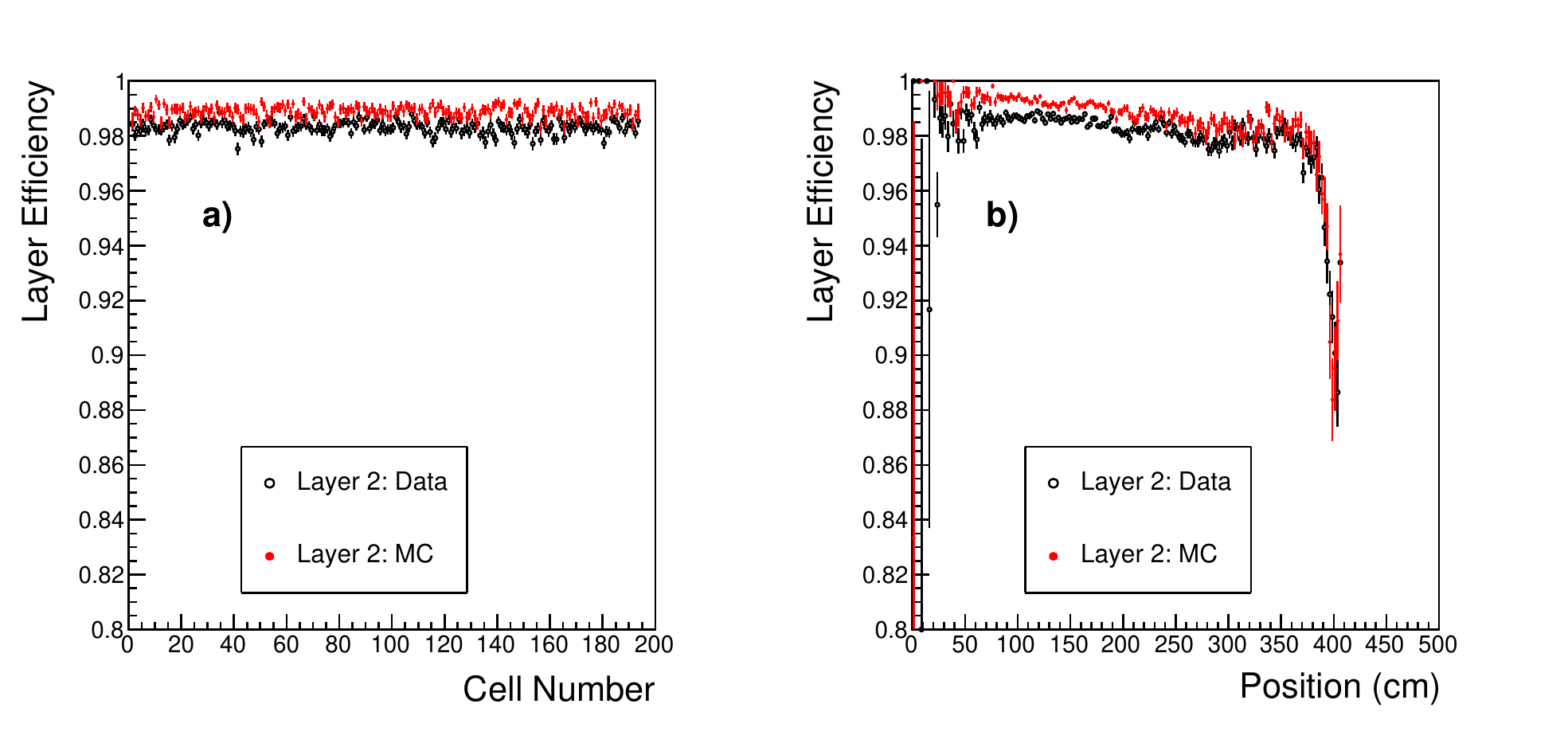}
\caption{\label{fig:NIM_hadronic_eff}
a) Layer 2 efficiency as a function of cell number. b) Layer 2 efficiency as a function of position along the BCAL module.
(Color online)   
  }   
\end{figure}

We note that this calculation does not give the efficiency for finding a charged particle signal. Rather, it gives the efficiency of finding a hit in a given layer when there is an expected
signal at that projected location. An example of the calculated efficiency
is shown in Fig.\,\ref{fig:NIM_hadronic_eff} for layer 2. The figure also shows the agreement with the expectation from Monte Carlo. 
A systematic source of inefficiency can be seen in the figure showing the efficiency as a function of the position along the length of the module, $z$. At the downstream end of the module there is a 
drop in efficiency due to showers that are not fully contained in the calorimeter. 
The averages for the calculated layer efficiencies are about 96.8\% for layers 1 and 3 and 98.4\% for layer 2. 

\section{Monitoring \label{sec:monitoring}}

\subsection{LED Pulser System}
\label{sec:ledpulsersystem}

\begin{figure}[thp]\centering
\includegraphics[width=0.5\textwidth]{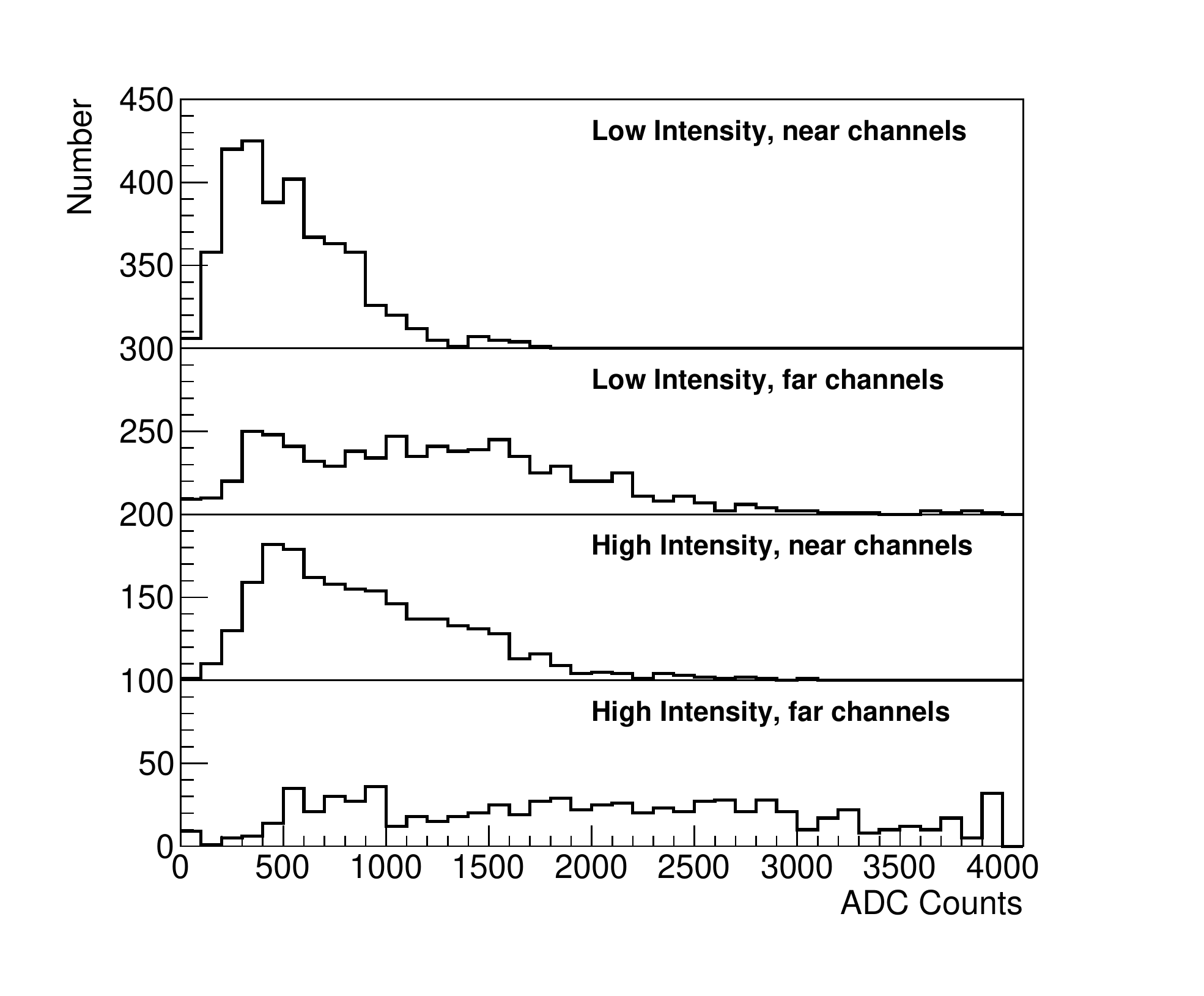}
\caption{\label{fig:bcalleddist_upstream}
Response of all cells to the upstream LEDs for two different bias settings. 
The distributions are offset vertically for clarity. The wide variation of the 
response from cell to cell is due to the high sensitivity of light collection to the precise placement of the LED onto the light guide. 
The response of each cell is tracked relative to its average over the run period.
  }  
\end{figure}

The relative gain of the BCAL photosensors is monitored using a modular LED-driver system whose design has been described elsewhere~\cite{Anassontzis201441}. In essence, 
all ten LEDs in a column are connected to a flexible bus or ``string," which pulses them at once; there are 384 such strings. The arrangement can be seen in  Fig.\,\ref{fig:ModPhotos}.   
The light intensities of the LEDs are controlled by a preset bias, which is established before issuing a trigger pulse. 
The illumination of each SiPM varies considerably from channel to channel due to its sensitivity to precise positioning of the LED onto the light guide. This is illustrated in Fig.\,\ref{fig:bcalleddist_upstream},
which displays the response of all cells to the two light intensities used during production. The response of sensors near the LED is shown separately from 
sensors located on the far side of the BCAL. 

\begin{figure}[tbp]
\centering
\includegraphics[width=0.49\textwidth]{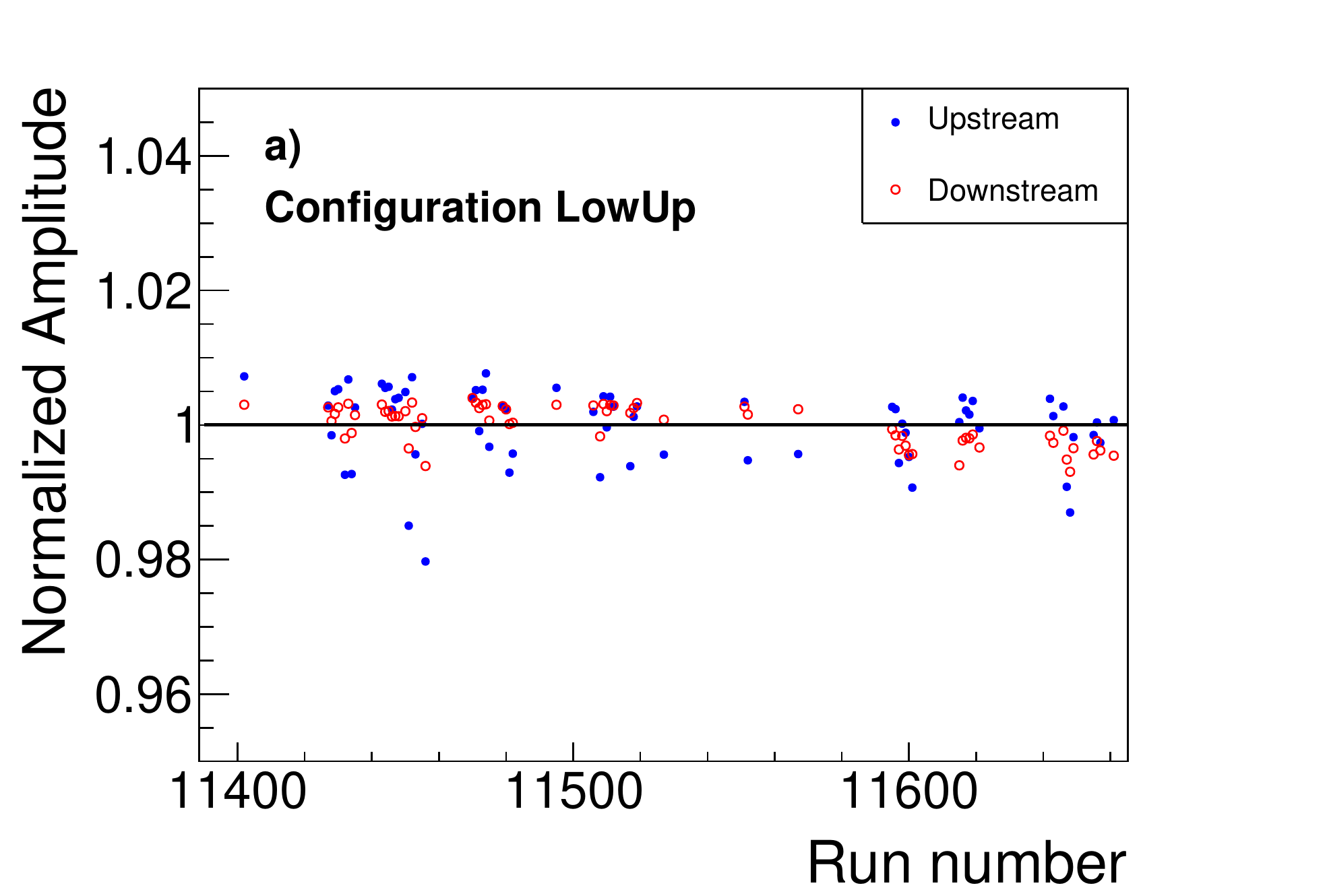}
\includegraphics[width=0.49\textwidth]{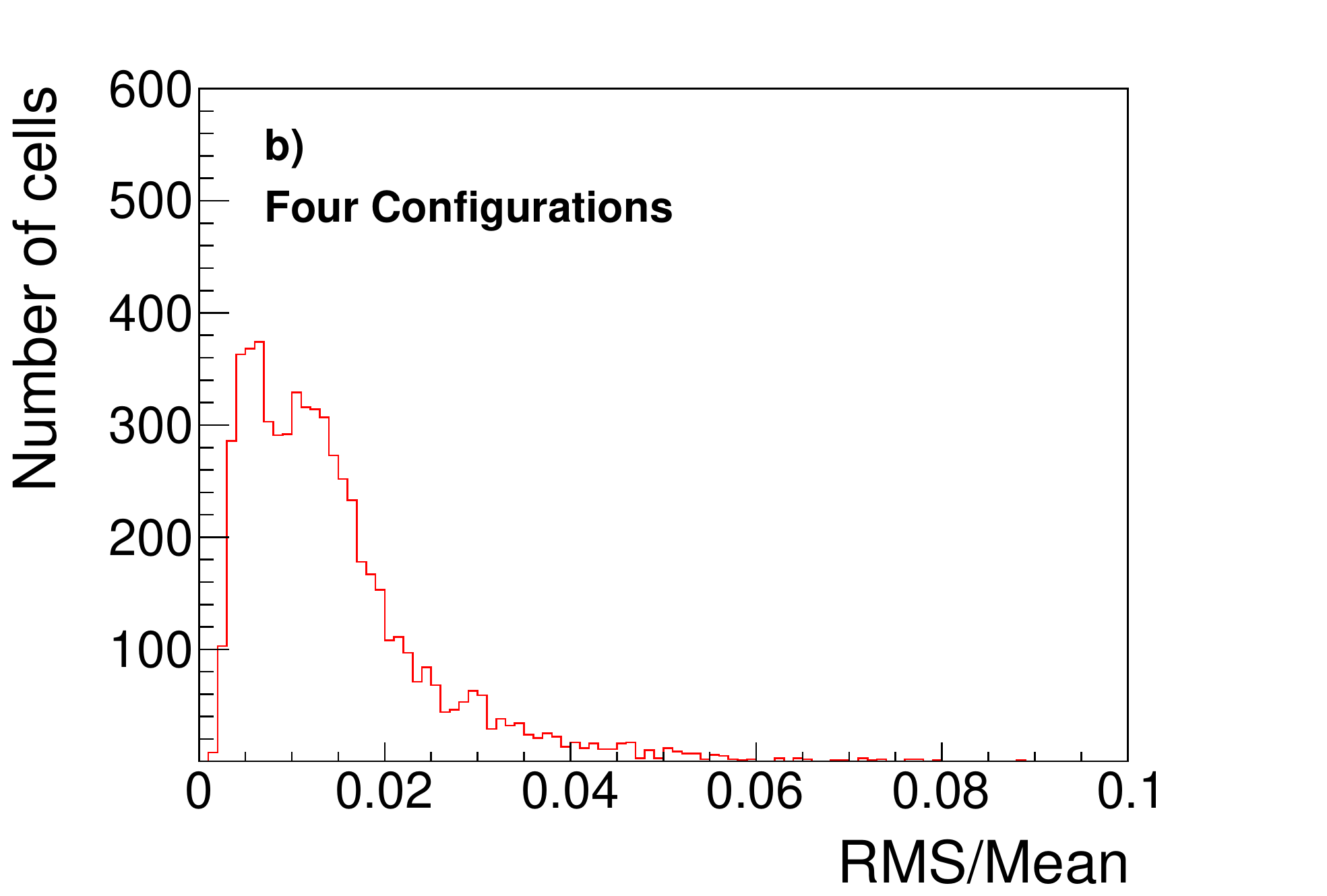}
\caption{\label{fig:LED_study}
a) Average normalized amplitude as a function of run number for all upstream and downstream cells in response to the upstream LEDs at low intensity.
b) Fractional root-mean-square deviation of the mean of each cell for one run from its average over the run period. All four pulsing configurations are included in the plot.  (Color online)
  }
\end{figure}

The LED monitoring system has two important applications.  First, it allows checking the response of individual SiPM sensors during special runs. In this application, individual sensors are enabled using the combined granularity of the LED pulsing system, which is column based, and the bias distribution system, which is arranged in rows. A particular combination of column and row selects a single sensor from a summed output, 
which normally includes more than one sensor. This test is conducted at the start of each data-taking period to verify that all sensors are operational. 

Second, during production runs ($\sim$ 1-2 h), all SiPM bias lines are enabled and we continuously monitor the response of the BCAL as physics data are being collected. In this mode we cycle through 16 different 
configurations,\footnote{Each cell is illuminated during four configurations.} each for one minute,
pulsing the LEDs at 10 Hz. The configurations differ by which LEDs fire (upstream or downstream), which columns are enabled (1-4), and the LED bias.
Two LED intensities were used, corresponding to low intensity (6.0 V) and high intensity (6.25 V). 
The large variation in response from cell to cell requires two different LED bias settings to ensure that all cells record signals in the central range of their FADC dynamic range. 
The LED signals are recorded by the light sensors on the near and far ends of the same cell, which can be subsequently compared.\footnote{There is some optical cross talk between 
cells ($\sim$ 2\% of the direct signal size) that results in cells of adjacent columns firing.}
Temperature drifts, which affect both LED light output and sensor response, can be tracked because the upstream and downstream sensors are hooked up to independent cooling systems.

In Fig.\,\ref{fig:LED_study}a we plot the normalized amplitude, averaged over all upstream and downstream cells as a function of run number. The amplitudes are normalized to their averages over the run period and
the plot is made for one of the four configurations. The average stability of the detector is better than 1\% spanning period of about ten days. 
The cells are studied by taking 
the fractional root-mean-square (RMS) deviations of the mean for each cell during a single run from the average over the run period are plotted in Fig.\,\ref{fig:LED_study}b. All four pulser
configurations are included in the plot. The fractional deviations are typically less than 2\%, which are due to a combination of changes in the monitoring system and the sensor gains.
To date we have not adjusted any gain constants using the feedback from the monitoring system, but studies 
are underway to evaluate how best to use these results in the analysis.

\section{Performance \label{sec:performance}}
In this section we discuss some BCAL reconstruction quantities during the data production running from 2016 and 2017.  There are no significant differences between the two run periods and in the discussion below data sets are 
combined when needed. The data were collected using a coherent bremsstrahlung beam \cite{AlGhoul:2017nbp} with the main coherent peak edge set at approximately 
9\,GeV\,\footnote{The electron beam energy
in 2016 was 12.1 GeV and in 2017 it was 11.6 GeV.} and a loose trigger programmed to select all hadronic interactions. 
Data were recorded with typical trigger rates of 20 KHz (2016), 25 KHz (low intensity in 2017)  and 50 KHz (high intensity 2017). Information from the
tagging spectrometer recorded incident photon energies in the range between 25\% and 95\% of the electron-beam energy.  The pulse structure of the machine is synchronized by the 250 MHz radio frequency (RF) clock.

\subsection{Time Resolution \label{sec:timeres}}

\begin{figure}[thp]\centering
 \includegraphics[width=0.45\textwidth,page=2]{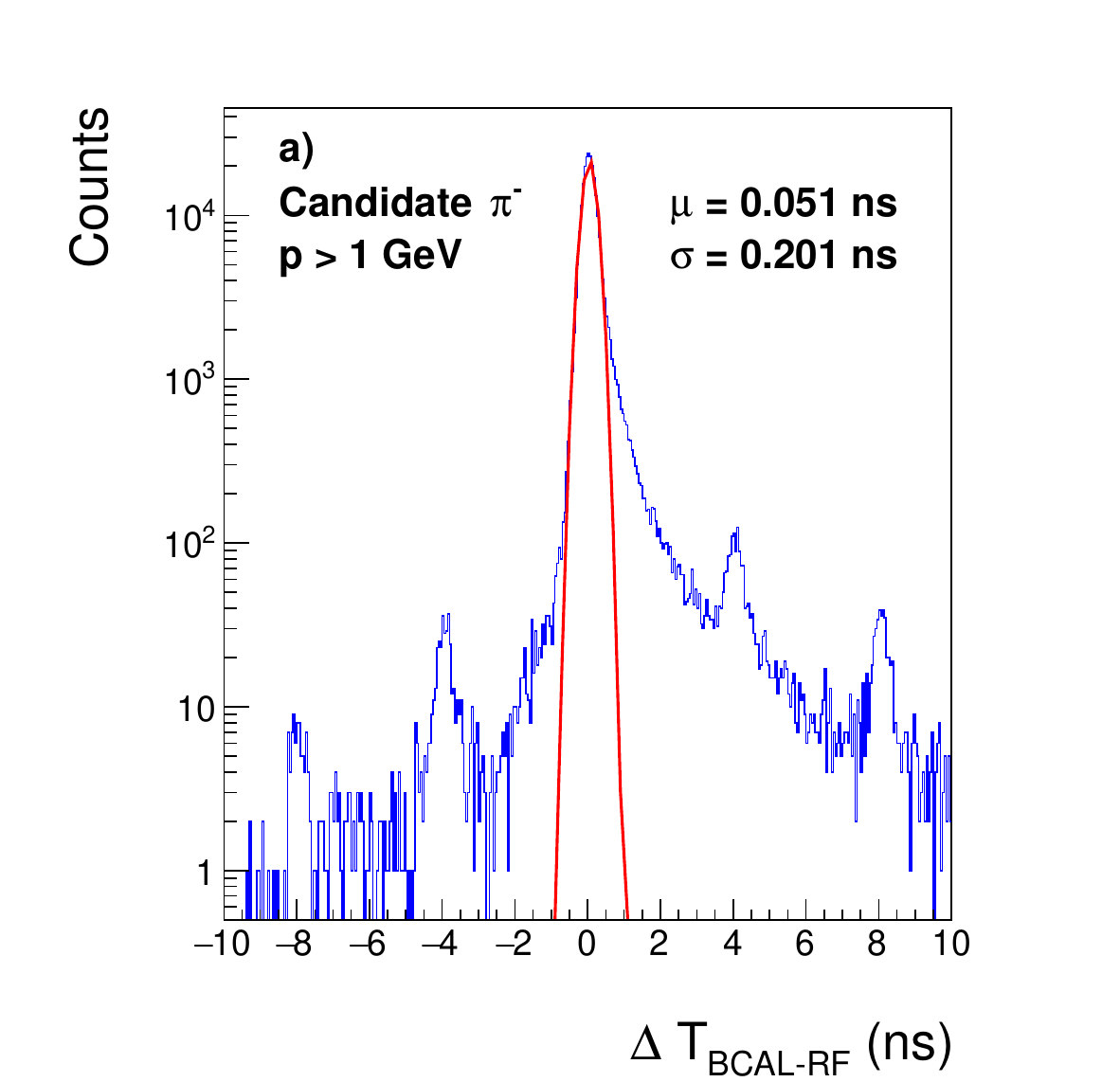}
 \includegraphics[width=0.45\textwidth,page=3]{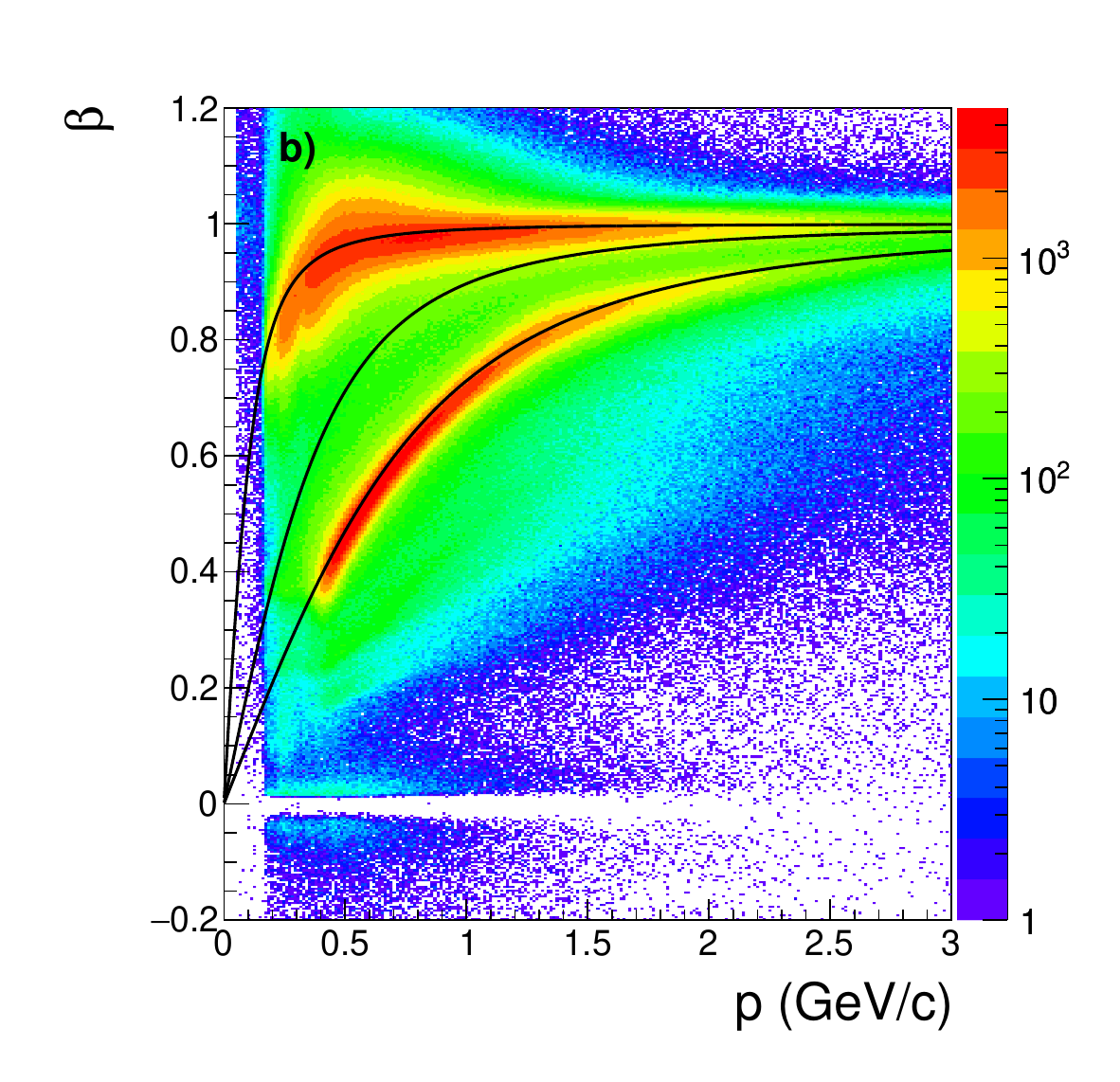}
\caption{ \label{fig:BCAL_pid}
a) Time difference between measured hits in the BCAL and the expectation based on the RF times for a sample of candidate negative pions with momentum greater than 1 GeV.  
The peaks at 4 ns intervals correspond to selecting the wrong RF bunch. 
b) Measured speed using BCAL hit times vs. the measured momentum for positive particles. The three curves correspond to the expected response of pions, kaons and protons.  (Color online)}
\end{figure}

The timing resolution directly affects the ability of the  detector to determine the velocity of charged particles, as shown in Fig.\,\ref{fig:BCAL_pid}. 
Fig.\,\ref{fig:BCAL_pid}a shows the difference from the BCAL shower time compared to the time calculated using the momentum and trajectories determined from tracking assuming they are 
pions. The distributions are plotted for a sample of candidate negative pions from a sample of $\gamma p \rightarrow p \pi^+ \pi^-$ events. The event time is determined by selecting the 
accelerator radio frequency (RF) bunch that best matches the track times
of all particles. At this stage of the reconstruction one finds that occasionally the wrong RF bunch was selected, which produces satellite peaks at 4 ns intervals. 
The fitted width of the central peak is $\sigma$\,=\,201 ps, which includes variations between modules as well as contributions from tracking. Using the flight time and the path length 
for positive tracks in the sample, one can compute
the particle speed and plot it vs momentum (Fig.\,\ref{fig:BCAL_pid}b). Clear bands corresponding to pions and protons are seen in the figure along with the expected curves for pions, kaons and protons.

\begin{figure}[thp]\centering
 \includegraphics[width=0.6\textwidth]{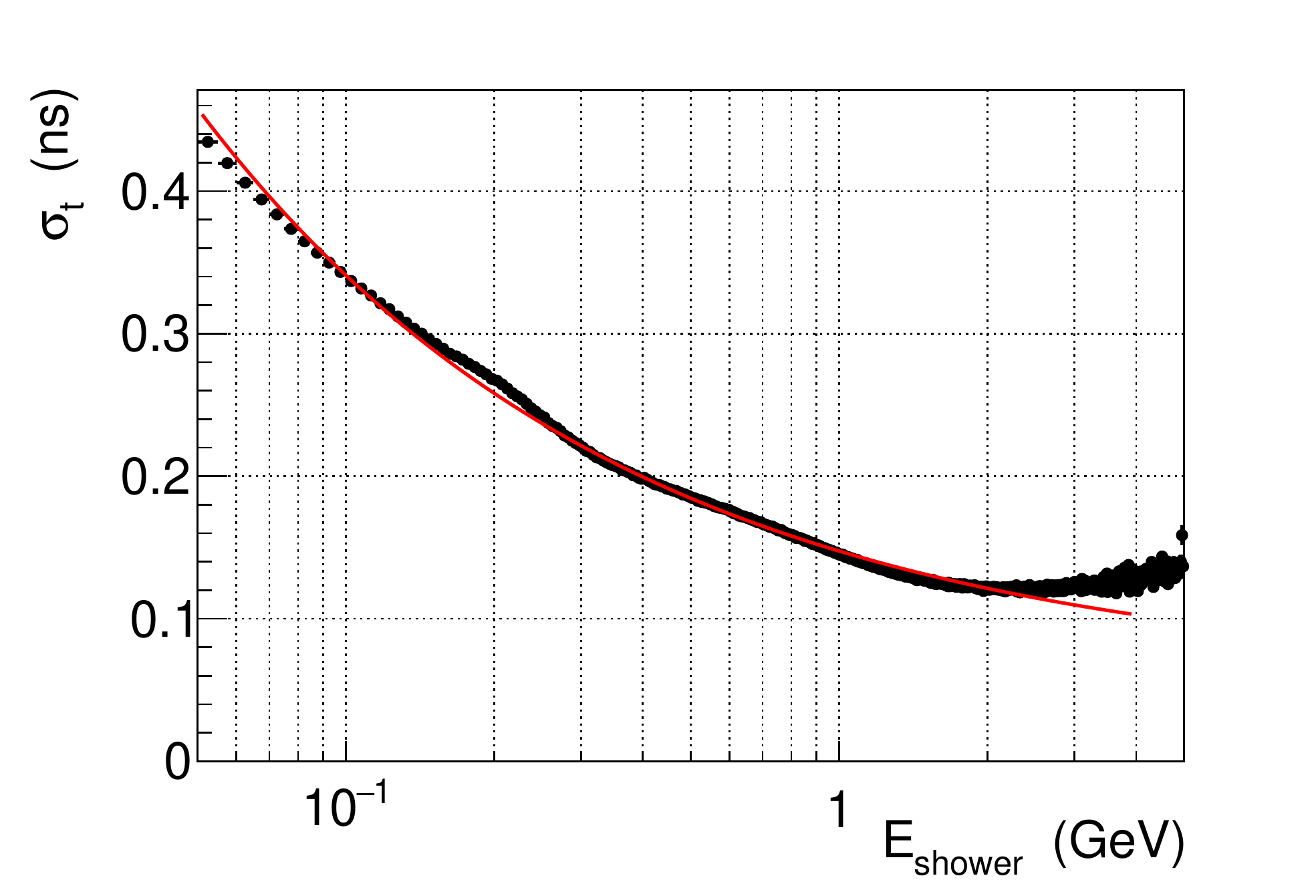}
\caption{ \label{fig:NIM_shower_sigma_Spring17_mod}
Time resolution for neutral showers as a function of shower energy. The curve is a fit to Eq.\,\ref{eq:Tresolution} with parameters c = 0.089 $\sqrt{\rm{GeV}}$ns and d = 0.058 ns, which gives a good qualitative description of
the data between 0.08 and 2 GeV. (Color online)
}
\end{figure}

The time resolution for neutral showers, i.e. photons, can be determined in a similar way for a sample of showers that are not matched to charged particles. 
We use a sample of events that contain at least two charged particles to determine the vertex of the event and additional showers that are not matched to any charged particle. 
For each unmatched shower, the BCAL shower time (Section \ref{sec:showerreconstruction})  is projected back to the event vertex assuming photons traveling in straight lines at the speed of light. 
The RF bunch for the event is chosen using all the charged tracks in an event.  The difference between the propagated BCAL time and the RF time is a measure of the time resolution for neutral showers. 
While this quantity  is dominated by 
the mean time resolution of the BCAL, it also includes other factors that affect the measurement of the shower position. 
The time resolution determined in this way is plotted as a function of the shower energy in Fig.\,\ref{fig:NIM_shower_sigma_Spring17_mod}. At 1 GeV,  $\sigma$\,=\,150\,ps.
The time resolution for photons  is somewhat better than for charged particles. 
This is partially due to the curvature of charged tracks that affects the time distribution of hits in the shower and to the lower deposited energy of non-interacting tracks ($\sim$ 0.23 GeV). 
Currently, the time of the shower is determined from the energy-weighted average time of points. This may not be the optimal strategy, especially for charged particles. 

\subsection{Mass Resolution}
\begin{figure}[thp]\centering
 \includegraphics[page=1,width=0.45\textwidth,clip=true]{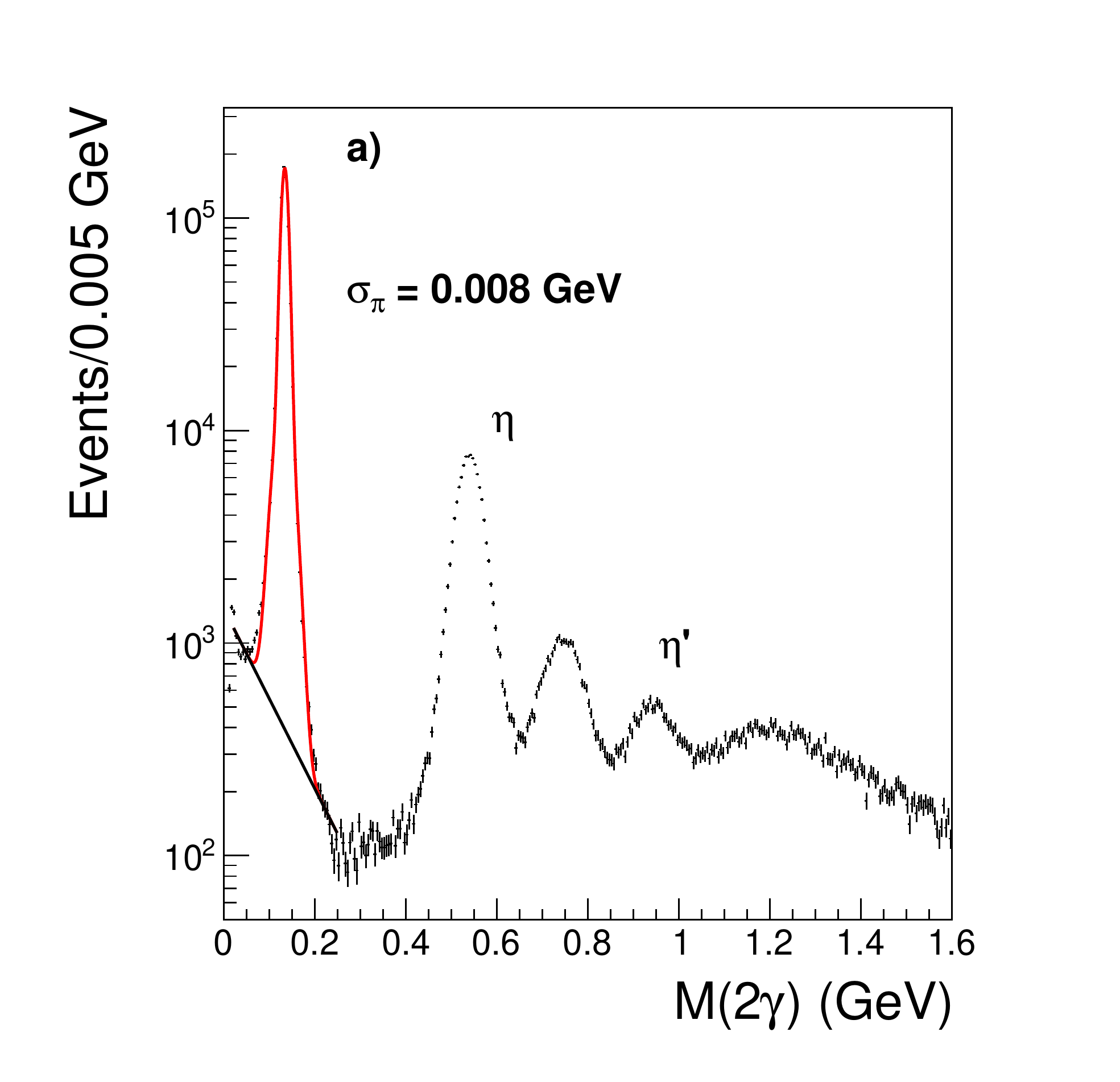}
 \includegraphics[page=3,width=0.45\textwidth,clip=true]{figures/fit_2gam_bcal.pdf}
 \caption{ \label{fig:fit_2gam_bcal}
 a)  Two-photon mass distribution for an exclusive sample of events $\gamma p \rightarrow p \gamma \gamma$ on a logarithmic scale over the full mass range. The fit to the $\pi^0$ peak and the assumed
 background are shown with solid curves. 
  b) Same distribution on a linear scale over the mass range
 of the $\eta$ and $\eta'$ mesons. The fit to the peaks and the assumed background are shown with solid curves. 
 The peak at about 0.75 GeV is due to mis-reconstructed $\omega$'s decaying to $\pi^0 \gamma$, where one low-energy photon is lost or the two photons from the $\pi^0$ decay merge into a single cluster.
 (Color online)
}
\end{figure}

The invariant mass resolution of narrow particles decaying into two photons is  directly related to the calorimeter energy resolution.   Figure\,\ref{fig:fit_2gam_bcal}
shows a typical experimental spectrum of the invariant mass of the two photons for the reaction $\gamma p \rightarrow p \gamma \gamma$ with peaks at the $\pi^0$, $\eta$ and $\eta'$. 
The event selection required four-momentum balance between the initial and final states and the detection of both photons in the BCAL.
The invariant mass resolution depends on both the position and energy resolution of the detector (see Eq.\,\ref{eq:invMass}). 
The importance of these two terms varies depending on kinematics and the particle mass, but generally the position resolution only affects the $\pi^0$ mass resolution.
A fit to the $\pi^0$ peak with a single Gaussian yields a width of $\sigma_{\pi}$\,=\,0.0088\,GeV, but does not describe the tails of the distribution very well. We use two Gaussians and an exponential background to 
fit the distribution including the tails (Fig.\ref{fig:fit_2gam_bcal}a), which yields a width of the main peak of $\sigma_{\pi}$\,=\,0.0075\,GeV. The $\eta$ and $\eta'$ 
peaks are fitted separately (Fig.\ref{fig:fit_2gam_bcal}b) over a polynomial background, yielding Gaussian widths of $\sigma_{\eta}$\,=\,0.028 GeV and $\sigma_{\eta'}$\,=\,0.030\,GeV, respectively. 

     \begin{figure}[tbh]\centering
\includegraphics[width=0.45\textwidth]{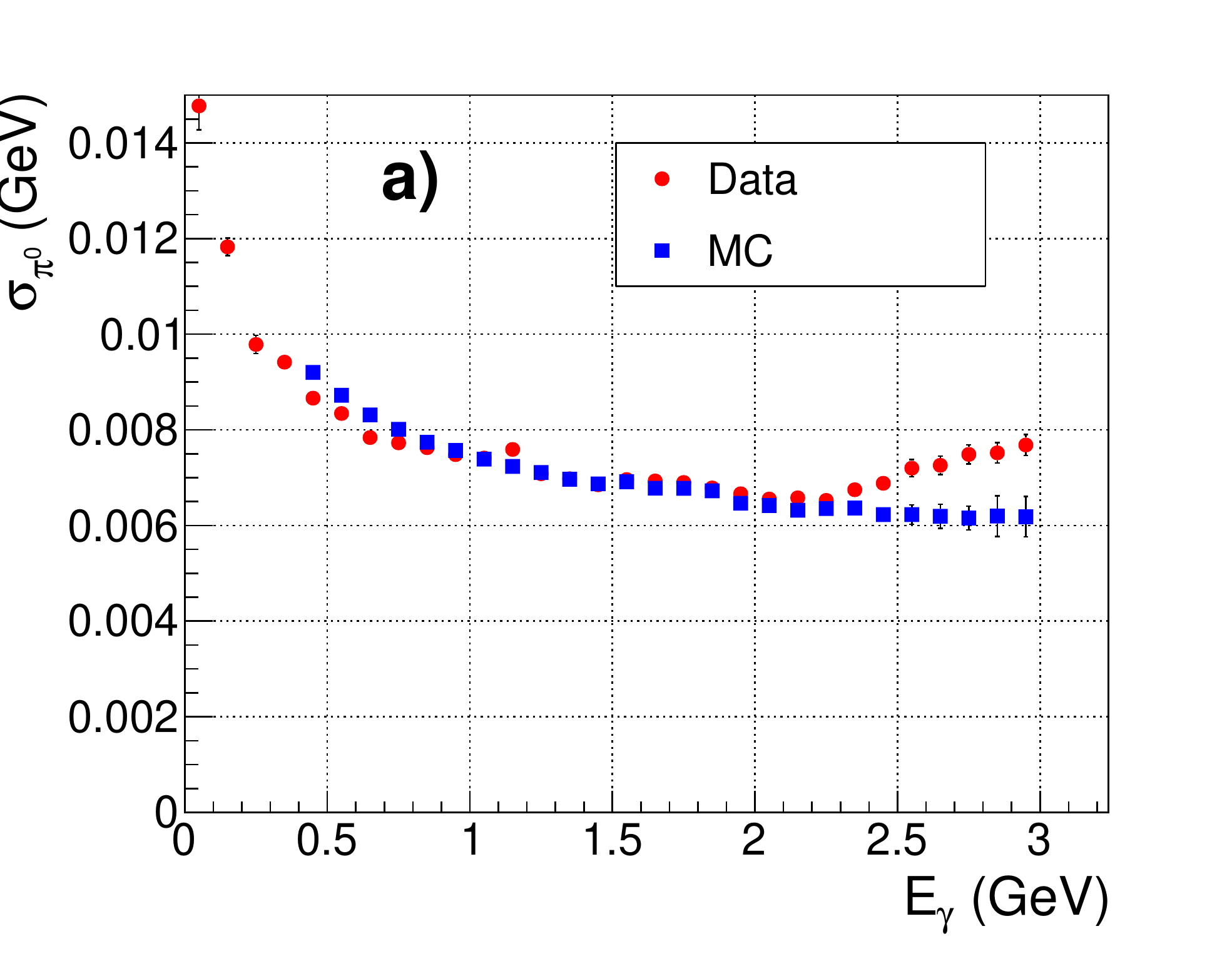}   
\includegraphics[width=0.45\textwidth]{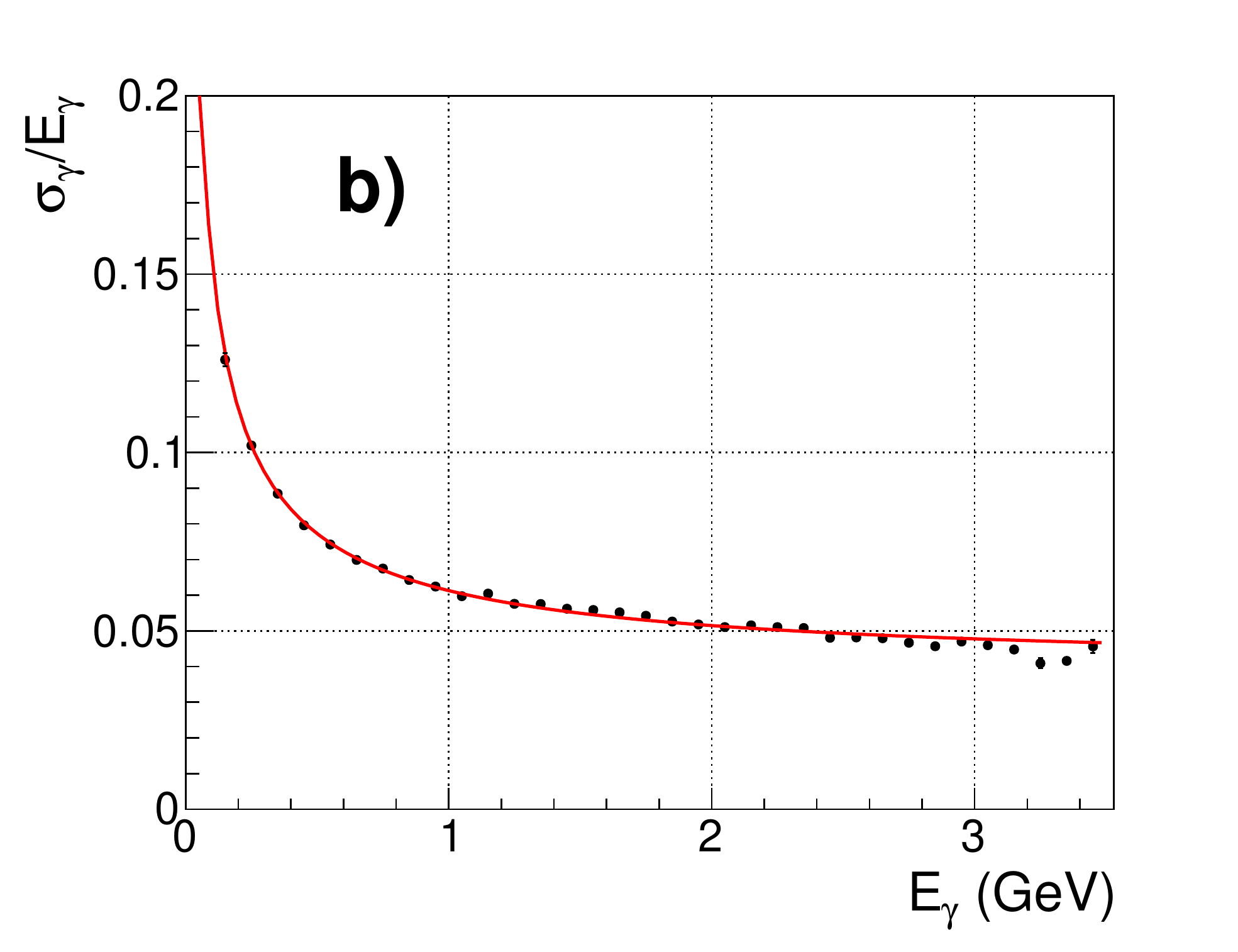}    
\caption{\label{fig:Energy_resolution}
a)  $\pi^{0}$ width as a function of energy for symmetric decays, where both photons are required to be within 0.1~GeV of each other. See text for discussion of the discrepancy at high energy. 
b) Fractional energy resolution used in the simulation 
to reproduce the $\pi^0$ width distributions in a).  The MC data points are fit to Eq.\,\ref{eq:resolutionfit} yielding $a$\,=\,4.70$\pm$0.03\%$\sqrt{\rm{GeV}}$ and $b$\,=\,3.93$\pm$0.03\%, which is shown as a
continuous curve.
(Color online)
 }
\end{figure}

\subsection{Energy Resolution \label{sec:energyres}}
The single-photon energy resolution is obtained from Monte Carlo (MC) because its direct determination using data-driven techniques results in larger uncertainties.
We use a subset of events selected for the gain calibration (Section\,\ref{sec:gaincalibration}) where the energies of the two photons are within 0.1\,GeV of each other.
The energy dependence of the $\pi^0 \rightarrow \gamma\gamma$ spectrum is shown in Fig.\,\ref{fig:Energy_resolution}a for data and Monte Carlo.
We remind the reader that the effective response of the BCAL matrix is modeled in our \textsc{hdgeant} MC (see Section\,\ref{sec:finegrainedMC}) assuming a uniform medium for the matrix. 
The energy flow and all geometrical effects including energy leakage are taken into account. However, the resolution is dominated by sampling fluctuations and these are included by smearing
the energy deposited in each cell using  Eq.\,\ref{eq:resolutionfit} with empirically determined parameters $a$\,=\,5.0$\sqrt{\rm{GeV}}$\,\% and $b$\,=\,3.7\% . 
The agreement between simulation and data is quite good below 2.5 GeV, which gives confidence in the simulated energy dependence of detector response. For higher energies, the uncertainty in position resolution
becomes important and points to the limitations of the current MC implementation. The single-photon energy resolution in the simulation is 
plotted as a function of energy in  Fig.\,\ref{fig:Energy_resolution}b by generating photons with the angular distribution of photons in the $\pi^0$ sample. 

\begin{figure}[tbh]\centering
\includegraphics[width=0.45\textwidth]{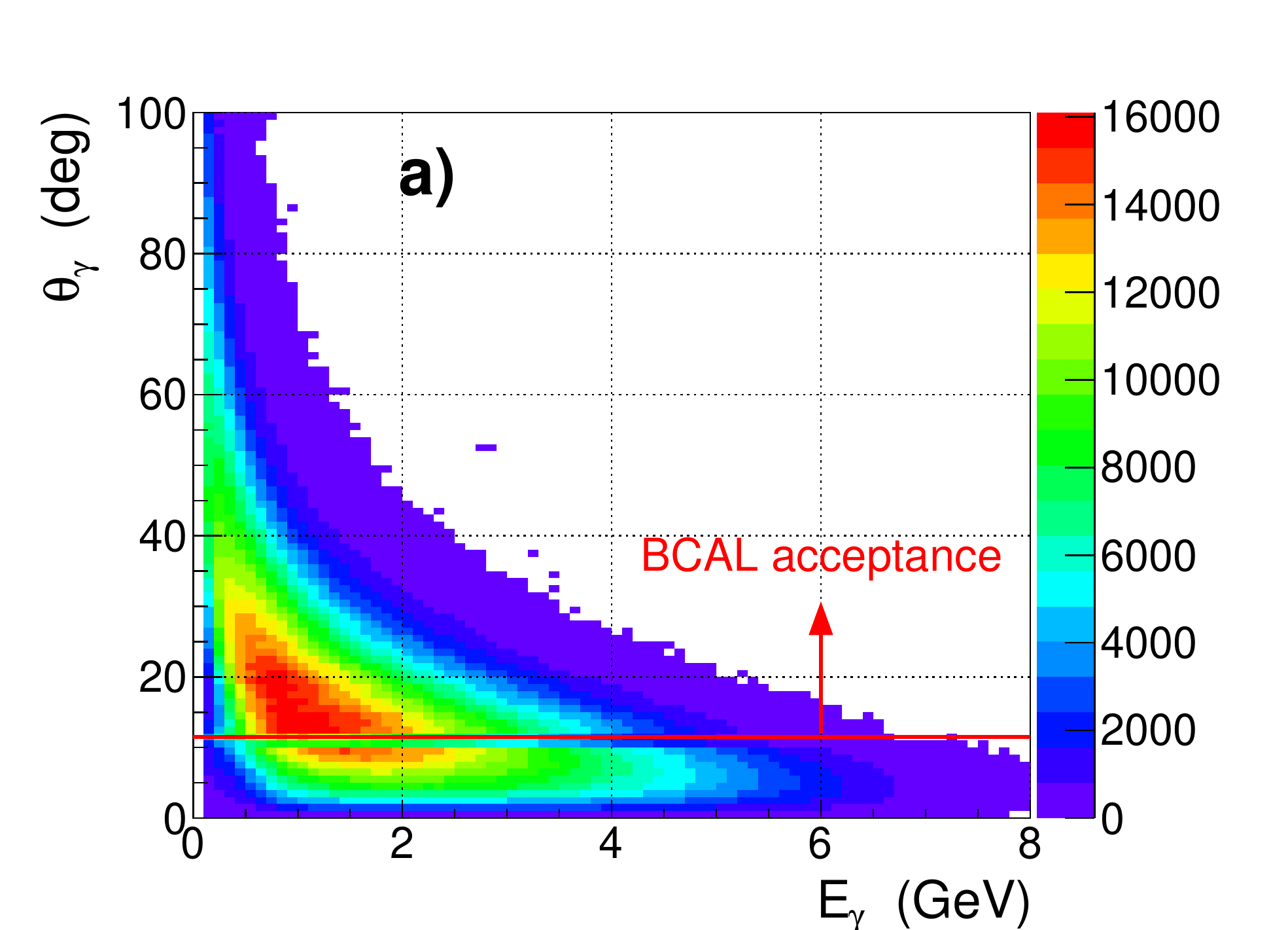}
\includegraphics[width=0.45\textwidth]{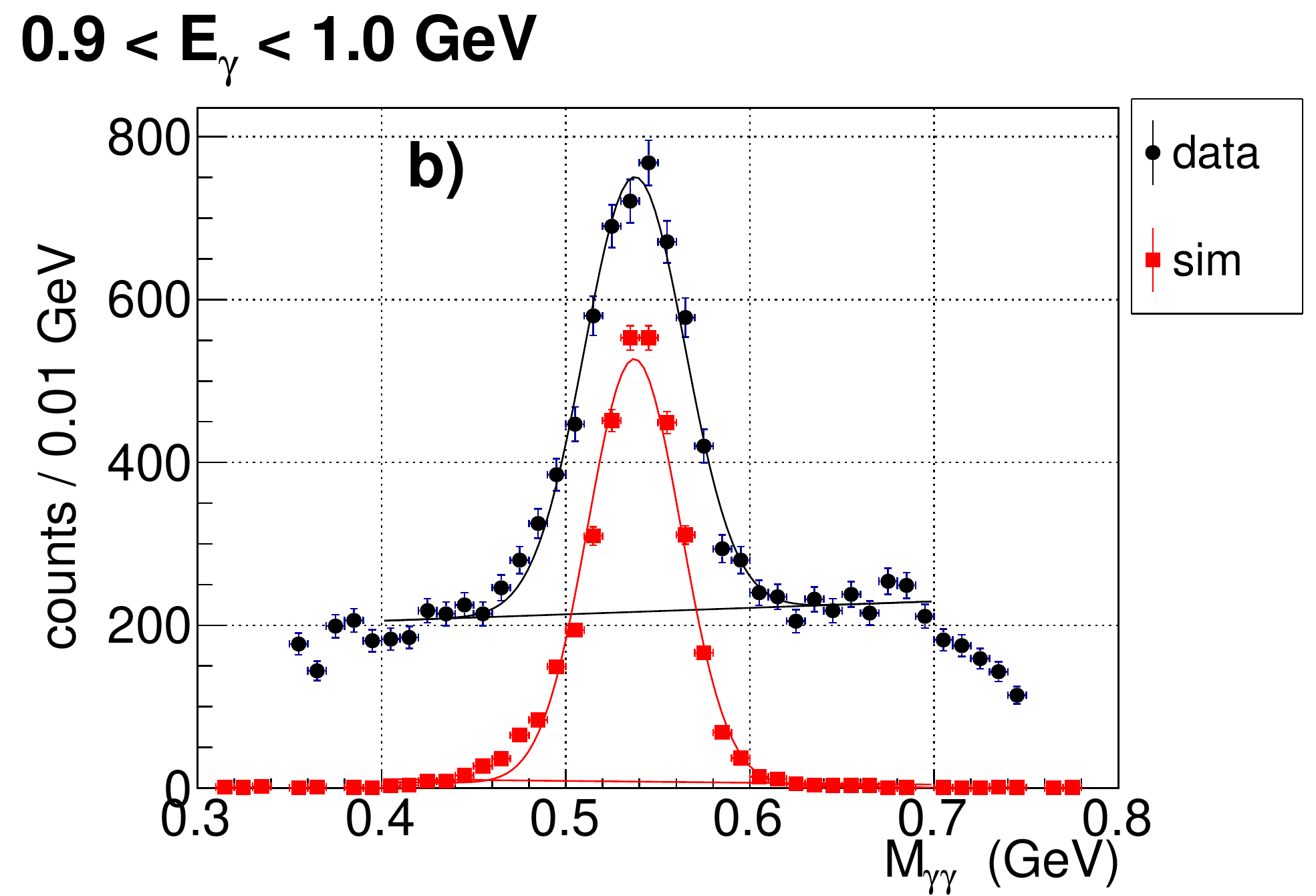}
\caption{\label{fig:NIM_gaus_fit_1GeV}
a) Angle vs. energy distribution for the $\eta$ sample. The BCAL covers the kinematic region above the line. The gap between the BCAL and FCAL is visible.
b) Example of the two-photon invariant mass integrated over angle in the $\eta$ mass range for simulation and data. Only signal is generated in the simulation. The curves show the fitted peak and the assumed background shape.
(Color online)
 }
\end{figure}

\begin{figure}[tbh]\centering
\includegraphics[width=0.45\textwidth]{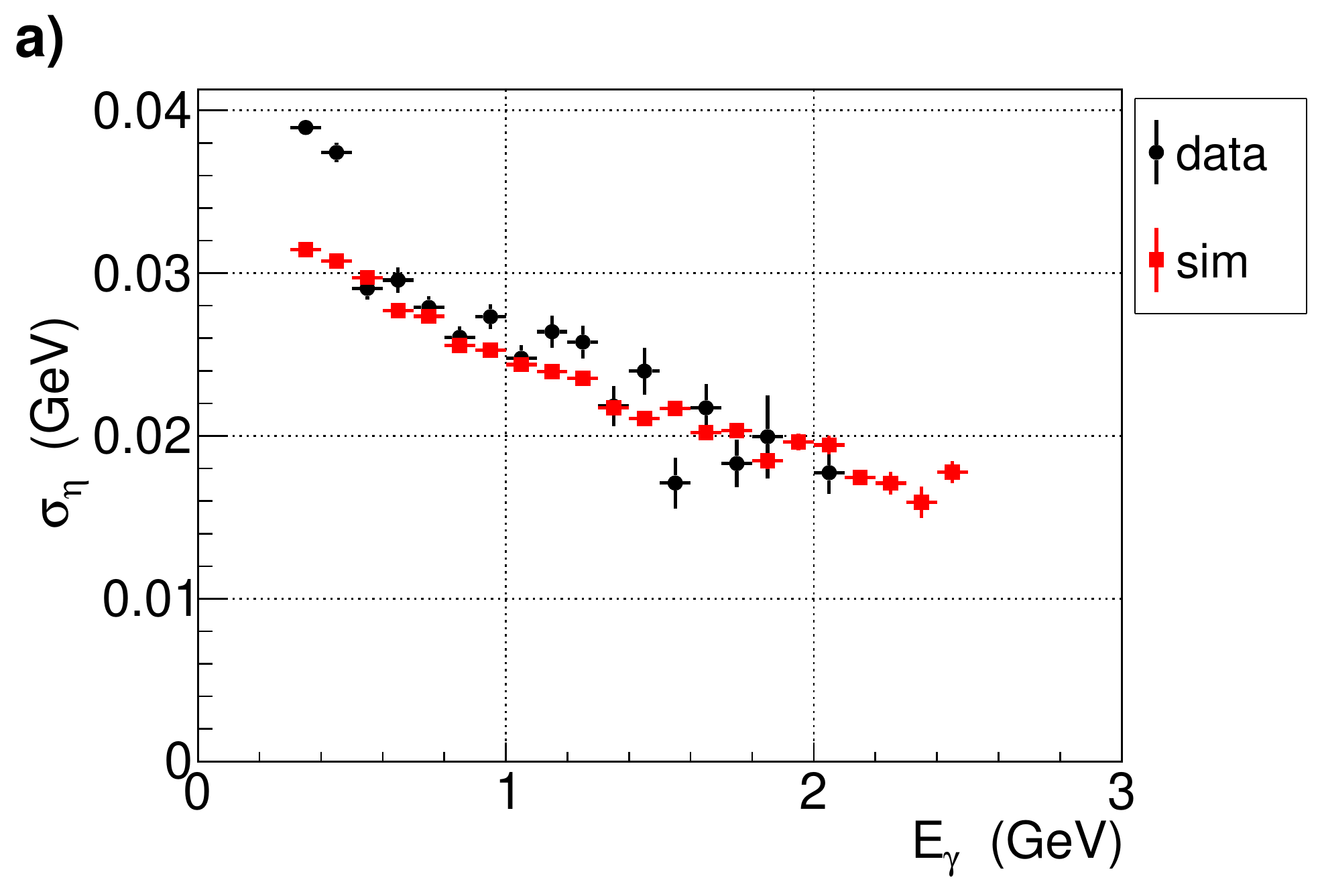}   
\includegraphics[width=0.45\textwidth]{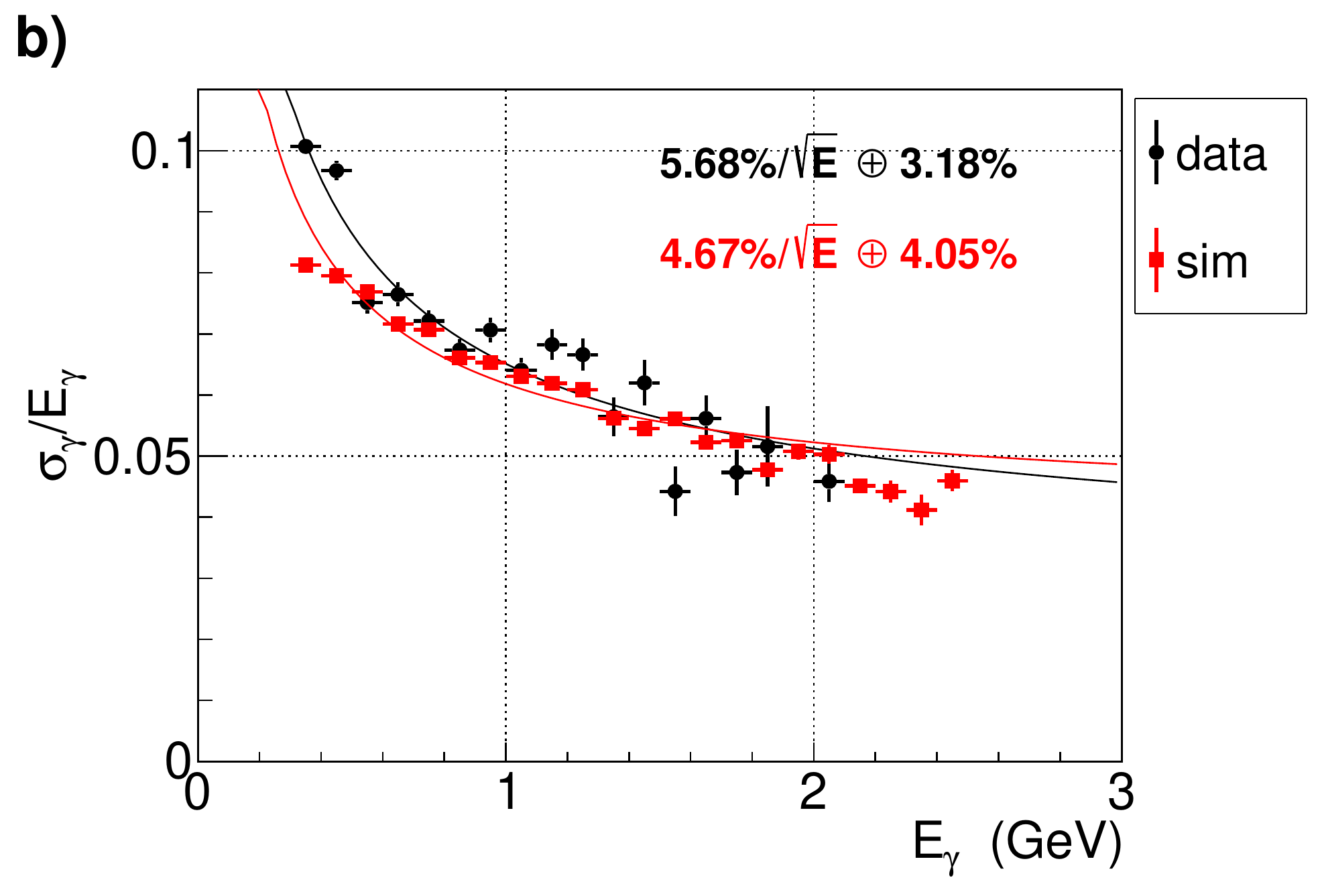}
\caption{\label{fig:eta_resolution}
a)  Measured and simulated  $\eta$ width as a function of energy for symmetric decays, where both photons are required to be within 0.1 GeV of each other. 
b) The energy resolution of single photons calculated under the assumption that only the energy resolution contributes to the $\eta$ width. The curves are fits based on  Eq.\,\ref{eq:resolutionfit}.
(Color online)
 }
\end{figure}

We also have a sufficient number of $\eta$'s to study the $\eta$ mass resolution as a function of energy.  Events were
selected using kinematic fits to $\gamma p \rightarrow p \pi^+ \pi^- \gamma \gamma$. 
This reaction provides a fairly clean sample of $\eta$'s where both photons are recorded in the BCAL.
The charged tracks were used to determine the event vertex needed to reconstruct the two-photon invariant mass.  The MC simulation 
generated $\gamma p \rightarrow p \pi^+ \pi^- \eta$ events, with $\eta\rightarrow \gamma\gamma$, where the kinematics were chosen to approximate the experimental distributions 
(Fig.\,\ref{fig:NIM_gaus_fit_1GeV}a). We further selected a restricted sample of events, where the energies of the two photons fell within 0.1\,GeV of each other. 
The two-photon mass distributions show a prominent peak at the position of the $\eta$ mass and the distributions were fit to a Gaussian with a linear background. An example of
the distribution is shown in Fig.\,\ref{fig:NIM_gaus_fit_1GeV}b for both MC and data. 
The fitted Gaussian widths are shown in Fig.\,\ref{fig:eta_resolution}a as a function of the photon energy, both for data and simulation. 
The single-photon energy resolution can be determined from the $\eta$ width by neglecting contributions from the opening angle.
This is shown in Fig.\,\ref{fig:eta_resolution}b, where one can see that the resolution extracted from the
data is only slightly above the MC.

\section{Summary and Discussion}
\label{sec:summary}  
We have designed, fabricated, assembled and installed the barrel calorimeter in the GlueX experiment in Hall D at Jefferson Lab. The BCAL was operated during the first two running periods of
the GlueX experiment and the calibration for these data has been completed. The BCAL is a scintillating fiber and lead calorimeter with 
a sampling fraction of about 9.45\%. The calorimeter records the time and energy deposited by charged and neutral particles created by a 9\,GeV photon beam. Particles impinge on the detector
over a wide range of angles, from normal incidence at 90 degrees down to 11.5 degrees, which define a geometry that is fairly unique among calorimeters. 
The scintillation light is collected at two ends of the barrel and detected using
3840 silicon photomultiplier arrays, the first large-scale application of this technology. The response of the calorimeter has been measured during a running experiment 
using samples of $\pi^0$ and $\eta$ decays and is shown to perform as expected for electromagnetic showers up to 2.5 GeV. 
The timing resolution of the BCAL has been measured to be about 150\,ps for 1\,GeV electromagnetic showers and 200\,ps for pions greater than 1\,GeV.

\begin{figure}[btp]\centering
\includegraphics[width=0.7\textwidth]{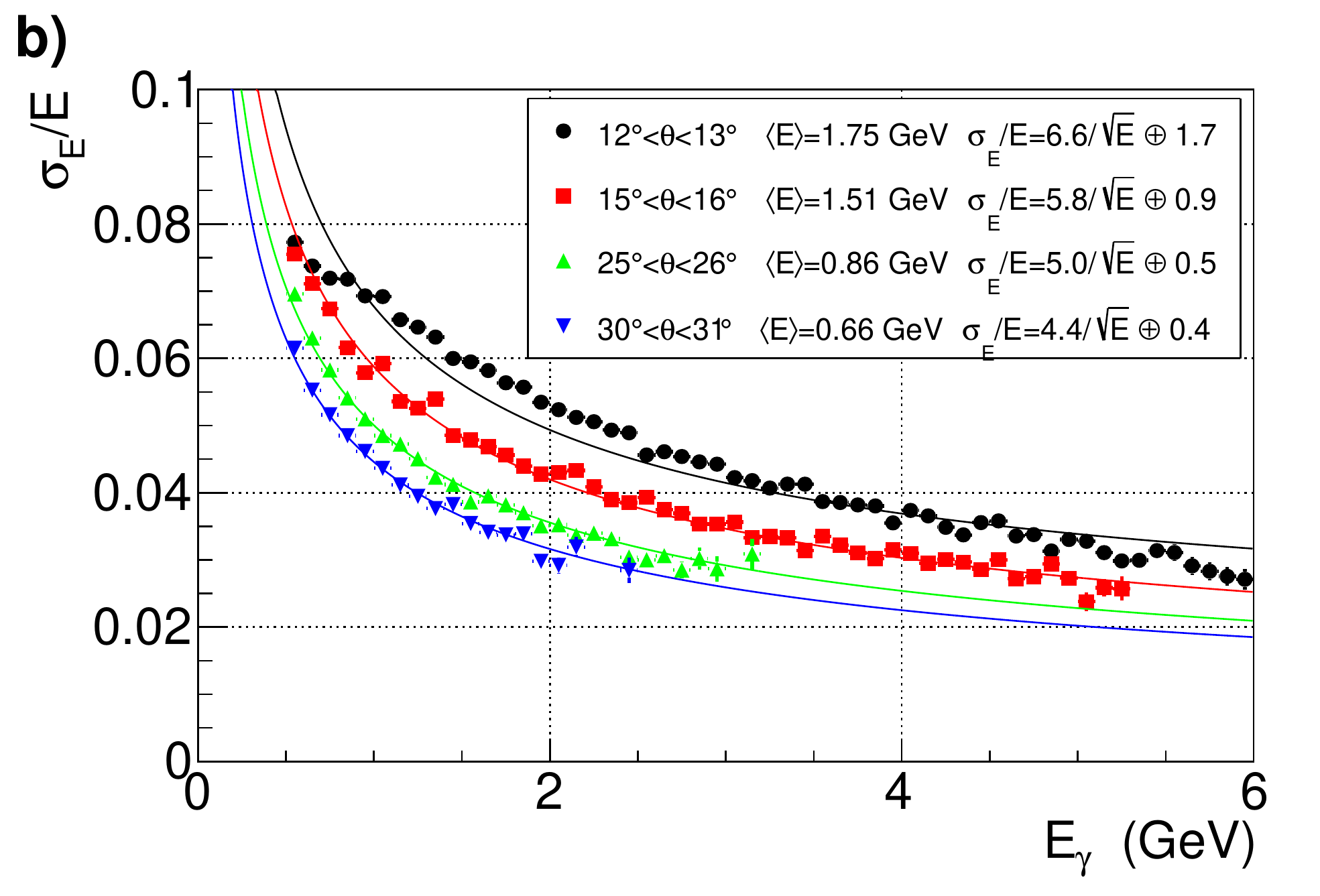}   
\caption{\label{fig:Z_resolution}
Simulated fractional energy resolution for data at different angles from the target for the distribution in Fig.\,\ref{fig:NIM_gaus_fit_1GeV}a.  
At shallow angles the stochastic term increases and the energy leakage generates a non-zero constant term.
The MC data points are fitted to Eq.\,\ref{eq:resolutionfit} with the results shown in the legend.
(Color online)  }
\end{figure}

The energy response of the BCAL to photons depends on their angle of incidence and therefore on the impact position along the 
detector. The energy resolution is plotted in Fig.\,\ref{fig:Z_resolution} for the simulated sample of $\eta$'s  in four angular bins, each fitted to Eq.\,\ref{eq:resolutionfit}. Due to the changing kinematics
from bin to bin, the highest energy photons correspond to shallow impact angles located at the downstream end of the barrel. 
Both the stochastic and the constant terms increase at small angles, reflecting the geometrical dependence of the energy containment. 
We note that the response
of the calorimeter averaged over its length, as done for the $\eta$ sample in Fig.\,\ref{fig:eta_resolution}, is not described well with Eq.\,\ref{eq:resolutionfit} 
and has a large correlation between the two parameters (-0.89). 
Nevertheless, in order to characterize the performance of the BCAL between 0.5 and 2.5 GeV, we take the fitted energy-resolution parameters integrated over the angular distributions for $\pi^0$ and $\eta$ production to obtain a typical energy resolution  for our detector of $\sigma_E/E$=5.2\%/$\sqrt{E(\rm{GeV})} \oplus$ 3.6\%.  
In order to estimate the resolution at high energy, we use the MC that describes  our data at low energy (Fig.\,\ref{fig:Z_resolution}) and results in a constant term of less than 1.7\%.
However, to verify this expectation, we would need additional data reaching to higher energy.


\section{Acknowledgments}

The design, fabrication, installation, calibration and operation of the barrel calorimeter was conducted within the GlueX collaboration, which provided essential feedback and guidance for the project.
We would like to acknowledge the outstanding efforts of the Hall D technical staff for exceptional work in the staging and installation of the detector.
This work was supported by the U.S. Department of Energy, Office of Nuclear Physics Division, under Contract No. DE-AC05-06OR23177, under which Jefferson Science Associates operate Jefferson Laboratory, by
the Natural Sciences and Engineering Research Council of Canada  grant SAPPJ-2015-00024, by
the Department of Energy  grants DE-FG02-05ER41374 and DE-FG02-87ER40315, by
the Chilean Comis\'on Nacional de Investigaci\'on Cient\'ifica y Tecnol\'ogica, and by 
the BASAL grant FB-0821 which supports the Centro Cient\'ifico Tecnol\'ogico de Valpara\'iso.

\newpage

\section*{References}

\bibliography{BCAL_nim}
\bibliographystyle{unsrt}

\clearpage

\end{document}